\documentclass[aps,prx,twocolumn,superscriptaddress,floatfix]{revtex4-2}

\usepackage{color}
\usepackage{soul}
\usepackage{dcolumn}
\usepackage{graphicx}
\usepackage{amsmath}
\usepackage{amssymb}
\usepackage{bm}
\usepackage{bbm}
\usepackage{tabularx}
\usepackage[hidelinks]{hyperref}
\usepackage{multirow}
\usepackage{adjustbox}

\newcommand{\showfontsize}{\f@size{} pt}

\newcommand{\q}[1]{|#1\rangle}
\newcommand{\qd}[1]{\langle#1|}

\begin{document}

\preprint{APS/123-QED}

\title{Multiplexed photon number measurement}

\author{Antoine Essig}
\thanks{Equal contribution author}
\affiliation{Univ Lyon, ENS de Lyon, CNRS, Laboratoire de Physique,F-69342 Lyon, France}

\author{Quentin Ficheux}
\thanks{Equal contribution author}
\affiliation{Univ Lyon, ENS de Lyon, CNRS, Laboratoire de Physique,F-69342 Lyon, France}

\author{Th\'eau Peronnin}
\affiliation{Univ Lyon, ENS de Lyon, CNRS, Laboratoire de Physique,F-69342 Lyon, France}

\author{Nathana\"el Cottet}
\affiliation{Univ Lyon, ENS de Lyon, CNRS, Laboratoire de Physique,F-69342 Lyon, France}

\author{Rapha\"el Lescanne}
\affiliation{Laboratoire de Physique de l'Ecole Normale Sup\'erieure, ENS, Universit\'e PSL, CNRS, Sorbonne Universit\'e, Universit\'e Paris-Diderot, Sorbonne Paris Cit\'e, Paris, France}
\affiliation{QUANTIC team, INRIA de Paris, 2 Rue Simone Iff, 75012 Paris, France}

\author{Alain Sarlette} 
\affiliation{Laboratoire de Physique de l'Ecole Normale Sup\'erieure, ENS, Universit\'e PSL, CNRS, Sorbonne Universit\'e, Universit\'e Paris-Diderot, Sorbonne Paris Cit\'e, Paris, France}
\affiliation{QUANTIC team, INRIA de Paris, 2 Rue Simone Iff, 75012 Paris, France}
\affiliation{Department of Electronics and Information Systems, Ghent University, Belgium}

\author{Pierre Rouchon}
\affiliation{Centre Automatique et Syst\`emes, Mines-ParisTech, PSL Research University, 60, bd Saint-Michel, 75006 Paris, France}
\affiliation{QUANTIC team, INRIA de Paris, 2 Rue Simone Iff, 75012 Paris, France}

\author{Zaki Leghtas}
\affiliation{Centre Automatique et Syst\`emes, Mines-ParisTech, PSL Research University, 60, bd Saint-Michel, 75006 Paris, France}
\affiliation{Laboratoire de Physique de l'Ecole Normale Sup\'erieure, ENS, Universit\'e PSL, CNRS, Sorbonne Universit\'e, Universit\'e Paris-Diderot, Sorbonne Paris Cit\'e, Paris, France}
\affiliation{QUANTIC team, INRIA de Paris, 2 Rue Simone Iff, 75012 Paris, France}

\author{Benjamin Huard}
\affiliation{Univ Lyon, ENS de Lyon, CNRS, Laboratoire de Physique,F-69342 Lyon, France}

\date{\today}

\begin{abstract}

When a two-level system -- a qubit -- is used as a probe of a larger system, it naturally leads to answering a single yes-no question about the system state.
Here we propose a method where a single qubit is 
able to extract, not a single, but many bits of information about the photon number of a microwave resonator using continuous measurement.
We realize a proof-of-principle experiment by recording the fluorescence emitted by a superconducting qubit reflecting a frequency comb, thus implementing multiplexed photon counting  where the information about each Fock state -- from 0 to 8 -- is simultaneously encoded in independent measurement channels.
Direct Wigner tomography of the quantum state of the resonator evidences the back-action of the measurement as well as the optimal information extraction parameters.
Our experiment
unleashes the full potential of quantum meters by replacing a
sequential
quantum measurements
with simultaneous and continuous measurements separated in the frequency domain.



\end{abstract}
\maketitle

\section{Introduction}
\label{sec:intro}

\begin{figure*}
\includegraphics[width=\textwidth]{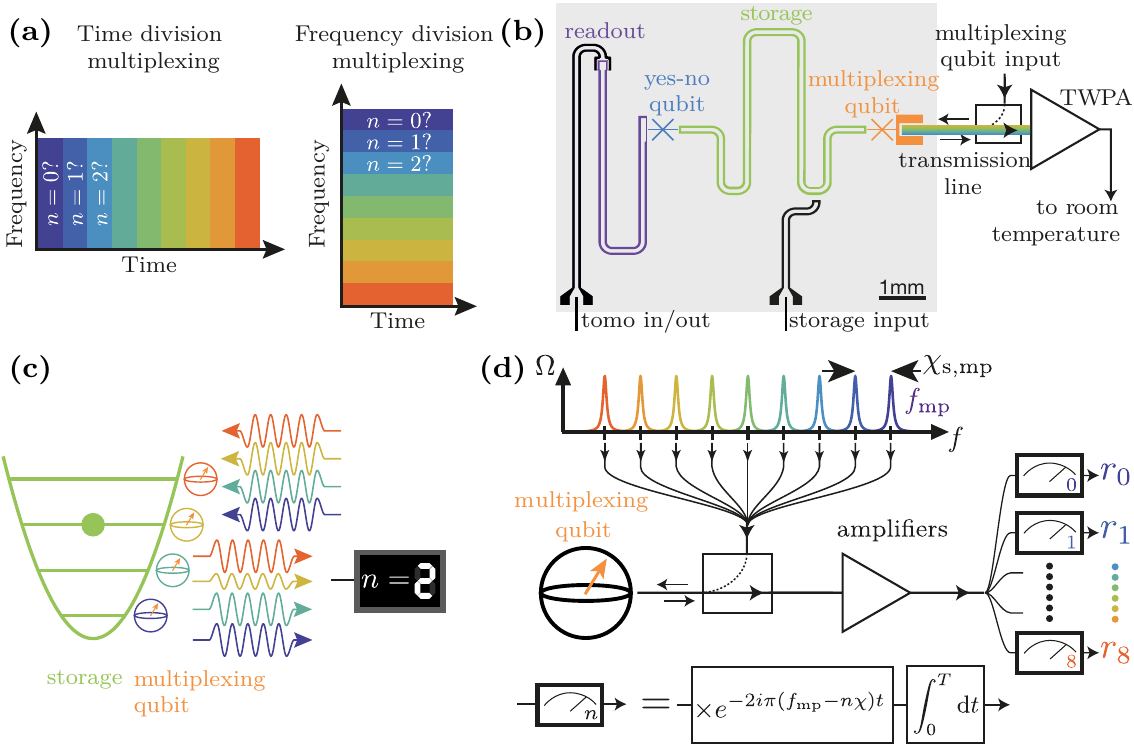}
\caption{\textbf{Multiplexed photon counting setup and protocol.} \textbf{a,} Time domain division multiplexing proceeds one binary question at a time. Frequency division multiplexing simultaneously retrieves multiple binary answers. \textbf{b,} Scheme of the device in coplanar waveguide architecture. The storage mode (green) is coupled to a transmon multiplexing qubit (orange), which is directly coupled to a transmission line (rainbow). A directional coupler and broadband Traveling Wave Parametric Amplifier (TWPA) allow us to probe the qubit in reflection. An additional transmon yes-no qubit (blue) and its readout resonator (purple) are used as a reference photon counter and for Wigner tomography (see Appendix \ref{sec:wigner_cal}). \textbf{c,} The frequency (color) of the multiplexing qubit encodes the storage photon number. The reduction in the reflection amplitude of the qubit at one of the frequencies reveals the number of photons in the storage mode -- e.g. here 2 photons. \textbf{d,} The qubit is 
probed by a frequency comb of amplitude $\Omega$. The reflected pulse is amplified and digitized before numerical demodulation at every frequency $f_\mathrm{mp}- k \chi_\mathrm{s,mp}$. This multiplexing-demultiplexing process leads to reflection coefficients $r_k$ that each encodes the probability that $k$ photons are stored.}
\label{fig1}
\end{figure*}

The most general measurement of a quantum system consists in using a quantum apparatus as a probe. The system interacts with the probe before the latter gets measured projectively. In the simplest case, the probe is a qubit whose readout answers a yes-no question about the system state. Identifying what is the state of a system thus comes down to playing a game of ``Guess Who?". A series of binary questions are asked iteratively to refine our knowledge about the state. Unlike the classical game, each answer disturbs the state of the system. To give a concrete example, in order to determine how many photons are stored in a cavity, one may ask ``is there an even number of photons?" or a series of binary questions such as ``are there $n$ photons?" for each integer $n$ (Fig.~\ref{fig1}a). 

Such experiments have been implemented with Rydberg atoms or superconducting circuits probing a microwave cavity~\cite{Guerlin2007,Johnson2010} with the possible refinement of choosing what binary question should be optimally asked conditioned on the previous answers~\cite{Peaudecerf2014}--
using a feedback loop~\cite{Haroche1992,Dassonneville2020} or advanced pulse shaping~\cite{Curtis2020,Wang2020}.
Determining an arbitrary number of photons in the cavity between $0$ and $2^m-1$ thus takes at least $m$ consecutive probe measurements since each answer provides at most one bit of information about the system state.
This limitation originates from the encoding of the extracted information into the quantum state of the qubit. But is it the best use of a qubit to determine an observable with many possible outcomes such as a photon number?

We propose to use a qubit as an \emph{encoder} of information about the cavity state into the many propagating modes of a transmission line. Assuming an ideal detector, we show that photon counting can then be implemented in a time independent of the number of photons. 
We demonstrate the practicality of this approach in an experiment where information about 9 possible photon numbers (more than 3 bits) in a microwave resonator is simultaneously extracted by a single superconducting qubit into 9 propagating modes of a transmission line.
Owing to dispersive interaction~\cite{Schuster2007,Gely2019}, each photon number corresponds to a unique qubit frequency. When driving the qubit at 9 test frequencies by multiplexing, the qubit simultaneously emits 9 microwave signals
that each reveals information about the photon number ranging from 0 to 8. 
Daring an analogy with communication protocols~\cite{ProakisSaheli2002}, previous measurement schemes with time series of binary questions used time division multiplexing while our experiment demonstrates the analogous of frequency division multiplexing, where the qubit alone acts as the frequency multiplexing transducer (Fig.~\ref{fig1}a).
This experiment directly benefits from the recent bandwidth improvements of near quantum limited amplifiers~\cite{Macklin2015}, which enabled us to bring the measurement process in the frequency domain. Current limitations in the cavity lifetime and detector efficiency prevented us from reaching single shot readout of the photon number in this proof-of-principle experiment, and hence from decoding 3 bits of information per experiment. However, unlike in sequential measurement schemes, a single run of our experiment does provide, in parallel, partial information about each bit of the photon number.
Besides we manage to observe the multiplexed measurement back-action on the resonator using direct Wigner tomography, which allowed us to measure the decoherence rate of the resonator induced by the measurement. We evidence an optimal qubit drive amplitude for information extraction, which matches the expected dynamics of a qubit under a multi-frequency drive.

The article is organized as follows, Sec. II 
demonstrates standard photon counting using a qubit dispersively coupled to a microwave resonator, and reviews the practical limitations of this technique. It provides a calibration of the photon number for the rest of the article.
Sec. III describes photon counting using the fluorescence of a second qubit. Sec. IV presents a gedanken experiment that shows how multiplexed photon number measurement can reach an outcome in a time that does not depend on the number of photons. Sec. V presents an actual experiment that implements a version of the gedanken experiment where the photo detectors are replaced by heterodyne detectors in each frequency band. In practice, it comes down to experimentally probing the qubit in fluorescence with a frequency comb, thus revealing how information about cavity photon numbers
is routed by the qubit
onto the transmission line modes.
Sec. VI evidences the measurement back-action of the multiplexed photon counting using direct Wigner tomography of the quantum state
and shows how to maximize information extraction. A detailed comparison between the efficiency of our approach compared to other measurement schemes is given in Appendix E. 

\section{Standard photocounting}
\label{sec:Stand_photoncounting}

The experiment implements both the standard approach and the multiplexed one in order to count the number of photons in a resonator dubbed the \emph{storage mode}, which resonates at $f_\mathrm{s}=4.558~\mathrm{GHz}$. Two off-resonant transmon qubits are coupled to the storage mode (see Fig.~\ref{fig1}b). The \emph{yes-no qubit} with a frequency $f_\mathrm{yn}=3.848~\mathrm{GHz}$ is used to ask 
standard binary questions about the photon number
or to perform storage mode tomography; while the \emph{multiplexing qubit} with a frequency $f_\mathrm{mp}=4.238~\mathrm{GHz}$
is used for fluorescence photon counting (Sec. III) and frequency multiplexed photon counting (Secs. V-VI). Both qubits are dispersively coupled to the resonator so that their frequency respectively redshifts by $\chi_\mathrm{s,yn}=1.4~\mathrm{MHz}$ and $\chi_\mathrm{s,mp}=4.9~\mathrm{MHz}$ per additional photon in the storage mode.

\begin{figure}
\includegraphics[width=\columnwidth]{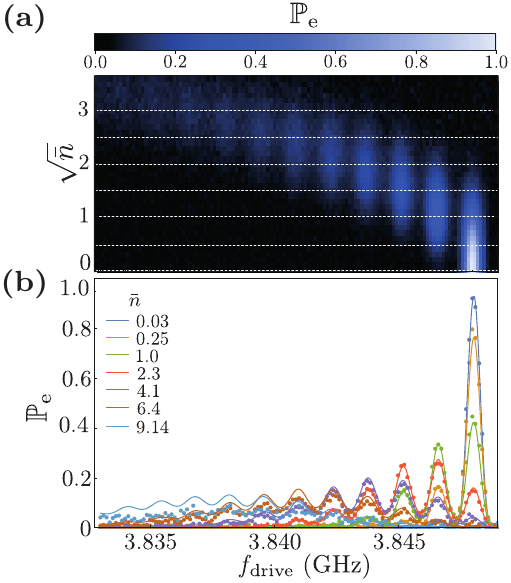}
\caption{\textbf{
Standard photon counting.} The storage mode is prepared in a coherent state with
an average photon number $\bar{n}$
using a microwave pulse at storage frequency and amplitude $V_s$. \textbf{a, b,} 
Measured probability $\mathbb{P}_\mathrm{e}$ that the yes-no qubit gets excited by a $\pi$-pulse at a frequency $f_\mathrm{drive}$.
Peaks appear at $f_\mathrm{yn}-k\chi_\mathrm{s,yn}$ and indicate the probability to store $k$ photons. The dots in \textbf{b} are cuts along the dashed lines in \textbf{a} and match the master equation model (solid lines), hence providing a calibration of $\bar{n}$ as a function of the drive amplitude $V_s$.
}
\label{fig2A}
\end{figure}

In the standard approach~\cite{Schuster2007,Johnson2010}, which probes whether there are $k$ photons, the probability to have $k$ photons is encoded as the probability $\mathbb{P}_\mathrm{e}$ to excite the yes-no qubit by driving it with a $\pi$-pulse at $f_\mathrm{drive}=f_\mathrm{yn}-k\chi_\mathrm{s,yn}$. The state of the yes-no qubit is read out using a dedicated resonator (Fig.~\ref{fig1}b). 
To demonstrate this photon counting ability, we use a microwave tone at $f_\mathrm{s}$ to prepare the storage mode in a coherent state $\left| \beta \right> = e^{- |\beta|^2/2} \sum_{n=0}^{+ \infty} \frac{\beta^n}{\sqrt{n!}} \left| n \right>$, which is a superposition of all Fock states with mean photon number $\bar{n}=|\beta|^2$. The probability $\mathbb{P}_\mathrm{e}$ is then measured and shows resolved peaks as a function of $f_\mathrm{drive}$ for every photon number up to about 7 (Fig.~\ref{fig2A}a,b). For the rest of the paper, we use this measurement as a calibration of the photon number in the storage mode. The linear relation between $\beta$ and the amplitude $V_\mathrm{s}$ of the tone at $f_\mathrm{s}$ is extracted using a master-equation based model (see Appendix Sec.~\ref{sec:photon_number}) reproducing the measured $\mathbb{P}_\mathrm{e}$ (solid lines in Fig.~\ref{fig2A}b).

With this approach the choice of binary question asked to the resonator is set by the frequency of the drive used to perform the conditional $\pi$-pulse on the qubit.
Finding the photon number $N$, by asking successively whether the resonator is in state $|k\rangle$ for $k = 0 \cdots N$, takes $N+1$ consecutive measurements and is
highly sensitive to measurement errors (see Appendix \ref{sec:comparison}). Each step reveals at most one bit of information about the photon number in the resonator since the qubit state encodes the information. The duration of this procedure can be reduced by adaptative measurement~\cite{Peaudecerf2014}. It reaches a minimal number of measurement steps using binary decimation at the expense of using a feedback loop to adjust the pulse sequence in real time~\cite{Haroche1992,Dassonneville2020} or numerical optimal control techniques such as Gradient Ascent Pulse Engineering (GRAPE)~\cite{Curtis2020,Wang2020}. Therefore, the best strategy using a qubit state as the probe of the photon number requires advanced control techniques and at least of the order of $\log_2(N)$ measurement steps that each requires a time of the order of $1/\chi_\mathrm{s,yn}$ at best (see Appendix \ref{sec:comparison}).


\section{Fluorescence photon counting}
\label{sec:fluorecence}

The intrinsic limitation of the standard approach is that measuring the qubit state can at most reveal one bit of information per step. It is possible to avoid this constraint by observing the qubit frequency 
directly instead of measuring its state. The multiplexing qubit is coupled to the transmission line so that when there are $k$ photons in the storage mode, the qubit emits a fluorescence signal into the mode of the transmission line that is centered around the qubit frequency $f_\mathrm{mp}- k \chi_\mathrm{s, mp}$. 
This encoding ability can be observed by driving the multiplexing qubit with a single microwave drive in reflection through the transmission line (Fig.~\ref{fig1}b)~\cite{Houck2007,Astafiev2010a,PhysRevLett.107.043604,PhysRevLett.112.180402}. The measured real part $\mathcal{R}e(r)$ of the reflection coefficient of a microwave pulse at frequency $f_\mathrm{probe}$ is reduced when the probe resonates with the qubit, hence revealing the photon number $k$ (Fig.~\ref{fig1}c)~\cite{Gely2019}. This reduction arises from the coherent emission by the qubit in phase opposition with the reflected drive~\cite{Cohen-Tannoudji2001en}. Therefore, on average, the distribution of photon numbers in the storage mode can be deduced from the relative amplitudes of the reduction of $\mathcal{R}e(r)$ at each frequency $f_\mathrm{mp}- l \chi_\mathrm{s, mp}$.



\begin{figure}
\includegraphics[width=\columnwidth]{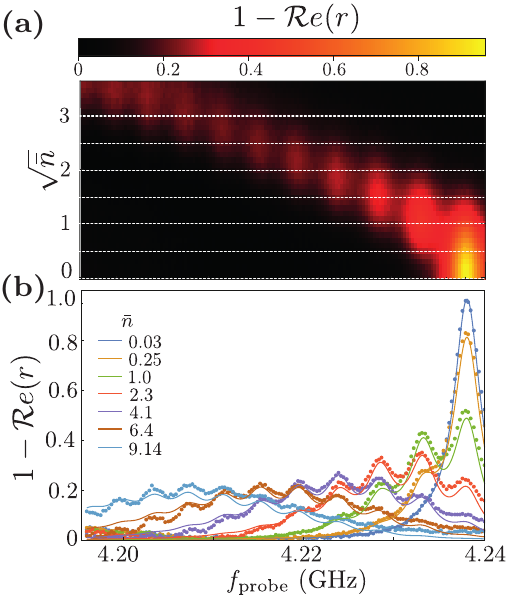}
\caption{\textbf{
Fluorescence photon counting
} \textbf{a, b,}
Measured emission coefficient $1-\mathcal{R}e(r)$ as a function of the probe frequency $f_\mathrm{probe}$ and the mean photon number $\bar{n}$ in the storage mode. The emission coefficient exhibits a resolved peak for each photon number. The dots in \textbf{b} are cuts along the dashed lines in \textbf{a} and are captured by a master equation model (solid lines).
}
\label{fig2B}
\end{figure}

In Figs.~\ref{fig2B} a,b, we show the measured qubit \emph{emission coefficient} $1-\mathcal{R}e(r)$ as a function of a single probe frequency $f_\mathrm{probe}$ and of the initial amplitude of the storage mode coherent state $\sqrt{\bar{n}}$. The measurement is performed using a drive strength $\Omega=\chi_\mathrm{s,mp}/4$ (expressed as the corresponding Rabi frequency) and pulse duration of $2~\mu\mathrm{s}$, which is smaller than the storage lifetime of $3.8~\mu\mathrm{s}$. Resolved peaks develop for every photon number up to at least 9. Using the former calibration of $\bar{n}$, a master-equation based model enables us to reproduce the measurement results (see Appendix \ref{sec:Photon_counting_sim}).

The observation of resolved peaks is due to our choice of parameters.
We designed the relaxation rate of the multiplexing qubit $\Gamma_\mathrm{1,mp}= (42 \mathrm{~ns})^{-1}$ so that the decoherence rate $\Gamma_\mathrm{2,mp}=\Gamma_\mathrm{1,mp}/2$ is smaller than the dispersive shift $2\pi\chi_\mathrm{s, mp}$. When peaks are separated, probing the qubit at one of its resonance frequencies $f_\mathrm{mp}- k \chi_\mathrm{s, mp}$ opens a communication channel with a maximal bandwidth $\Gamma_\mathrm{2,mp}$ carrying information only about Fock state $|k\rangle$. We maximize the bandwidth of each channel by designing $\Gamma_\mathrm{1,mp}$ as large as possible by adjusting the direct coupling to the transmission line, under the constraint of keeping the peaks resolved (see Appendix \ref{sec:parameters}).

Therefore we have shown that both the fluorescence photon counting and the standard photon counting
(Figs.~\ref{fig2A} and \ref{fig2B})
allow us to ask questions of the kind ``are there $k$ photons?".
The important difference between both techniques is that only the fluorescence photon counting can be multiplexed. Indeed, for the standard technique, one needs to read out and reset the qubit at the end of each step. The readout step cannot be multiplexed as it always occurs at the readout mode frequency. In contrast, with the fluorescence readout, information about a given photon number $k$ is constantly extracted through the frequency mode $f_\mathrm{mp} - k \chi_\mathrm{s, mp}$ of the transmission line. It thus enables the key ingredient of our approach: the multiplexing measurement of reflection at every frequency $f_\mathrm{mp} - k \chi_\mathrm{s, mp}$. The qubit thus acts as an encoder of the state of the storage mode into the many modes of the transmission line at frequencies $\{f_\mathrm{mp}- l \chi_\mathrm{s, mp}\}_l$, which can collectively host much more than a single bit of information.





\begin{figure*}
    \centering
    \includegraphics[width=\textwidth]{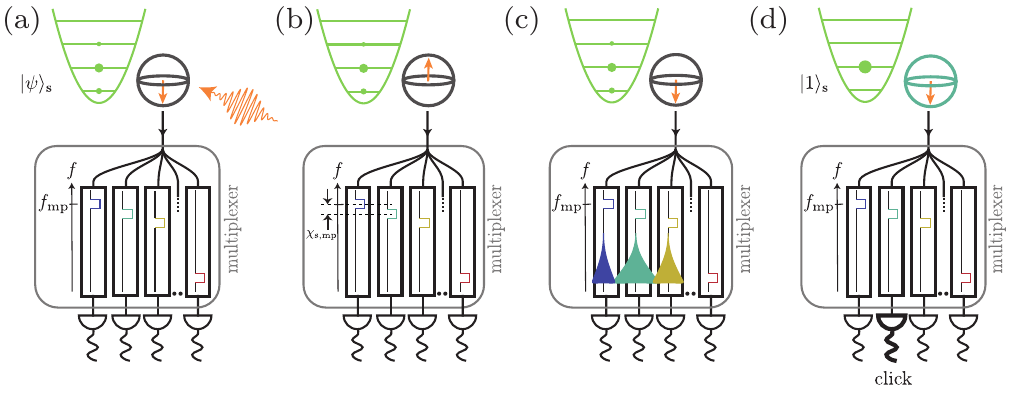}
    \caption{\textbf{Gedanken multiplexing experiment.} 
    a) An unconditional $\pi$ pulse is applied to the multiplexing qubit while the cavity is prepared in state $|\psi\rangle_s$. b) The qubit is prepared in the excited state. c) The qubit spontaneously emits a photon into the transmission line, where a multiplexer sorts the emitted radiation according to its frequency. Each port $k$ of the multiplexer is bandpass filtered around frequency $f_\mathrm{mp}-k \chi_\mathrm{s,mp}$ by a rectangular component displaying the frequency band.
d) Eventually a single photodetector (detector $k = 1$ in the figure) clicks with probability $|\langle k | \psi \rangle|^2$, allowing us to deduce the photon number. The storage mode is projected on the corresponding Fock state (here $|1 \rangle_\mathrm{s}$) in a typical time $T_\mathrm{1,mp}$ that does not depend on the average number of photons in the storage mode.}
    \label{fig:photondetector}
\end{figure*}

\section{Gedanken multiplexed experiment}
\label{sec:constantmeastime}

In this section, we show how, despite using a single qubit as well, multiplexed measurements are able to determine the photon number in a constant time in contrast with the standard approach. We consider an ideal detector for the propagating modes in order to better illustrate the power of multiplexing. The ideal detector is made of a frequency multiplexer followed by a perfect photodetector on each of its outputs (Fig.~\ref{fig:photondetector}). The multiplexer is made of a parallel ensemble of bandpass filters that are each centered on the frequency $f_\mathrm{mp}-k \chi_\mathrm{s,mp}$ with a bandwidth $\chi_\mathrm{s,mp}$. The protocol proceeds in three steps to count the number of photons in the storage mode starting in state $|\psi\rangle_\mathrm{s}$, as detailed in Fig.~\ref{fig:photondetector}. First, the multiplexing qubit is excited with a $\pi$-pulse that is short enough so that it prepares the qubit in the excited state irrespective on the number of photons. Second, the qubit decays in the transmission line converting its excitation into a single photon contained in a propagating wavepacket whose envelope decays at a rate $\Gamma_\mathrm{1,mp}$. In the limit where $\Gamma_\mathrm{1,mp}\ll \chi_\mathrm{s,mp}$, and without pure dephasing of the qubit, the photon emission produces an entangled state between the storage mode and the propagating modes of the line
\[
\sum_k \langle k |\psi\rangle_\mathrm{s} \bigotimes_j|\delta_{k,j}\rangle_j\otimes|k\rangle_\mathrm{s},\]
\noindent where $|\cdot\rangle_j$ represents the quantum state of the propagating mode going through the multiplexer on branch $j$ corresponding to frequencies in the band $[f_\mathrm{mp}-(j+1/2) \chi_\mathrm{s,mp},f_\mathrm{mp}-(j-1/2) \chi_\mathrm{s,mp}[$ (see Fig.~\ref{fig:photondetector}c). Matching the temporal envelope of the modes to the exponential decay at a rate $\Gamma_1$~\cite{Besse2018}, the mode is occupied by either $|0\rangle_j$ or $|1\rangle_j$ depending on the storage photon number, hence the notation $|\delta_{k,j}\rangle_j$. Finally, a single photodetector clicks and reveals the number of photons $k$ with probability $|\langle k|\psi\rangle|^2$ (Fig.~\ref{fig:photondetector}d). In case of ideal detectors with zero false positives, the click detects the associated propagating mode in $|1\rangle_k$, and therefore, as the line is entangled with the storage mode, the measurement backaction projects the storage mode in Fock state $|k\rangle$. The total measurement time, a few $1/\Gamma_\mathrm{1,mp}$, corresponds to the time it takes for one photodetector to click. The time is thus \emph{independent on the number of photons} stored in the storage mode. Note that in order to avoid spectral leakage into other ports, $\Gamma_\mathrm{1,mp}$ is limited by $\chi_\mathrm{s,mp}$ so that the shortest measurement time is limited to a few $1/\chi_\mathrm{s,mp}$.



In contrast to sequential measurements for which increasing the maximal number of photons that can be detected requires additional temporal resources (of the order of $\log_2(N)/\chi_\mathrm{s,mp}$), this gedanken experiment shows that the multiplexed measurement is able to operate in a constant time at the expense of additional spectral resources.




\section{Multiplexed photon counting}
\label{sec:multiplexing}





\begin{figure}
\includegraphics[width=\columnwidth]{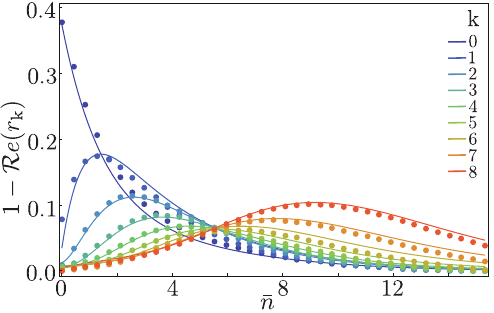}
\caption{\textbf{Multiplexed photon counting.} Dots: simultaneously measured average emission coefficients corresponding to every photon number $k$ from 0 to 8 as a function of the initial mean photon number $\bar{n}$ in the storage mode. $r_k$ is here the reflection coefficient at $f_\mathrm{mp}-k\chi_\mathrm{s,mp}$. Solid lines: prediction based on a master equation without free parameters (see Appendix \ref{sec:sim_multiplexing}).
}
\label{fig2C}
\end{figure}

\begin{figure*}
\centering
\includegraphics[width=0.7\textwidth]{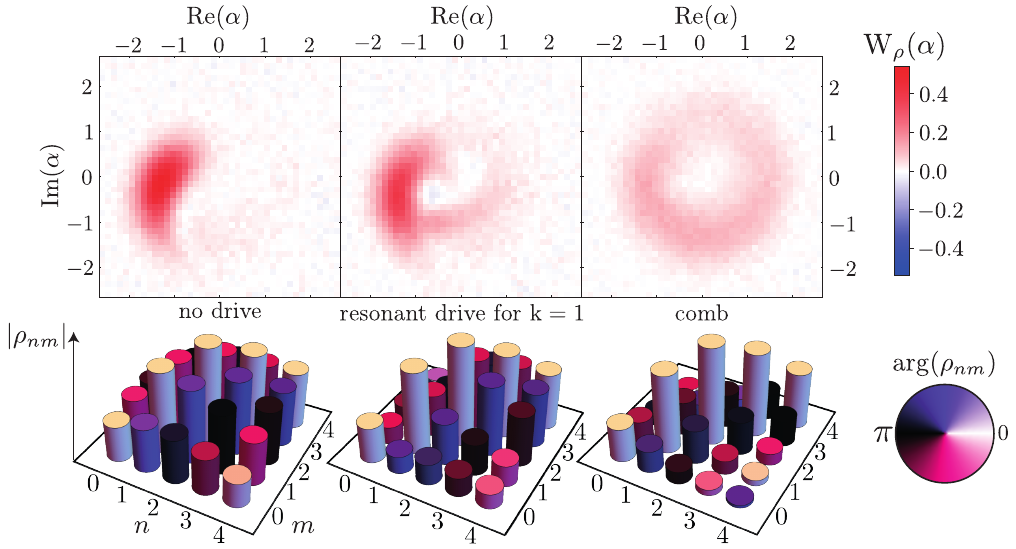}
\caption{\textbf{Measurement back-action.} Direct Wigner tomography of the storage mode at time $t=0.5~\mu\mathrm{s}$ after initialization in a coherent state with amplitude $\beta=-1.55$. Left: free evolution without driving. Middle: evolution with a single tone at $f_\mathrm{mp}- \chi_\mathrm{s,mp}$ that probes whether there is 1 photon in the storage mode. Right: evolution in presence of the frequency comb $\{f_\mathrm{mp} - k \chi_\mathrm{s,mp}\}_{0\leq k\leq 8}$. Each bottom panel represents the density matrix $\{\rho_{nm}\}_{n,m}$ that is calculated from the Wigner function of its top panel. The cylinder heights represent $|\rho_{nm}|$ and the color encodes $\mathrm{arg}(\rho_{nm})$.
}
\label{fig3A}
\end{figure*}

In practice, building such an array of frequency sensitive photodetectors remains an open challenge in the microwave domain, despite encouraging recent progress towards this goal~\cite{Chen2011,Inomata2016,Narla2016,Besse2018,Kono2018,Lescanne2019,Dassonneville2020}. In this section, we demonstrate an actual experiment that implements a continuous version of the multiplexing measurement using heterodyne detectors instead of photodetectors. We multiplex the fluorescence photon counting of Sec. III
by sending a pulse containing a comb with $9$ frequencies corresponding to photon numbers from 0 to 8.
We then demultiplex the reflected pulse at the same 9 frequencies $\{f_\mathrm{mp}-k\chi_\mathrm{s,mp}\}_{0\leq k\leq 8}$ and extract a reflection coefficient $r_k$ for each of them (Fig.~\ref{fig1}d). The measurement consists in simultaneously measuring the emission coefficients $1-\mathcal{R}e(r_k)$ for each peak in Fig.~\ref{fig2B}a,b, which is
much faster than measuring them one at a time. Figure \ref{fig2C} shows these emission coefficients as a function of the average initial photon number for a drive strength $\Omega=\chi_\mathrm{s,mp}/2$ and a measurement duration of $2~\mu\mathrm{s}$.
For a given $\bar{n}$, every measurement channel $k$ gives an average signal that is proportional to the probability of having $k$ photons in the storage mode. As $\bar{n}$ is varied, the shape of the average signal of channel $k$ reproduces a Poisson distribution distorted by relaxation processes and channel cross-talk that increases with driving strength (see Appendix \ref{sec:sim_multiplexing}). This multiplexed photon counting signal can be reproduced using a master equation approach (solid lines in Fig.~\ref{fig2C}) using the photon number calibration of the standard photon counting approach.
This result thus demonstrates the applicability of our approach to photon counting by simultaneously probing information about the presence of 9 possible photon numbers in the resonator.

So how does this proof-of-principle experiment compare with standard photocounting? Each method has its own advantages and drawbacks. The multiplexed photon counting scheme trades off the temporal constraint and complexity of optimal control of the standard approach for the need of an efficient quantum measurement on a large frequency bandwidth. Indeed, the efficient measurement of the reflected pulse requires the use of a near-quantum limited amplifier with a dynamical bandwidth of at least a dozen of $\chi_\mathrm{s,mp}$ which is now possible using a TWPA~\cite{Macklin2015}. It is a comparable technical requirement to the recently demonstrated high-efficiency multiplexed readout of as many as 6 qubits coupled to a single feed line~\cite{Schmitt2014,Heinsoo2018,Kundu2019,Arute2019}. The number independent measurement time in the case of ideal photodetectors (Sec. IV) relies on the absence of noise when measuring a mode in the vaccum state. Instead, heterodyne detectors produce at least vacuum fluctuations in each frequency band, as Heisenberg uncertainty relations command. For this reason, the measurement time is expected to scale as the logarithm of the photon number similarly to state-of-the-art sequential measurement schemes. The two main advantages of our multiplexing technique over the standard approaches is that the prefactor is not limited by feedback latency or optimal control duration (see Appendix \ref{sec:multiplexed_time}) and that it is a continuous measurement that does not require any subtle temporal control.

\section{Measurement back-action}
\label{sec:backaction}

The measurement strength of the multiplexing measurement can be characterized using the yes-no qubit to observe the dynamics of the cavity state under the action of the continuous multiplexed measurement. The advantage of this method is that it does not require a single shot measurement of the photon number, which we could not reach owing to the limited efficiency of our amplifier, and the too short lifetime of the storage mode.
In the reciprocal case of measuring a qubit using a cavity as a probe, the measurement rate is bounded by the dephasing rate of the qubit, which grows as the square of the cavity driving strength~\cite{Clerk2008,Boissonneault2009}. Thus, characterizing the measurement rate of our multiplexed photon counting can 
be done by observing how the storage mode dephases for a given driving strength $\Omega$. Indeed, owing to the inherent quantum backaction of the photon number measurement, the measurement rate is bounded by how fast the conjugated operator, here the mode phase, diffuses. As the probe is based on a qubit driven by a frequency comb, one expects a different dependence of the measurement rate on $\Omega$ than for standard dispersive qubit readout using a single tone driving a probe cavity.

In order to measure this dephasing rate, we use the yes-no qubit to perform a direct Wigner tomography~\cite{Lutterbach1997,Bertet2002,Vlastakis2013a} of the storage mode at various times $t$. It provides a representation of the state $\rho$ in the  phase space of the mode and can be expressed as
\begin{equation}
W(\alpha) = \frac{2}{\pi} \mathrm{Tr}(D^\dagger(\alpha) \rho D(\alpha) \mathcal{P}).
\end{equation}
Here $D(\alpha) = e^{\alpha a_s^\dagger-\alpha^* a_s}$ is the storage displacement operator, $\mathcal{P} = e^{i \pi a_s^\dagger a_s}$ is the photon number parity operator, and $a_s$ is the canonical annihilation operator of the storage mode. Preparing the storage mode in a coherent state $|\beta=-1.55\rangle$, the Wigner function starts as a Gaussian distribution centered at $\alpha=\beta$. On the left of Fig.~\ref{fig3A}, one can see how the bare dephasing rate and the self-Kerr effect of the storage mode ($0.02~\mathrm{MHz}$ frequency shift per photon) distort the Gaussian distribution towards a torus with no phase when time increases even without any photon counting drives. Using a single drive with $\Omega = \chi_\mathrm{s,mp}/2$ to measure whether there is 1 photon, the phase diffuses faster and the Wigner function exhibits negativities in middle of Fig.~\ref{fig3A}. As seen in the corresponding density matrix, a tone at $f_\mathrm{probe}=f_\mathrm{mp} - \chi_\mathrm{s, mp}$ notably induces dephasing between states $|1\rangle$ and all other states $|m \neq 1\rangle$ (see density matrix as a function of drive frequency in Appendix \ref{sec:density_vs_frequency}). The phase diffusion is more intense when all the tones of the multiplexed readout are turned on than for a single tone with the same drive strength $\Omega$ (right of Fig.~\ref{fig3A}). Likewise all off-diagonal elements of the density matrix are then reduced.

\begin{figure}
\centering
\includegraphics[width=\columnwidth]{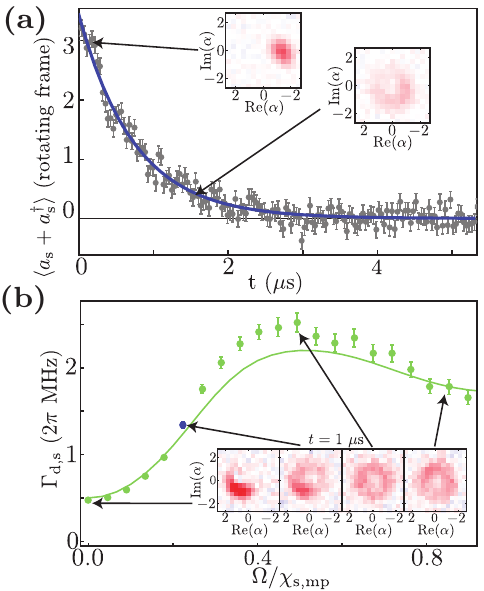}
\caption{\textbf{Measurement induced dephasing of the multiplexing measurement.} \textbf{a,} Measured exponential decay of the average storage mode quadrature $\langle a_\mathrm{s}+a_\mathrm{s}^\dagger\rangle$ in the case of a driving comb of strength $\Omega=0.23\chi_\mathrm{s,mp}$, and in a frame rotating at the storage mode resonant frequency. Insets show Wigner tomography of the storage mode for two values of $t$. \textbf{b,} Measured dephasing rate induced on the storage mode as a function of the drive strength $\Omega$ of the frequency comb. Insets show Wigner tomographies of the storage mode at $t=1~\mu\mathrm{s}$ for four values of $\Omega$.
}
\label{fig3B}
\end{figure}

To be more quantitative, the dephasing rate $\Gamma_\mathrm{d,s}$ of the cavity is accessed through the decay of the mean quadrature of the storage mode~\cite{Campagne-Ibarcq2020} (see Appendix \ref{sec:Ramsey})
\begin{equation}
\langle a_s+a_s^\dagger\rangle=\int 2 x W(x+iy)\mathrm{d}x\mathrm{d}y .
\end{equation}
In Fig.~\ref{fig3B}a, we show $\langle a_s+a_s^\dagger\rangle$ as a function of time under the multiplexed drive with a strength $\Omega=0.23\chi_\mathrm{s,mp}$. Repeating this experiment for various values of the multiplexed driving strength $\Omega$ allows us to determine how the latter affects the dephasing rate $\Gamma_\mathrm{d,s}$, and thus the measurement rate. The dephasing rate is non monotonic in the drive strength (Fig.~\ref{fig3B}b). Noticeably, it reaches a maximum when $\Omega = \chi_\mathrm{s, mp}/2$ for which information is extracted at a rate approximately 5 times larger than the natural dephasing rate. It is possible to understand this behavior by considering a model system where the comb has an infinite number of Dirac peaks $\{f_\mathrm{mp}+k\chi_\mathrm{s,mp}\}_{k\in\mathbb{Z}}$ (see Appendix \ref{sec:theory}). The Fourier transform of a comb being a comb, the drive performs sudden rotations by an angle $2\pi\Omega/\chi_\mathrm{s,mp}$ of the Bloch vector of the qubit every time step $1/\chi_\mathrm{s,mp}$. When $\Omega/\chi_\mathrm{s,mp}$ is integer, the comb does not affect the qubit and thus $\Gamma_\mathrm{d,s}$ vanishes. Conversely, the maximum measurement rate corresponds to half-integer $\Omega/\chi_\mathrm{s,mp}$ for which the effect of the comb on the qubit dynamics is maximum and leads to the strongest qubit emission. With the finite comb used in the experiment, this maximum persists and is reproduced by a model based on a master equation without any free parameter (line in Fig.~\ref{fig3B}b).

\section{Conclusion}
\label{sec:Discussion}

We have experimentally demonstrated that a single qubit can be used to continuously probe a multidimensional system by encoding information about its quantum state in the frequency domain. Improving further the detection efficiency $\eta$, the dispersive shift $\chi_\mathrm{s,mp}$ and the coupling rate $\Gamma_\mathrm{1,mp}$ between the qubit and the transmission line (while protecting the storage lifetime with a notch filter at its frequency), should enable single shot photon counting by multiplexing. To be more accurate, if the parameter $\eta \Gamma_\mathrm{1,mp}/\Gamma_\mathrm{1,s}=17$ increases by an order of magnitude, single shot measurements would be possible. Interestingly, assuming perfect detectors, our gedanken experiment shows how the multiplexed measurement can determine a photon number $N$ in a time that is $\log(N)$ faster than the best standard sequential approach. 
Our continuous measurement opens new possibilities in terms of feedback control of the quantum state of a cavity. It can readily be applied to stabilize quantum states by feedback control~\cite{Wiseman2009}, probe quantum trajectories of microwave modes~\cite{Haroche2006},  observe quantum Zeno dynamics~\cite{Bretheau2015}, or engineer desired decoherence channels by varying in time the amplitude of the probe tones. This measurement scheme enables the future implementation of a large class of measurement operators that would be 
useful to stabilize bosonic codes~\cite{Cai2021}, to stabilize a Fock state parity by autonomous feedback~\cite{Gertler2020}, or to extend the reach of simultaneous probing of a single quantum system by multiple observers~\cite{Hacohen-Gourgy2016,FICHEUX2018} to larger systems and arbitrarily many observers.
Our photocounter for stationnary modes can also be converted into a photocounter for propagating modes using a catch and count protocol~\cite{Dassonneville2020}. Moving further, one could extend this frequency domain measurement to more complex probes than a single qubit and many possible physical systems beyond superconducting circuits.

\begin{acknowledgments} 
This work was made possible through the support of Fondation Del Duca, IDEX Lyon (Contract No. ANR-16-IDEX-0005) and ANR JCJC-HAMROQS. We acknowledge IARPA and Lincoln Labs for providing a Josephson Traveling-Wave Parametric Amplifier. The device was fabricated in the cleanrooms of Coll\`ege
de France, ENS Paris, CEA Saclay, and Observatoire de Paris. We are grateful to R\'emi Azouit, Michel Devoret, and Mazyar Mirrahimi for discussions.
\end{acknowledgments}

\appendix

\section{Device and measurement setup}
\subsection{Design \label{sec:design}}
\begin{figure}[h!]
    \centering
    \includegraphics[width=\columnwidth]{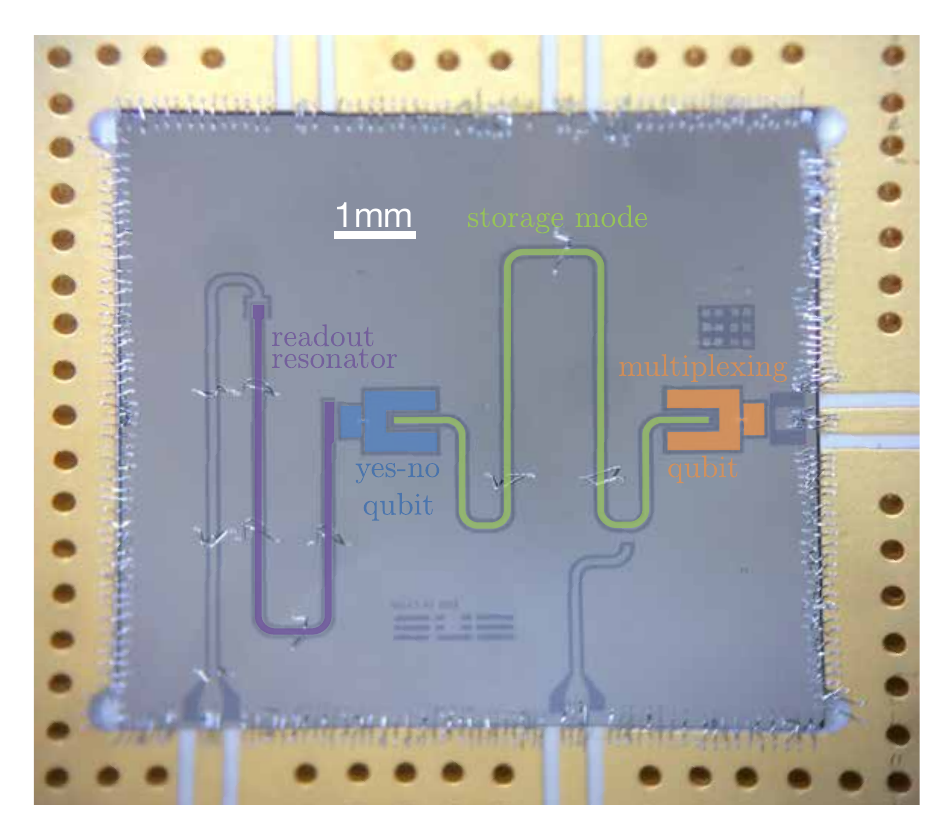}
    \caption{\textbf{Optical image of the circuit.} The readout resonator is colored in purple, storage mode in green, yes-no qubit in blue and multiplexing qubit in orange. All dark grey areas are silicon, grey areas are niobium on silicon and Josephson junctions are made of Al/AlOx/Al.}
    \label{fig::photo_circuit}
\end{figure}

The circuit  is composed of 4 electromagnetic modes whose parameters can be found in Methods. A high-Q harmonic oscillator, called storage mode, is composed of a $\mathrm{\lambda/2}$ coplanar waveguide (CPW) resonator (green in Fig.~\ref{fig::photo_circuit}). The storage resonator is capacitively coupled to two transmon qubits. The multiplexing qubit (orange) has a high spontaneous photon emission rate $\Gamma_\mathrm{1,mp} = (44 \mathrm{~ns})^{-1}$ into a transmission line compared to other modes. In contrast, the yes-no qubit is capacitively coupled to a low-Q readout resonator and has a long coherence time $T_\mathrm{2,yn} =27~\mu\mathrm{s}$. As required by Wigner tomography, the yes-no qubit coherence time and the lifetime of storage mode are larger than the
time needed to measure the parity of storage photon number
$1/2\chi_\mathrm{s,yn} \ll T\mathrm{_{1,s}},T_\mathrm{2,yn}$. As we use the multiplexing qubit to count the photon number in the storage mode, we need it to be photon number resolved~\cite{Schuster2007} otherwise each record of the multiplexing measurement could not be associated to a single specific photon number. This photon number resolved constraint imposes that the multiplexing qubit decoherence rate must be smaller than the cross-Kerr rate between the multiplexing qubit and the storage mode $\mathrm{\Gamma_{2,mp}<2\pi\chi_{s,mp}}$. This resolution constraint is not critical, as in fact a finite amount of photon number information can be extracted as soon as $\chi_\mathrm{s,mp}$ is nonzero, but the decoding is much simpler if we can reason in terms of well-separated resonance peaks.

\subsection{Device fabrication.}
\label{sec:fabrication}
The length of the readout resonator and storage mode was designed to obtain the resonant frequencies $f_\mathrm{ro} \sim 7~\mathrm{GHz}$ and $f_\mathrm{s} \sim  4.5~\mathrm{GHz}$. 
The circuit consists of a sputtered 120 nm-thick Niobium film deposited on a 280 $\mu$m-thick undoped silicon wafer. The resonators and feed lines are dry etched after optical lithography. After an ion milling step, the Josephson junctions are made out of e-beam evaporated Al/AlOx/Al through a PMMA/MAA resist mask patterned in a distinct e-beam lithography step. For each transmon qubit a single Dolan bridge is used to make the junctions.

\subsection{Measurement setup}
\label{sec:setup}

The readout resonator, the yes-no qubit and the multiplexing qubit are driven by pulses that are generated using a Tektronix\textregistered ~Arbitrary Waveform Generator (AWG) AWG5014C with a sample rate of 1 GS/s. Storage mode pulses are generated using a Zurich Instruments\textregistered ~UHFLI with a sample rate of 1.8 GS/s. The UHFLI allows us to change the pulse amplitude and phase without recompiling the sequence. This feature decreases the time needed for Wigner tomography compared to a standard AWG and makes the pulse sequence simple with the drawback of having to synchronize the two AWGs. AWG pulses are modulated at a frequency 25 MHz for readout, 100 MHz for yes-no qubit, and 75 MHz for storage and multiplexing qubit. They are up-converted using single sideband mixers for readout resonator and multiplexing qubit and regular mixers for the storage resonator and yes-no qubit, with continuous microwave tones produced respectively by AnaPico\textregistered ~APSIN12G, Agilent\textregistered ~E8257D, WindFreak\textregistered ~SynthHD, and AnaPico\textregistered ~APSIN20G sources that are set at the frequencies $f_\mathrm{ro}+25$ MHz, $f_\mathrm{mp}+75$ MHz, $f_\mathrm{s}+75$ MHz and $f_\mathrm{yn}+100$ MHz.

\begin{figure}[h!]
    \centering
    \includegraphics[width=\columnwidth]{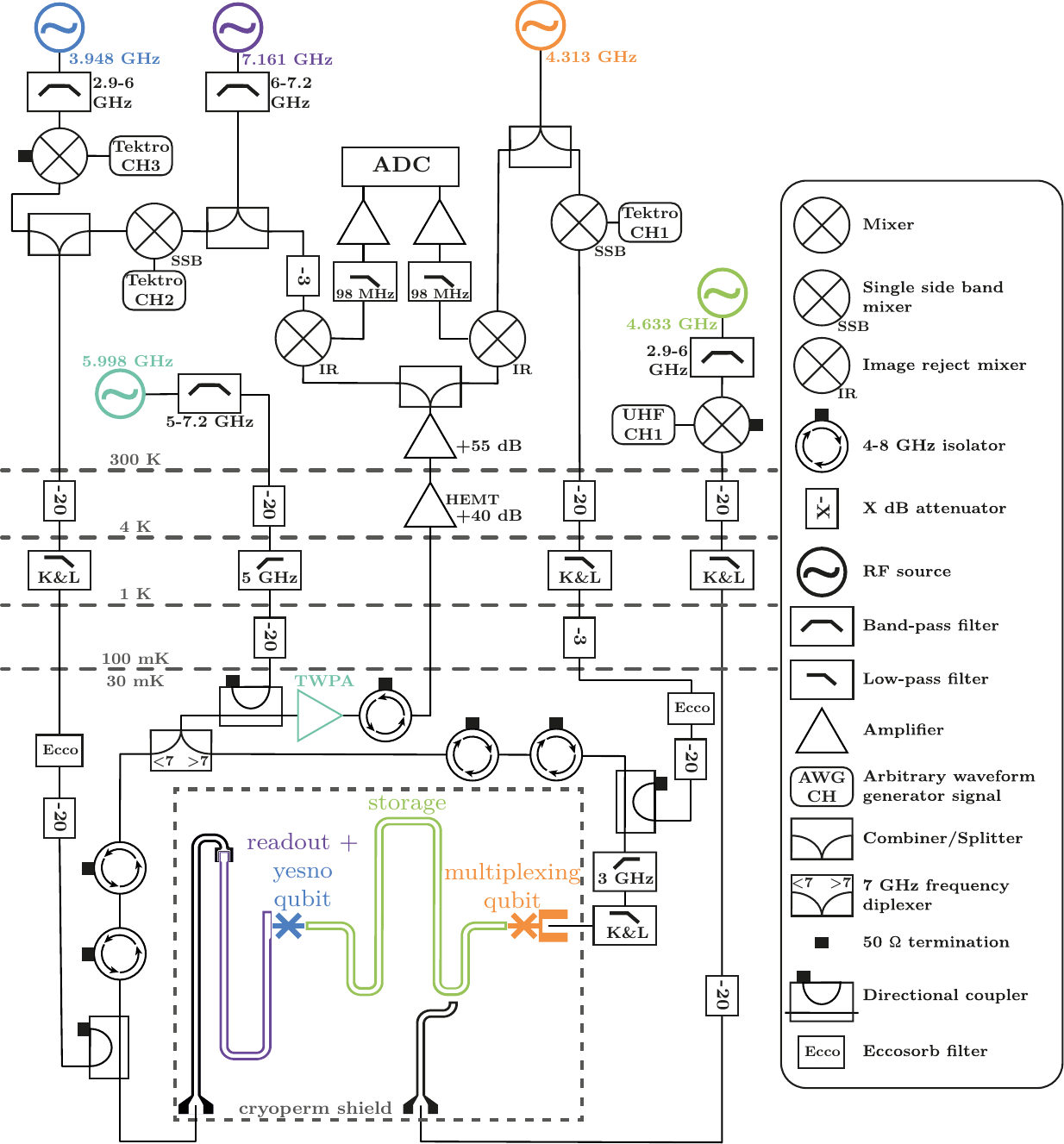}
    \caption{\textbf{Schematic of the setup.} Each electromagnetic mode of the experiment is driven by a RF generator detuned by the modulation frequency and whose color matches the color of the corresponding mode in Fig.~\ref{fig::photo_circuit}. Room temperature isolators are not represented for the sake of clarity.}
    \label{fig::cablage}
\end{figure}

The two reflected signals from the readout and multiplexing qubit are combined with a diplexer and then amplified with a Travelling Wave Parametric Amplifier (TWPA) provided by Lincoln Labs. We tuned the pump frequency ($f_\mathrm{TWPA}$ = 5.998 GHz) and power in order to reach a gain of 20.7 dB at 7.138 GHz and 18.2 dB at 4.238 GHz. The quantum efficiency of the yes-no readout signal was measured to be $18.7 \pm 0.4\%$, and should be close to the efficiency $\eta$ of the multiplexing detection. We estimate that this efficiency is the product of the efficiency of the microwave components before the TWPA (25 to 60\%), the efficiency of the TWPA itself (33\% to 83\%) and the (90 to 95\%) efficiency coming from what is above the HEMT amplifier. The follow-up amplification is performed by a High Electron Mobility Transistor (HEMT) amplifier from Low Noise Factory (LNF\textregistered)~at 4 K and by two room temperature amplifiers. The two signals are down-converted using image reject mixers before digitization by an Alazar\textregistered~acquisition board and numerical demodulation. Actually for the multiplexed signal, nine demodulation operations are performed at each of the down-converted frequencies $75\mathrm{~MHz} + k\chi_\mathrm{s,mp}$ for $0\le k \le 8$. The full setup is shown in Fig.~\ref{fig::cablage}. The Tektronix\textregistered ~~AWG is used as the master that triggers the UHFLI and the Alazar\textregistered ~board.

The frequency comb that is used for the multiplexing measurement is generated and demodulated using the following method. Nine cosine functions at frequencies $\{75\mathrm{~MHz} + k\chi_\mathrm{s,mp}\}_{0\le k \le 8}$ are summed and multiplied by a Gaussian envelop numerically with a sampling rate of $1\mathrm{~GHz/s}$ over the duration of the pulse. A waveform is then generated by the AWG following this list of values. This method  ensures a good phase coherence between all the comb frequencies. The AWG output is up-converted using a single side band mixer whose LO port is driven at frequency $f_\mathrm{mp}+75$ MHz. 

\subsection{Circuit parameters and master equation.}
\label{sec:parameters}

\begin{table}[t!]
    \resizebox{\columnwidth}{!}{
    \begin{tabular}{|l|c|c|c|}
    \hline
     circuit parameters& symbol& Hamiltonian term & value\\
     \hline
     readout resonator frequency &$f_\mathrm{ro}$& $h f_\mathrm{ro} \hat{n}_\mathrm{ro}$ & $7.138~ \mathrm{GHz}$\\
     storage mode frequency &$f_\mathrm{s}$ & $h f_\mathrm{s} \hat{n}_\mathrm{s}$ & $4.558 ~\mathrm{GHz}$\\
     yes-no qubit frequency &$f_\mathrm{yn}$ & $h f_\mathrm{yn} \hat{n}_\mathrm{yn}$ & $3.848 ~\mathrm{GHz}$\\
     multiplexing qubit frequency &$f_\mathrm{mp}$ & $h f_\mathrm{mp} \hat{n}_\mathrm{mp}$ & $4.238~ \mathrm{GHz}$\\
     readout/yes-no qubit cross-Kerr rate &$\chi_\mathrm{ro,yn}$ & $-h \chi_\mathrm{ro,yn} \hat{n}_\mathrm{ro}\hat{n}_\mathrm{yn}$ & $0.4~ \mathrm{MHz}$\\
     storage/yes-no qubit cross-Kerr rate &$\chi_\mathrm{s,yn}$ & $-h \chi_\mathrm{s,yn} \hat{n}_\mathrm{s}\hat{n}_\mathrm{yn}$ & $1.4~ \mathrm{MHz}$\\
     storage/multiplexing qubit cross-Kerr rate &$\chi_\mathrm{s,mp}$ & $-h \chi_\mathrm{s,mp} \hat{n}_\mathrm{s}\hat{n}_\mathrm{mp}$ & $4.9~ \mathrm{MHz}$\\
     yes-no qubit anharmonicity &$\chi_\mathrm{yn,yn}$ & $-h \chi_\mathrm{yn,yn} \hat{n}_\mathrm{yn}(\hat{n}_\mathrm{yn}-1)$ & $160~ \mathrm{MHz}$\\
     multiplexing qubit anharmonicity &$\chi_\mathrm{mp,mp}$ & $-h \chi_\mathrm{mp,mp} \hat{n}_\mathrm{mp}(\hat{n}_\mathrm{mp}-1)$ & $116~ \mathrm{MHz}$\\
     \hline
    \hline
     circuit parameters & symbol & jump operator & value\\
     \hline
     readout decay rate & $\Gamma_\mathrm{ro}$& $\Gamma_\mathrm{ro}\mathcal{L}(\hat{a}_\mathrm{ro})\rho$  & $(40~\mathrm{ns})^{-1}$\\
     storage decay rate & $\Gamma_\mathrm{1,s}$& $\Gamma_\mathrm{1,s}\mathcal{L}(\hat{a}_\mathrm{s})\rho$& $(3.8~ \mathrm{\mu s})^{-1}$\\
     storage decoherence rate & $\Gamma_\mathrm{2,s}$&$2\Gamma_\mathrm{\phi,s}\mathcal{L}(\hat{n}_\mathrm{ro})\rho$ & $(2~ \mathrm{\mu s})^{-1}$\\
     yes-no decay rate & $\Gamma_\mathrm{1,yn}$& $\Gamma_\mathrm{1,yn}\mathcal{L}(\hat{a}_\mathrm{yn})\rho$& $(20~ \mathrm{\mu s})^{-1}$\\
     yes-no decoherence rate & $\Gamma_\mathrm{2,yn}$& $2\Gamma_\mathrm{\phi,yn}\mathcal{L}(\hat{n}_\mathrm{yn})\rho$& $(27~ \mathrm{\mu s})^{-1}$\\
     multiplexing decay rate & $\Gamma_\mathrm{1,mp}$& $\Gamma_\mathrm{1,mp}\mathcal{L}(\hat{a}_\mathrm{mp})\rho$& $(42~ \mathrm{ns})^{-1}$\\
     multiplexing decoherence rate & $\Gamma_\mathrm{2,mp}$& $2\Gamma_\mathrm{\phi,mp}\mathcal{L}(\hat{n}_\mathrm{mp})\rho$& $(84~ \mathrm{ns})^{-1}$\\
     \hline
    \end{tabular}
    }
    \caption{\label{tab:lindblad_para}\textbf{Table of circuit parameters.} 
    }
\end{table}
All parameters of the 4 modes can be measured using standard circuit-QED measurement (see Table \ref{tab:lindblad_para}). Frequencies of the readout mode and multiplexing qubit are measured by spectroscopy. Frequencies of storage mode and yes-no qubit are measured using two-tone spectroscopy with the readout mode. Yes-no qubit decay and decoherence rate are measured with the time evolution of the probability to find the qubit excited after a $\pi$ pulse and using Ramsey oscillations. Readout mode decay rate and cross-Kerr rate between readout mode and yes-no qubit are measured using the measurement induced dephasing rate by the readout mode on the yes-no qubit. Cross-Kerr rate between the storage mode and the two qubits are measured using qubit spectroscopy with the storage state initialized in various coherent states. Anharmonicities are measured using spectroscopy of the qubit excited state. Decay and decoherence rates of the storage mode are measured with the time evolution of the probability to have $0$ photon in the storage mode after a displacement and storage Ramsey interferometry experiment. Multiplexing qubit decay and decoherence rates are measured by fitting the qubit spectroscopy for various drive amplitudes. All those parameters enable us to write a master equation model based on the Lindblad equation with the Hamiltonian
\begin{equation}
\label{eq:hamiltonian}
\begin{array}{rcl}
  \hat{H} &=& h f_{\mathrm{ro}} \hat{n}_{\mathrm{ro}} +h f_{\mathrm{s}} \hat{n}_{\mathrm{s}} + h f_{\mathrm{yn}} \hat{n}_{\mathrm{yn}} + h f_{\mathrm{mp}} \hat{n}_\mathrm{{mp}} \\&&
    - h \chi_\mathrm{ro,yn} \hat{n}_\mathrm{ro}\hat{n}_\mathrm{yn} - h \chi_\mathrm{s,yn} \hat{n}_\mathrm{s}\hat{n}_\mathrm{yn} - h \chi_\mathrm{s,mp} \hat{n}_\mathrm{s}\hat{n}_\mathrm{mp} \\&&
    - h \chi_\mathrm{yn,yn} \hat{n}_\mathrm{yn}(\hat{n}_\mathrm{yn}-1) - h \chi_\mathrm{mp,mp} \hat{n}_\mathrm{mp}(\hat{n}_\mathrm{mp}-1),
\end{array}
\end{equation}
where $\hat{n}_\mathrm{{ro}}$, $\hat{n}_\mathrm{s}$, $\hat{n}_\mathrm{{yn}}$, and $\hat{n}_\mathrm{{mp}}$ are the photon number operators respectively for the readout, storage, yes-no qubit and multiplexing qubit. $\chi_{a,b}$ is the cross-Kerr rate between modes $a$ and $b$. $\chi_{a,a}$ is the anharmonicity of the mode $a$.
The master equation on the system density matrix $\rho$ reads

\begin{equation}
\begin{array}{rcl}
    \dot{\rho} &=& - \dfrac{i}{\hbar} [\hat{H},\rho] + \Gamma_\mathrm{ro}\mathcal{L}(\hat{a}_\mathrm{ro})\rho+ 2 \Gamma_\mathrm{\phi,s}\mathcal{L}(\hat{n}_\mathrm{s})\rho+ \Gamma_\mathrm{1,s}\mathcal{L}(\hat{a}_\mathrm{s})\rho \\&&+ 2 \Gamma_\mathrm{\phi,yn}\mathcal{L}(\hat{n}_\mathrm{yn})\rho+  \Gamma_\mathrm{1,yn}\mathcal{L}(\hat{a}_\mathrm{yn})\rho + 2 \Gamma_\mathrm{\phi,mp}\mathcal{L}(\hat{n}_\mathrm{mp})\rho \\&& + \Gamma_\mathrm{1,mp}\mathcal{L}(\hat{a}_\mathrm{mp})\rho,
\end{array}
\label{eq::Lindblad}
\end{equation}where $\mathcal{L}$ is the Lindblad superoperator defined as $\mathcal{L}(\hat{L})\rho = \hat{L}\rho\hat{L}^\dag-\left\{\hat{L}^\dag\hat{L},\rho \right\}/2$ and $\hat{a}_b$  is the annihilation operator of mode $b$. For a qubit mode $b$, the dephasing rate $\Gamma_\mathrm{\phi,b}$ is linked to the decoherence rate by $\Gamma_\mathrm{2,b}= \Gamma_\mathrm{1,b}/2+\Gamma_\mathrm{\phi,b}$. 

\section{Calibrations}
\label{sec:calibration}

\subsection{Calibration of the storage mode displacement amplitude}
\label{sec:photon_number}
\begin{figure}[h!]
    \centering
    \includegraphics[width=0.9\columnwidth]{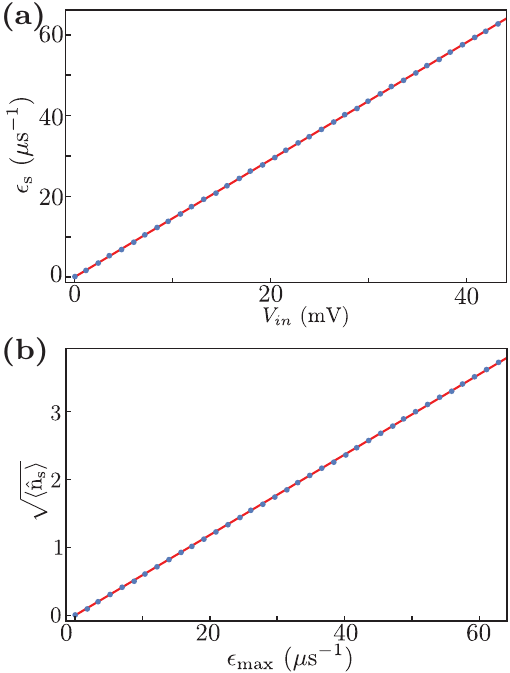}
    \caption{\textbf{Calibration of the average number of photons $\langle n_\mathrm{s}\rangle$ in the storage mode as a function of the displacement amplitude. a.} Evolution of the displacement amplitude $\epsilon_s$ with the pulse envelope $V_{s}$. The calibration is obtained by comparing the results of a photon counting experiment using the yes-no qubit with a master equation simulation (see section \ref{sec:Photon_counting_sim}). \textbf{b.} Square root of the average photon number $\langle n_\mathrm{s}\rangle$ in the storage mode as a function of the drive amplitude. The storage is displaced by 100~ns long Gaussian pulse with a width of 25~ns. The same pulse shape is used in the simulation. From the two linear fits we extract the evolution of the mean number of photons with the amplitude of the pulse $\sqrt{\langle n_\mathrm{s}\rangle} = \left(85.9~\mathrm{V}^{-1}\right) \, V_\mathrm{max,s}$. }
    \label{fig:photon_number}
\end{figure}

The storage mode can be displaced by driving it on resonance with a voltage $V_\mathrm{s}(t)\cos(2\pi f_\mathrm{s} t+\phi_\mathrm{s})$, where $V_\mathrm{s}(t)$ is the pulse envelope.
The driving Hamiltonian of the storage mode reads $\hbar (\epsilon_\mathrm{s} (t) \hat{a}_\mathrm{s}^\dag +\epsilon_\mathrm{s}^\ast(t) \hat{a}_\mathrm{s})$ where $\epsilon_\mathrm{s} (t) = \mu V_\mathrm{s} (t) e^{i\phi_\mathrm{s}}$. The scaling factor $\mu = 1.45~\mathrm{(mV~\mu s)^{-1}}$ is calibrated by fitting the photocounting measurement results obtained using the yes-no qubit with the master equation simulation (see section \ref{sec:Photon_counting_sim}). Fig.~\ref{fig:photon_number}a shows the evolution of $\epsilon_s$ with $V_\mathrm{s}$. For every experiment, the storage mode displacements are realized using a Gaussian pulse shape $\epsilon_\mathrm{s}(t)=\lambda(t)\epsilon_\mathrm{max}$ with a maximum amplitude $\epsilon_\mathrm{max}$, a width  $25~\mathrm{~ns}$ and a duration  $100~\mathrm{~ns}$. We simulated the dynamics of the storage mode under this Gaussian displacement taking into account the couplings, relaxation and decoherence rates (see section \ref{sec:photon_number_sim}) for various amplitudes $\epsilon_\mathrm{max}$. We then computed the expectation value of the number of photon operator $\langle \hat{n}_\mathrm{s}\rangle$ at the end of the pulse. Fig.~\ref{fig:photon_number}b shows the square root of $\langle \hat{n}_\mathrm{s}\rangle$ as a function of $\epsilon_\mathrm{max}$. Fitting with a linear function, we find that $\sqrt{\langle \hat{n}_\mathrm{s}\rangle} = 59.1 \epsilon_\mathrm{max}$. As $\epsilon_s$ increases linearly with $V_{\mathrm{s}}$, $\epsilon_\mathrm{max}$ increases linearly with the maximum voltage amplitude $V_\mathrm{max,s}$ of the Gaussian pulse $V_{\mathrm{s}}(t)=\lambda(t)V_\mathrm{max,s}$. Using the two linear regressions,  we can express the photon number of the storage mode as $\sqrt{\langle n_\mathrm{s}\rangle} = (85.9~\mathrm{V}^{-1}) \, V_\mathrm{max,s}$.

\subsection{Wigner tomography calibration}
 \label{sec:wigner_cal}

\begin{figure}[h!]
    \centering
    \includegraphics[width=0.85\columnwidth]{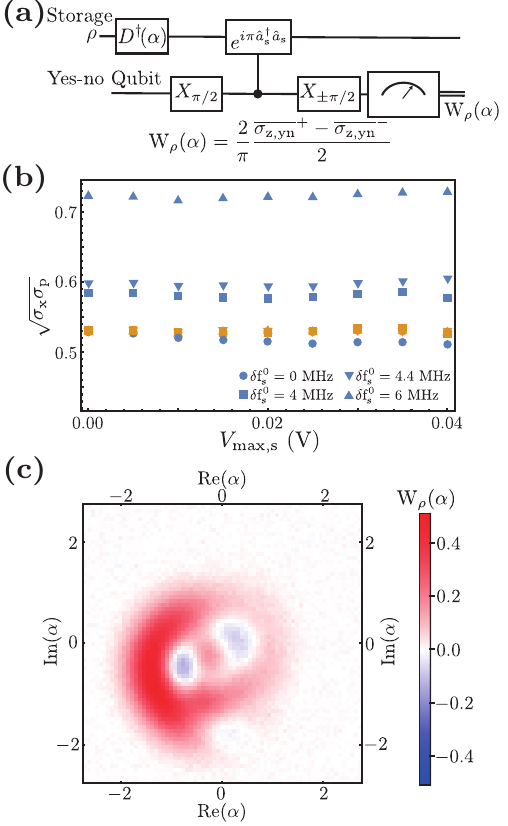}
    \caption{\textbf{Direct Wigner tomography of the storage mode a.} Circuit diagram of a Wigner tomography using a parity measurement based on dispersive interaction. After a 100~ns displacement pulse on the storage mode, an unconditional $\pi/2$ pulse is applied to the yes-no qubit. The qubit evolves freely during a time $\Delta\tau=337 ~\mathrm{ns}\approx 1/2\chi_\mathrm{s,yn}$ before a new $\pm\pi/2$ pulse is sent and the state of the yes-no qubit is measured using the readout resonator. \textbf{b.} Calibration of the quadrature axes for Wigner tomography. Blue dots represent the standard deviation of the quadratures of the displaced thermal equilibrium state of the storage mode as a function of  drive amplitude for various detuning using only the photon number calibration (see section \ref{sec:photon_number}). In contrast, yellow dots show the same standard deviation with the noise based quadrature calibration. 
    \textbf{c.} Wigner tomography of the storage mode. Here, the mode was prepared in two steps. First, the storage mode is displaced by a pulse with an amplitude 1.7 and then the multiplexing qubit is driven at a single tone at $f_\mathrm{mp}-1.4\chi_\mathrm{s,mp}$ during 750~ns with an amplitude $\Omega = \chi_\mathrm{s,mp}/2$. The appearance of  negative values in the Wigner function demonstrate that one can prepare non-classical states in the storage mode using the multiplexing qubit backaction alone.}
    \label{fig::Wigner_ex}
\end{figure}

The Wigner function of a harmonic oscillator with density matrix $\rho$ is defined as $\mathrm{W(\alpha) = 2\,\mathrm{Tr}(D^\dag(\alpha)\rho D(\alpha)\mathcal{P})}/\pi$ where $\mathrm{D}(\alpha)=e^{\alpha \hat{a}_\mathrm{s}^\dagger-\alpha^* \hat{a}_\mathrm{s}}$ is the displacement operator of the storage mode by a coherent field $\alpha$ and $\mathcal{P} = e^{i\pi \hat{a}_\mathrm{s}^\dag\hat{a}_\mathrm{s}}$ is the photon number parity operator. A Wigner function is the expectation value of the parity after a displacement by an amplitude $\alpha$.
The Wigner tomography sequence is represented on Fig.~\ref{fig::Wigner_ex}a . It starts by realizing a displacement on the storage mode with a 100 ns long Gaussian pulse at frequency $f_\mathrm{s}$ (or detuned for Ramsey interferometry of the storage mode, see section \ref{sec:Ramsey}) with a width of 25 ns. Then two successive $\pi/2$ Gaussian pulses  of 18 ns with a width of 4.5 ns are sent to the yes-no qubit at $f_\mathrm{yn}$ and are separated by a waiting time $\Delta\tau=337 ~\mathrm{ns}\approx 1/2\chi_\mathrm{s,yn}$. It implements a parity measurement and maps the parity of the storage mode onto the $z$-axis on the yes-no qubit~\cite{Lutterbach1997,Bertet2002,Vlastakis2013a}. The sequence terminates by a 2 $\mu$s long square pulse on the readout resonator to read out the state of the yes-no qubit.
For high amplitude $\alpha$, higher order Kerr terms distort the Wigner function. To mitigate this effect, we interleave two sequences with a final pulse of phase either $+\pi/2$ or $-\pi/2$.(see Fig.~\ref{fig::Wigner_ex}a). The difference between the two signals gives us the Wigner function without the distortion due to the storage mode anharmonicity and enables us to remove low frequency noise.
The $z$ axis of Fig.~\ref{fig::Wigner_ex}b is calibrated using the yes-no qubit Rabi oscillation amplitude to express the signal using Pauli operators. Multiplying the result by $2/\pi$ yields the Wigner function W($\alpha$) in Fig.~\ref{fig::Wigner_ex}c. 

\begin{figure*}[t]
    \centering
    \includegraphics[width=0.9\textwidth]{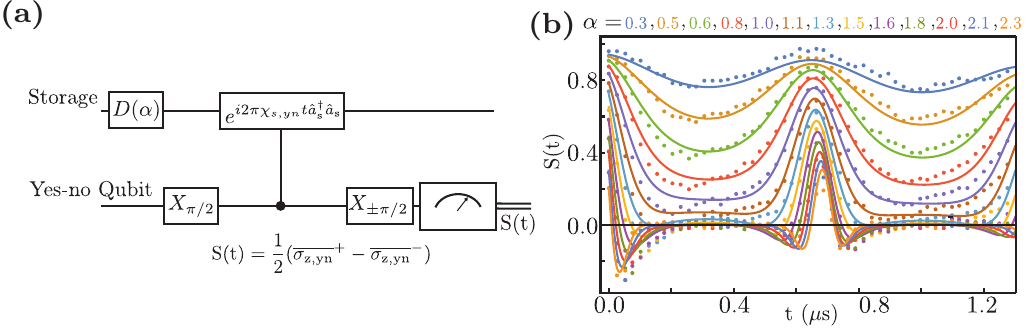}
    \caption{\textbf{Revival of the Ramsey interferometry on the yes-no qubit a.} Circuit diagram for Ramsey interferometry in the presence of storage photons. After a 100~ns displacement pulse at the storage frequency, an unconditional $\pi/2$ pulse is applied to the yes-no qubit. We then let the qubit evolve freely during a time $t$ before doing a new $\pm\pi/2$ pulse and measure the state of the yes-no qubit. The signal S(t) is half the difference between the average outcomes of the two sequences. \textbf{b.} Measured (dots) and predicted (lines) signal S as a function of waiting time $t$. Predicted signal is computed from Eq.~(\ref{eq:S(t)}). yes-no qubit revivals occur every $1/\chi_\mathrm{s,yn}\approx 0.7 \mu \mathrm{s}$.}
    \label{fig::yes-no_revival}
\end{figure*}

The axes of the phase space ${x,p}$ are calibrated using the same pulse sequence. The photon number calibration realized before (see section \ref{sec:photon_number}) cannot be used here for two reasons. First Ramsey oscillations of the storage mode impose to play the Wigner sequence with displacement pulses detuned from the storage mode frequency, while the photon number calibration is only valid for resonant pulses. Second high order Kerr interaction affects the calibration when the storage mode hosts a large number of photons. We decided to use the width of the Wigner function when the storage mode is in the thermal equilibrium state to calibrate the phase space axes. For a thermal state with a thermal photon number $n_\mathrm{th}$ the Wigner function is a 2D Gaussian function with a width $\sqrt{n_\mathrm{th}+1/2}$~\cite{Haroche2006}
\begin{equation}
    W_{\rho(n_\mathrm{th})}(\alpha = x+ip) = \dfrac{2}{\pi}\dfrac{1}{2 n_\mathrm{th}+1}e^{-2|\alpha|^2/(2 n_\mathrm{th}+1)}.
    \label{eq:Wigner_thermal}
\end{equation}
For a thermal state displaced by an amplitude $\beta$ the Wigner function is still a 2D Gaussian function with a width $\sqrt{n_\mathrm{th}+1/2}$ but centered on $\beta$. In thermal equilibrium, the storage mode has an average photon number $n_\mathrm{th}=0.03$, which is measured using the photon counting experiment. We calibrated the quadrature axes in order to get the expected geometrical mean $\sqrt{\sigma_x \sigma_p} =0.53$ of the spread along the quadratures $x$ and $p$  when the storage mode is at thermal equilibrium. To take into account high order Kerr effects, we displace the storage mode equilibrium state and measure its Wigner function. We adjust the calibration to still find a spread of $\sqrt{\sigma_x \sigma_p}=0.53$. The function used for the calibration is a third order polynomial function which gives $|\alpha|$ as a function of the pulse amplitude $V_\mathrm{max,s}$. We repeat this protocol for 3 detuning values $\delta\!f_\mathrm{s}$ between the displacement pulse and storage mode frequencies. Fig.~\ref{fig::Wigner_ex}b shows the mean quadrature spread of the displaced storage mode thermal state Wigner function as a function of drive amplitude $V_\mathrm{max,s}$ for the photon number calibration and the Wigner phase space calibration. For example, the polynomial function for a detuning of 4 MHz reads $\alpha=x+ip = e^{i\phi_\mathrm{s}}(77.3 V_\mathrm{max,s} + 86.7 V_\mathrm{max,s}^2-1343 V_\mathrm{max,s}^3)$ where  $V_\mathrm{max,s}$ is expressed in Volts and $\phi_\mathrm{s}$ is the phase of the pulse. For a typical value $V_\mathrm{max,s}=20 \mathrm{~mV}$, the second order term is a correction of about 2\% and the third one is a correction of about 0.07\%.

The duration $\Delta\tau$ is calibrated using qubit state revival during Ramsey interferometry~(see supplementary material of Ref.~\cite{Bretheau2015}). We used a Ramsey interferometry sequence (Fig.~\ref{fig::yes-no_revival}a) for the yes-no qubit at its resonance frequency for various coherent states in the storage mode. Revivals happen every $1/\chi_\mathrm{s,yn}$ which allows us to set $\Delta\tau$ as half the revival time in Fig.~\ref{fig::yes-no_revival}b. The signal difference between the final $-\pi/2$ and $+\pi/2$ pulses can be expressed as
\begin{equation}
\begin{aligned}
    S(t) & =\dfrac{\overline{\sigma_\mathrm{z,yn}}^+-\overline{\sigma_\mathrm{z,yn}}^-}{2} \\& = e^{|\alpha|^2 ( \mathrm{cos}(2 \pi \chi_\mathrm{s,yn} t)-1)}\\& \times \mathrm{cos}(|\alpha|^2 \mathrm{sin}(2 \pi \chi_\mathrm{s,yn} t)) e^{-t \Gamma_\mathrm{2,yn}-\gamma |\alpha|^2 t}.
\end{aligned}
    \label{eq:S(t)}
\end{equation}

This expression is derived in the supplementary material of Ref.~\cite{Bretheau2015}. The last exponential decay factor was added to take into account the intrinsic decoherence of the yes-no qubit and the measurement induced dephasing rate of the storage mode on the yes-no qubit during the waiting time. We also take into account a second order Kerr correction that shifts the revival time with the amplitude of the coherent state~\cite{Bretheau2015}. At first order this shift is given by 

\begin{equation}
    t_\mathrm{revival} = 2 \Delta\tau \left( 1+2|\alpha|^2\chi_\mathrm{s,s,yn}\Delta\tau\right).
\end{equation}
By adjusting the above parameters to allow the model to match the measured signal shown in Fig.~\ref{fig::yes-no_revival}b, we find $\Delta\tau=337 ~\mathrm{ns}$, $\gamma=0.23~\mathrm{\mu s^{-1}}$ and $\chi_\mathrm{s,s,yn} = 14$ kHz. However, this simple expression does not take into account the finite lifetime of the storage mode and we prefer not to consider these values as accurate enough compared to what we obtain with the other methods presented in this work. 

\subsection{Measuring the mean quadratures of the resonator field}
\label{sec:Ramsey}

The two mean quadratures of the storage mode are computed from the Wigner tomography as follows. For any operator $\hat{O}$, one can apply the Wigner transform to obtain the operator Wigner map $ W_{\hat{O}}$~\cite{Haroche2006} as
\begin{equation}
\begin{aligned}
         W_{\hat{O}}(\alpha = x+ip) &= W_{\hat{O}}(x,p) \\&= \dfrac{1}{\pi}  \int \mathrm{d}y\, e^{-2ipy}\langle x+ y/2|\hat{O}|x- y/2\rangle \\&= \frac{2}\pi \, \mathrm{Tr}(D^\dag(\alpha)\hat{O} D(\alpha)\mathcal{P})
\end{aligned}
\end{equation}
where $\{|x\rangle\}$ is the eigenbasis of the quadrature operator $\hat{X}$. With this tool, the Wigner function of a state $|\Psi\rangle$ (respectively a density matrix $\rho$) is simply given by $W_{|\Psi\rangle\langle\Psi|}(\alpha)$ (respectively $W_{\rho}(\alpha)$). The mean value of an operator $\hat{O}$ can be derived from the integral over the phase-space of the product of the two Wigner distributions multiplied by $\pi$,
\begin{widetext}
\begin{equation}
\begin{array}{llll}
     \pi \int\mathrm{d}x \int\mathrm{d}p\, W_{\rho}(x,p) W_{\hat{O}}(x,p) \\= \frac{1}{\pi}\int\mathrm{d}x \int\mathrm{d}p \int \mathrm{d}y\int \mathrm{d}y\prime \,e^{-2ip(y+y\prime)} \langle x+y/2|\rho|x-y/2\rangle \langle x+y\prime/2|\hat{O}|x-y\prime/2\rangle \\
     = \int\mathrm{d}x\int \mathrm{d}y\int \mathrm{d}y\prime \,\delta (y+y\prime)\langle x+y/2|\rho|x-y/2\rangle \langle x+y\prime/2|\hat{O}|x-y\prime/2\rangle \\
     = \int\mathrm{d}x\int \mathrm{d}y \,\langle x+y/2|\rho|x-y/2\rangle \langle x-y/2|\hat{O}|x+y/2\rangle \\= \int\mathrm{d}u\int \mathrm{d}v \,\langle u|\rho|v\rangle \langle v|\hat{O}|u\rangle \\
     = \mathrm{Tr}(\rho\hat{O}) = \langle\hat{O}\rangle_\rho
\end{array}.
\label{eq:Wigner_expect_value}
\end{equation}
\end{widetext}


In the case  of $\hat{X}$ and $\hat{P}$ operators, Wigner maps take a simple expression
\begin{equation}
    \begin{array}{ll}
         W_{\hat{X}}(\alpha = x+ip) =x/\pi \\
         W_{\hat{P}}(\alpha = x+ip) = p/\pi
    \end{array}.
\end{equation}
For any density matrix $\rho$, one can extract $\langle\hat{X}\rangle=\mathrm{Tr}(\hat{X}\rho)$ and $\langle\hat{P}\rangle=\mathrm{Tr}(\hat{P}\rho)$ from the Wigner function $W\equiv W_\rho$ as
 \begin{equation}
    \begin{array}{ll}
         \langle\hat{X}\rangle =\int\mathrm{d}x \int\mathrm{d}p\, W(x,p) x  \\
         \langle\hat{P}\rangle = \int\mathrm{d}x \int\mathrm{d}p\, W(x,p) p
    \end{array}.
\end{equation}

\subsection{Measuring the decoherence of the storage mode.}
\label{sec:decoherence_storage}
For a qubit, Ramsey oscillations correspond to the evolution of the real part of the coherence between the $|g\rangle$ and $|e\rangle$ states. A typical sequence starts by a $\pi/2$ pulse detuned from resonance by $\delta\!f$ to create a coherent superposition of $|g\rangle$ and $|e\rangle$ states. Then the qubit evolves freely before its state tomography. Both $\sigma_\mathrm{x}$ and $\sigma_\mathrm{y}$ oscillate at $\delta\!f$ while decaying at the decoherence rate $\Gamma_2$.

\begin{figure}[h!]
    \centering
    \includegraphics[width=\columnwidth]{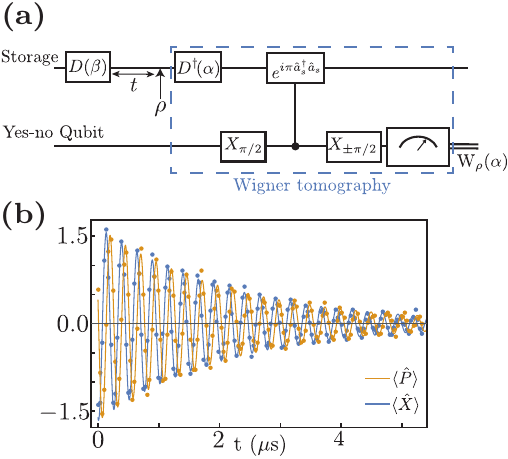}
    \caption{\textbf{Ramsey oscillations of the storage mode a,} Circuit diagram for Ramsey oscillations of an harmonic oscillator. All storage displacement pulses are performed in 100 ns with a Gaussian envelope of 25 ns width. In this experiment, the amplitude of the prepared coherent state $\beta$ is set to -1.55. The detuning between displacement pulse and bare storage frequencies is $\delta\!f^0_\mathrm{s}=3.96$ MHz. \textbf{b,} Measured (dots) and expected (lines) signals for $\langle\hat{X}\rangle$ (blue) and $\langle\hat{P}\rangle$ (orange). The expected signals are matched to the experiment using Eq.~(\ref{eq:XP_th}) with a frequency detuning of $\delta\!f_\mathrm{s}=3.96$ MHz and a decay rate $\Gamma_\mathrm{2,s} = (2~\mu \mathrm{s})^{-1}$.}
    \label{fig:Ramsey}
\end{figure}

We decided to realize an analogous sequence based on the same idea for a harmonic oscillator (a similar sequence was recently performed in Ref.~\cite{Campagne-Ibarcq2020}). The first $\pi/2$ pulse is replaced by a detuned displacement pulse $D(\beta)$ on the storage mode. The field then evolves during a time $t$ (during which the multiplexing measurement could be applied) before a Wigner tomography is realized (see Fig.~\ref{fig:Ramsey}a). The expectation value of $\hat{X} = (\hat{a}_\mathrm{s}+\hat{a}^\dag_\mathrm{s})/2$ and $\hat{P} = (\hat{a}_\mathrm{s}-\hat{a}^\dag_\mathrm{s})/2i$ quadratures are computed from the Wigner tomography (see Sec.~\ref{sec:Ramsey}). The time trace of $\langle\hat{X}\rangle$ and $\langle\hat{P}\rangle$ is what we call the Ramsey oscillations for the storage mode. As in the qubit case, the frequency of the oscillations is set by the detuning $\delta\!f_\mathrm{s}$ between the drive and the resonant frequency of the mode, which allows us to extract the frequency of the storage mode. At this point, a distinction has to be made between the detuning $\delta\!f^0_\mathrm{s}=f_\mathrm{drive}-f_s$ between the drive and the bare storage mode frequency (the resonant frequency when the multiplexed qubit and the storage are undriven) and the detuning $\delta\!f_\mathrm{s}$ between the drive and the resonant frequency of the storage mode, which depends on the multiplexed measurement strength in perfect analogy with the AC-Stark effect for a qubit readout. Note that the Wigner tomography sequence uses the same detuned frequency $\delta\!f_\mathrm{s}$ for its displacement pulse $D^\dag(\alpha)$ in order to keep the same phase reference. The measurement of Ramsey oscillations of a harmonic oscillator takes longer than the ones of a qubit because we fully determine the quantum state of an oscillator at each time instead of a simple Bloch vector. From Eq.~(\ref{eq::Lindblad}), one finds that $\langle\hat{X}\rangle$ and $\langle\hat{P}\rangle$ evolve as 
\begin{equation}
\begin{array}{ll}
    \langle\hat{X}\rangle = |\beta| \mathrm{cos}(2 \pi  \delta\!f_\mathrm{s} t+\phi) e^{-t \Gamma_\mathrm{d,s}} \\
    \langle\hat{P}\rangle = |\beta| \mathrm{sin}(2 \pi \delta\!f_\mathrm{s} t+\phi) e^{-t \Gamma_\mathrm{d,s}}
    \end{array}
    \label{eq:XP_th}
\end{equation}where $\beta = |\beta| e^{i\phi}=\langle \hat{a}_s\rangle(t=0)$. For each time $t$, we computed $\langle\hat{X}\rangle$ and $\langle\hat{P}\rangle$ and defined the storage mode dephasing rate as $\Gamma_\mathrm{d,s}$ which contains the intrinsic decoherence rate $\Gamma_\mathrm{2,s}$. Data Extended Fig.~\ref{fig:Ramsey}b shows an example of measured Ramsey oscillations. In the main text, Fig.~3B does not exhibit oscillations because it is the mean value $\langle \hat{a}_\mathrm{s} +\hat{a}_\mathrm{s}^\dag \rangle$ in the frame rotating at the resonant frequency of the storage mode. In practice, we measured them with a detuning 
and numerically computed the non-oscillating quantity
$2\mathcal{R}e((\langle\hat{X}\rangle+ i \langle\hat{P}\rangle)\mathrm{exp}(-2i\pi \delta\!f_\mathrm{s} t))$. 

\subsection{Storage mode frequency shift and induced dephasing rate by driving the multiplexing qubit with a comb}
\label{sec:MID}

\begin{figure*}[t]
    \centering
    \includegraphics[width=\textwidth]{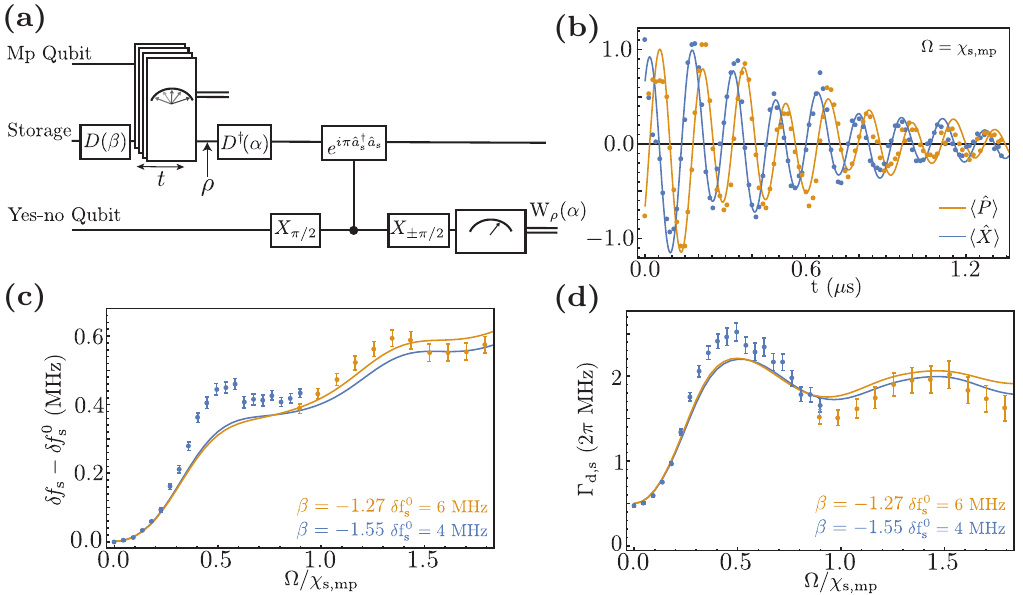}
    \caption{\textbf{Frequency shift and dephasing rate of the storage mode induced by the multiplexed photocounting measurement. a.} Circuit diagram of the protocol used to determine the dephasing rate and frequency shift of the storage mode induced by the multiplexed photocounting measurement. The amplitude $\beta$ of the initial displacement is set at -1.55 for small measurement amplitude $\Omega/\chi_\mathrm{s,mp}<0.9$ and at -1.27 for large measurement amplitude $\Omega/\chi_\mathrm{s,mp}>0.9$. The blocks linking the multiplexing qubit and the storage mode represent the multiplexed measurement during a time $t$ made by the qubit on the storage mode. This measurement is realized by driving the qubit with a frequency comb $[f_\mathrm{mp},f_\mathrm{mp}-\chi_\mathrm{s,mp},...,f_\mathrm{mp}-8\chi_\mathrm{s,mp}]$ within a Gaussian envelope. \textbf{b.} Ramsey oscillations of the storage mode for ``large" measurement amplitude $\Omega/\chi_\mathrm{s,mp}=1$. One can observe that the dynamics $\langle\hat{X}\rangle$ and $\langle\hat{P}\rangle$ are not governed by a simple decaying sine function. The theory does not reproduce quantitatively the measurement when using the naive version of the model Eq.~(\ref{eq:XPoftime}). we use the simple model Eq.~(\ref{eq:XPlargeM}) to capture this modulation.
    \textbf{c.} and \textbf{d.} ac-Stark shift and measurement induced dephasing rate measured (dots) and simulated (line) as a function of multiplexing qubit drive amplitude $\Omega$ in units of $\chi_\mathrm{s,mp}$. The evolution of the detuning and dephasing rate are strongly non linear with drive amplitude.}
    \label{fig:MID_Storage_cal}
\end{figure*}

In analogy with the ac-Stark shift of the frequency of a qubit coupled to a driven resonator, we also call ac-Stark shift the frequency shift of the storage mode induced by driving the multiplexing qubit. 
In order to measure this frequency shift and the dephasing rate that is induced by the multiplexing qubit on the storage mode, we realize the reciprocal protocol for a qubit measured by a cavity.
We use a Ramsey interferometry sequence on the storage mode during which the multiplexing qubit is driven with a frequency comb
(see Fig.~\ref{fig:MID_Storage_cal}a). The drive amplitude is given by the sum of nine sine functions at the frequencies $[f_\mathrm{mp},f_\mathrm{mp}-\chi_\mathrm{s,mp},...,f_\mathrm{mp}-8\chi_\mathrm{s,mp}]$ multiplied by a Gaussian envelope of duration $t$ and width $t/4$.

For small measurement strength $\Omega/\chi_\mathrm{s,mp}<0.9$, we generated the Ramsey sequence with a displacement pulse of amplitude $\beta = -1.55$ detuned from the base storage mode by $\delta\!f^0_\mathrm{s}=3.96$ MHz.
We fit the time evolution of $\langle\hat{X}\rangle$ and $\langle\hat{P}\rangle$ using the damped sine function
\begin{equation}
    \begin{array}{ll}
    \langle\hat{X}\rangle = A \mathrm{cos}(2 \pi \delta\!f_\mathrm{s} t+\phi) e^{-t \Gamma_\mathrm{d,s}} \\
    \langle\hat{P}\rangle = A \mathrm{sin}(2 \pi \delta\!f_\mathrm{s} t+\phi) e^{-t \Gamma_\mathrm{d,s}}
\end{array}.
\label{eq:XPoftime}
\end{equation}
The parameters $A$, $\delta\!f_\mathrm{s}$, $\phi$, and $\Gamma_\mathrm{d,s}$ are determined altogether by fitting the model to the measured oscillations. $\Gamma_\mathrm{d,s}$ is the sum of the intrinsic storage dephasing rate $\Gamma_\mathrm{2,s}$ and of the measurement induced dephasing rate. $\delta\!f_\mathrm{s}$ is the sum of the detuning from the bare storage mode frequency $\delta\!f^0_\mathrm{s}$ and of the ac-Stark shift of the storage mode. Both parameters are shown in Fig.~\ref{fig:MID_Storage_cal}c,d (blue dots) as a function of $\Omega/\chi_\mathrm{s,mp}$.

For larger measurement strength $\Omega/\chi_\mathrm{s,mp}>0.9$, we generated the Ramsey sequence with a displacement pulse detuning of $\delta\!f^0_\mathrm{s}=5.96$ MHz, an amplitude of $\beta=-1.27$,  and we model the time evolution of $\langle\hat{X}\rangle$ and $\langle\hat{P}\rangle$ by the sum of two sine functions with an exponential decay
\begin{equation}
    \begin{array}{ll}
    \langle\hat{X}\rangle = A (\mathrm{cos}(2 \pi \delta\!f_\mathrm{s} t+\phi) +\zeta \mathrm{cos}(2 \pi \nu t+\psi_\mathrm{X})) e^{-t \Gamma_\mathrm{d,s}} \\
    \langle\hat{P}\rangle = A (\mathrm{sin}(2 \pi \delta\!f_\mathrm{s} t+\phi)+\zeta \mathrm{sin}(2 \pi \nu t+\psi_\mathrm{P})) e^{-t \Gamma_\mathrm{d,s}}
\end{array}.
\label{eq:XPlargeM}
\end{equation}
This empirical model originates from three ideas. The first term is identical to the simple model in Eq.~(\ref{eq:XPoftime}).  Second, the measured Ramsey oscillations seem to show a small modulation in amplitude, which we try to capture with a second sine function. Third we try to keep the model as simple as possible.

Fig.~\ref{fig:MID_Storage_cal}b shows an example of Ramsey oscillations of the storage mode with a large amplitude of measurement. The two signals are fitted simultaneously to extract the parameters A,  $\delta\!f_\mathrm{s}$, $\nu$, $\phi$, $\psi_\mathrm{X}$, $\psi_\mathrm{P}$, and $\Gamma_\mathrm{d,s}$. The frequency $\nu$ varies from 2.15 MHz to 2.5 MHz. The parameter $\zeta$ is roughly constant, it varies between 0.2 to 0.27.
Fig.~\ref{fig:MID_Storage_cal}c shows measurement induced detuning as a function of measurement drive amplitude.

\subsection{Rabi frequency calibration}
\label{sec:calibration_Omega}

\begin{figure}[h!]
    \centering
    \includegraphics[width=\columnwidth]{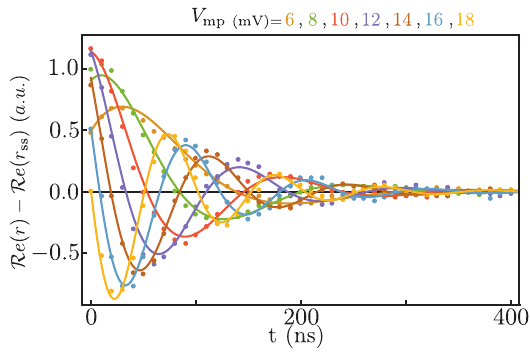}
    \caption{\textbf{Multiplexing qubit Rabi oscillations for various driving amplitude.} The measured Rabi oscillations observed in the reflection coefficient (dots) is reproduced by theory (solid line from Eq.~(\ref{eq:OmegaRabi})). The vertical axis represents the deviation of the real part of the reflection coefficient to its steady state value. This calibration allows us to extract the scaling parameter $\xi$ such that $\Omega =\xi V_\mathrm{mp} = (0.543\mathrm{~GHz.V^{-1}~}) V_\mathrm{mp}$. 
}
    \label{fig:fit_omega}
\end{figure}

We observe Rabi oscillations of the multiplexing qubit by applying a $1\mathrm{~\mu s}$-long square pulse at $f_\mathrm{mp}$ with a varying amplitude $V_\mathrm{mp}$. The reflected signal, demodulated by time steps of $10 \mathrm{~ns}$, displays damped oscillations given by~\cite{Ficheux2018thesis}
\begin{equation}
\begin{aligned}
    &\mathcal{R}e\left(r(t)\right)-\mathcal{R}e\left(r_\mathrm{ss}\right)= \\& A\cos \left[ \sqrt{(2 \pi\xi V_\mathrm{mp})^2-\left(\dfrac{\Gamma_\mathrm{1,mp}-2\Gamma_\mathrm{\phi,mp}}{16}\right)^2} (t - t_0)+ \phi\right]  \\&~~~~~~~~~~~~~~~ \times e^{-(t - t_0)/T},
    \label{eq:OmegaRabi}
\end{aligned}
\end{equation}
where $r_\mathrm{ss}$ is the value of the reflection coefficient  in the steady state. We obtain $\xi=0.543\mathrm{~GHz.V^{-1}~}$ so that the Rabi frequency is calibrated as  $\Omega =\xi V_\mathrm{mp} = (0.543\mathrm{~GHz.V^{-1}~})V_\mathrm{mp}$ (see Fig.~\ref{fig:fit_omega}).

  \section{Maximal measurement strength}
 \label{sec:theory}

The dephasing rate $\Gamma_\mathrm{d,s}$ depends on the driving amplitude $\Omega$, the dispersive shift $\chi_\mathrm{s,mp}$ and on the relaxation rate $\Gamma_\mathrm{1,mp}$. To gain insight in its dependence and learn how to maximize the measurement strength, we start by considering the case of an isolated single qubit before extending the model to a bipartite qubit-resonator system.

\subsection{Dynamics of a qubit driven by a comb}
 \label{sec:qubit_dynamics}

\subsubsection{Hamiltonian evolution}

\begin{figure}
\includegraphics[width=\columnwidth]{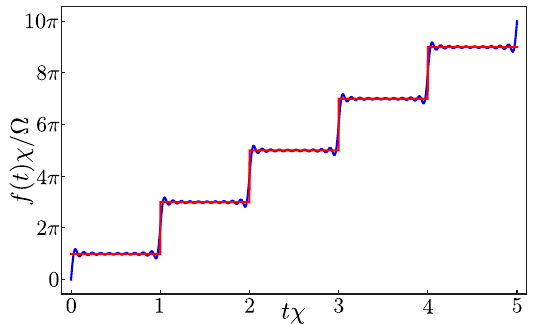}
\caption{Exact rotation angle $f(t)$ in blue and its staircase approximation $\bar{f}(t)$ in red. The duration of each step is equal to  $1/\chi$ and its height to $\frac{2\pi\Omega}{\chi}$. The quality of the approximation improves as the number of peaks  $2p+1$  in the comb gets larger. The fact that the trajectory starts with a half-jump is a particularity of having assumed that all comb components have the same phase at $t=0$. Random initial phases of the signals ( e.g. due to initializing the qubit after its photon emission into the measuring transmission line at a random time ) would most often position $t=0$ on a flat portion of the stairs. }\label{fig:ft}
\end{figure}

We consider a single qubit driven by a frequency comb with $2p+1$ frequency peaks at every $f_\mathrm{mp}+k\chi$ for $-p\leq k\leq p$. In the frame rotating at the qubit frequency, the Hamiltonian reads
\begin{equation}
   H(t) = \pi \Omega\left(\sum_{k=-p}^p \cos(2 \pi k\chi t)\right) \, \sigma_x \; . 
\end{equation}
After a time $t$, the qubit state will thus be rotated around the $x$-axis of the Bloch sphere by an angle $f(t)$ with
\begin{equation}
    f(t) = 2 \pi \Omega t \sum_{k=-p}^p \text{sinc}(2 \pi k  \chi t) \, ,
\end{equation}
where $\text{sinc}(x) = \sin( x)/( x)$. For large integers $p$, we can approximate the sum as
$\sum_{k=-\infty}^{+\infty} \text{sinc}(\pi k T) = \frac{1}{T}$, which is valid for $0<T<2$. Note that for a small number of peaks $2p+1$, this approximation is invalid close to $T=0$ or $2$. The expression allows us to approximate $f(t)$ for $0<\chi t<1$. It is then simple to derive $f(t)$ at any time $t$ since it is periodic up to the term in $k=0$.
With this we get
\begin{equation}
    f(t) \approx \pi\dfrac{\Omega}{\chi}+ 2\pi\dfrac{\Omega}{\chi}\lfloor t\chi \rfloor=: \bar{f}(t),
\end{equation}
where $\lfloor x \rfloor$ is the integer part of $x$. 
Therefore the rotation angle $f(t)$ evolves by steps. A comparison of the actual $f(t)$ and of the staircase approximation for a comb with 21 frequencies ($p=10$) is shown in Figure \ref{fig:ft}. To put it simply, the action of the comb consists in performing a Rabi rotation on the qubit by discrete steps instead of a continuous evolution as is the case for a single driving frequency. At each period $1/ \chi$, the qubit rotates almost instantaneously by an angle $2\pi \frac{\Omega}{\chi}$.

Without decoherence, if the qubit starts in state $|g\rangle$ at time $t_0$, the qubit state after a time $t$ reads
\begin{equation}
    \begin{array}{c}
\q{\psi(t)}= \cos\left(\frac{f(t)-f(t_0)}2\right) \q{g} + i \sin\left(\frac{f(t) - f(t_0)}2\right) \q{e} \\ \approx  \cos \left( \frac{\Omega \pi}{\chi} \lfloor(t-t_0) \chi\rfloor \right) \q{g} + i \sin \left( \frac{\Omega \pi}{\chi} \lfloor(t-t_0) \chi\rfloor \right) \q{e} .
\end{array}
\end{equation}

Let us focus on some particular values of $\frac{\Omega}{\chi}$.
\begin{itemize}
\item If $\frac{\Omega}{\chi}$ is integer, the staircase approximation with $\bar{f}(t)$ keeps the qubit in $\q{g}$ at all times, just performing a full rotation on the Bloch sphere at each Rabi pulse. In presence of relaxation, a photon loss can only happen during the short duration of the Rabi pulse, which decreases as $1/(p+1)$. The qubit remaining in the ground state cannot encode information on the resonator. This explains why there are minima of measurement induced dephasing rate when $\Omega/\chi$ is integer (see Fig.~\ref{fig:MID_Storage_cal}d). 
\item If $\frac{\Omega}{\chi}$ is half-integer, the staircase approximation with $f=\bar{f}$ makes the qubit state jump periodically between $\q{g}$ and $\q{e}$. This maximal extent of the evolution on the Bloch sphere intuitively explains the maximal measurement strength at the driving amplitude. In the following, we explain why this situation corresponds to a larger extraction of information than any steps with a different rotation angle.

\end{itemize}

\subsubsection{Integrated qubit dynamics in the presence of relaxation}

In the following, we use the infinite comb approximation $f(t) \approx \bar{f}(t)$. This allows us to integrate the qubit dynamics exactly. The continuous photon decay at rate $\Gamma_1$ is interrupted by discrete Rabi rotations at discrete times. The qubit state is confined in the $y-z$ plane of the Bloch sphere. Under this approximation, after the Rabi pulse number $k+1$, the qubit state is given by
\begin{equation}
\begin{array}{ll}
\left(\begin{array}{c} y(k\,T+T) \\ z(k\,T+T) \end{array} \right) =\\ \left( \begin{array}{cc} \cos\theta & -\sin\theta \\ \sin\theta & \cos\theta \end{array} \right) \; 
\left(\begin{array}{c} e^{-\Gamma_1 T/2}\, y(k\,T) \\ e^{-\Gamma_1 T} z(k\,T) + (e^{-\Gamma_1 T}-1)\end{array} \right)
\end{array}
\end{equation}
where $T=1/\chi$ is the period, $\theta=\frac{2 \pi \Omega}{\chi}$ is the angle spanned in the Bloch sphere during a discrete jump. The origin of time $t=0$ is chosen to start just after a Rabi jump.

The permanent solution of this discrete-time map (Poincar\'e map) right after $k$ steps is
\begin{equation}
\begin{aligned}
&\bar{y}(k\,T) = \frac{e^{\Gamma_1 T/4} \sinh(\Gamma_1 T/2) \sin\theta}{\cosh(3\Gamma_1 T/4) - \cos\theta\cosh(\Gamma_1 T/4)} \; , \\ \quad &\bar{z}(k\, T) = \frac{\sinh(\Gamma_1 T/2) (e^{-\Gamma_1 T/4} - e^{\Gamma_1 T/4} \cos\theta)}{\cosh(3\Gamma_1 T/4) - \cos\theta\cosh(\Gamma_1 T/4)} \; . \label{eq:ybar}
\end{aligned}
\end{equation}
The average qubit excited state population over one period in the permanent regime reads
\begin{equation}
  P(\Omega/\chi) := \frac{1}{T} \int_0^T \langle e \vert \rho(t) \q{e} = \frac{1-e^{-\Gamma_1 T}}{\Gamma_1 T} \frac{(1+\bar{z}(k\, T))}2 \; .  
\end{equation}
When using the parametrization $c\equiv \cos(\theta)$, one easily checks that $P$ is a strictly decreasing function on $c \in [-1,1]$. As a function of $\theta$, it has maxima for $\theta=(2k+1)\pi$ and minima for $\theta=2k\pi$. The latter are no surprise and give $P=0$ as the Rabi pulse takes the state from $\q{g}$ back to $\q{g}$; for finite number of peaks in the comb $2p+1$, the Rabi pulse is not instantaneous and $P>0$ at these minima. The value of the maxima would be $P=\frac{\tanh(\Gamma_1 T/2)}{\Gamma_1 T}$ with the infinite-comb approximation.

According to this approximation, the average rate of photon emission, which is linked to the measurement strength (each photon reveals information about the qubit frequency and hence the photon number), is thus $P\Gamma_1=\tanh(\Gamma_1 T/2)/T$ with the optimal choice of $\Omega=\chi/2+k\chi$, where $k$ is integer. 
\begin{itemize}
\item At fixed $T=1/\chi$, the emission rate increases with $\Gamma_1$ and converges towards $\chi$. For  $\Gamma_1 \gg \chi$, the qubit has the time to fully relax during one period. Therefore, in simple terms, at each period in the stepwise evolution, the qubit is excited and then releases deterministically a single photon into the output transmission line.
\item Likewise, for a fixed $\Gamma_1/\chi$ (thus fixed probability $P$ to emit a photon during a period), the average emission rate increases when $T$ decreases. Therefore the average emission rate increases as $\chi$.
\item For a fixed $\Gamma_1$, the largest average emission rate is obtained for $\chi=1/T$ as large as possible, but it saturates at
$P\Gamma_1=\Gamma_1/2$. This is consistent with the fact that $\Gamma_1$ is a hard limit on the photon emission rate.
\end{itemize}


\subsection{Dynamics of the qubit-resonator bipartite system driven by comb}
 \label{sec:system_dynamic}

For the sake of simplicity, we consider a lossless storage mode in this section. In this case, each 2 by 2 sub-matrix $\rho_{n_1,n_2} =\qd{n_1} \rho \q{n_2}$, with $\q{n_i}$ the $n_i$ Fock state, evolves independently of the others $\rho_{m_1,m_2}$ similarly to a collection of qubit-like system. The sub-matrix $\rho_{n_1,n_2}$ is non-normalized because it is an off diagonal sub-matrix of the storage-qubit system. We have
\begin{eqnarray}
\begin{array}{rcl}
    \tfrac{d}{dt} \rho_{n_1,n_2} &=& -i 2 \pi \chi \tfrac{n_1+n_2}{2} [\frac{\sigma_z}{2}, \rho_{n_1,n_2}]\\
& &-i 2 \pi \chi  \tfrac{n_1-n_2}{2} \{\frac{\sigma_z}{2}, \rho_{n_1,n_2}\} \\ 
& &- \frac{i}{\hbar} [H(t), \rho_{n_1,n_2}]  + \Gamma_1 \mathcal{L}(\sigma_-)\rho_{n_1,n_2} \; ,
\end{array}
\end{eqnarray}
where $H(t)$ is the frequency comb drive from the previous section.

\subsubsection{Computing the decoherence rate of the $n_1,n_2$ component}

The infinite comb approximation again helps. We view $H(t)$ as applying a Rabi pulse of angle $\theta=\frac{2 \pi \Omega}{\chi}$ at each period $T=1/\chi$, without any effect for the rest of the time. Over one period, we thus have
\begin{equation}
   \rho_{n_1,n_2}(kT+T) = \mathcal{K}_0 \circ \mathcal{K}_1 \rho_{n_1,n_2}(kT)\, , 
\end{equation}
where $\mathcal{K}_0$ applies the Rabi pulse, while $\mathcal{K}_1$ contains dynamics associated to the dispersive coupling and to qubit decay. During each period between Rabi jumps, denoting $\rho_{n_1,n_2} = \frac{x \sigma_x + y \sigma_y + z \sigma_z + \eta I}{2}$, the dynamics $\mathcal{K}_1$ corresponds to the integration of the set of equations
\begin{eqnarray*}
\tfrac{d}{dt} x &=& -\tfrac{\Gamma_1}{2} x - \tfrac{ 2 \pi \chi(n_1+n_2)}{2} y \\
\tfrac{d}{dt} y &=& -\tfrac{\Gamma_1}{2} y + \tfrac{2 \pi \chi(n_1+n_2)}{2} x \\
\tfrac{d}{dt} z &=& -\Gamma_1 (z+\eta) - i \tfrac{2 \pi \chi(n_1-n_2)}{2} \eta \\
\tfrac{d}{dt} \eta &=& - i \tfrac{2 \pi \chi(n_1-n_2)}{2} z \; .
\end{eqnarray*}
After one period $T$, since the peaks in the comb are exactly separated by the dispersive shift $\chi$, the effect of the precession at a frequency $2\pi\chi(n_1+n_2)/2$ is canceled out (modulo a possible phase flip every period when $n_1+n_2$ is odd). Note that the infinite comb approximation differs from the usual rotating wave approximation that would lead to a similar disabling of the precession  for $2 \pi \chi \gg \Gamma_1$. We then obtain, in the above coordinates, the matrix expression 
\begin{equation}
\begin{array}{ll}
\mathcal{K}_1 = 
(-1)^{n_1+n_2} \times \\ 
\left( \begin{array}{cccc}
e^{-\Gamma_1 T/2} & 0 & 0 & 0 \\
0 & e^{-\Gamma_1 T/2} & 0 & 0 \\
0 & 0 &  (e^{-\Gamma_1 T} - G) &   (e^{-\Gamma_1 T} -1 - G) \\
0 & 0 & G & (G+1)
\end{array}\right)
\end{array}
\end{equation}

with $\;\; G=\frac{-i \pi \chi(n_1-n_2)(1-e^{-\Gamma_1T})}{\Gamma_1+i 2 \pi \chi(n_1-n_2)} \; .$
Besides, the Rabi rotation corresponds to
\begin{equation}
  \mathcal{K}_0 = 
\left( \begin{array}{cccc}
1 & 0 & 0 & 0 \\
0 & \cos\theta & -\sin\theta & 0 \\
0 & \sin\theta & \cos\theta & 0 \\
0 & 0 & 0 & 1
\end{array}\right) \; .  
\end{equation}
This expression further simplifies for the two drive strengths $\Omega=\chi/2+k\chi$ or $\Omega=k\chi$, which respectively lead to $\theta=0$ (no qubit emission) and $\theta =\pi$ (maximal qubit emission), since the $(z,\eta)$ variables of interest decouple from $(x,y)$.
\begin{itemize}
\item For $\theta=0$, we tend to a stationary regime $(z,\eta)_{kT + T} = (z,\eta)_{kT}$. Note that these are the values associated to a coherence between Fock states $n_1$ and $n_2$ of the storage mode. This steady value thus confirms the intuition developed in the single qubit case: there is no change in the coherences between Fock states in the resonator, which means that no measurement is performed (minima in $\Gamma_\mathrm{d,s}$ in Fig.~\ref{fig:MID_Storage_cal}d).
\item For $\theta=\pi$, we can compute an analytical expression for the factor $R$ by which the trace $\eta$ decreases every period $T$ in the permanent regime \footnote{We find two solutions where $z/\eta$ is stationary, one stable and one unstable. Considering the stable solution only, we compute $R$ such that $\eta(kT+T)=R\, \eta(kT)$. The computation boils down to solving order-two polynomials.}. Thus the average decay rate is given by $-\log(|R|)/T$ with
\begin{widetext}
\begin{equation}
\begin{aligned}
R & =  \frac{1}{2(\Gamma_1 + i 2 \pi \chi(n_1-n_2))} \left( \Gamma_1(1-e^{-\Gamma_1\,T}) + 2 \sqrt{B} \left({\textstyle\sqrt{\sqrt{1+A^2}+A} + i \sqrt{\sqrt{1+A^2}-A}} \right) \right) \\
&  \text{ where } A=\frac{\Gamma_1^2(1+e^{-\Gamma_1 T})^2 - 4(2 \pi\chi)^2 (n_1-n_2)^2 e^{-\Gamma_1 T}}{8 B}
\\& \phantom{\text{ where }}  B= 2 \pi \chi (n_1-n_2) \Gamma_1 e^{-\Gamma_1 T} \\
& \text{ and we recall } T = 1 / \chi \; .
\end{aligned}
\end{equation}
\end{widetext}

We now analyze this last expression.
\end{itemize}

\subsubsection{Optimal decoherence rate of the $n_1,n_2$ component}

\begin{figure}
\includegraphics[width=\columnwidth]{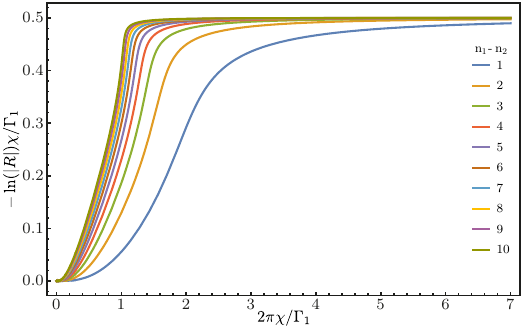}
\caption{Decoherence rate between Fock states $|n_1\rangle$ and $|n_2\rangle$ in units of $\Gamma_1$ for a driving amplitude $\Omega=\chi/2$ as a function of $2\pi\chi/\Gamma_1$ and for various values of $n_1-n_2$.}\label{fig:Opti3}
\end{figure}

The decoherence rate $-\log(|R|)\chi$ can be calculated for any value of the dispersive shift $\chi$ and of the photon numbers $n_1$ and $n_2$. In Fig.~\ref{fig:Opti3} are shown the rates corresponding to several values of $n_1-n_2$ for a driving amplitude $\Omega=\chi/2$ (maximal measurement strength). As expected, the decoherence is stronger when the photon numbers are further apart. Besides, for $2\pi\chi|n_1-n_2|\gg \Gamma_1$ the rate saturates to $\Gamma_1/2$ similarly to the emission rate of the single qubit.


\subsection{Upper bound on the total measurement rate as a function of photon number}
 \label{sec:best_MID}

Beyond the determination of the decoherence rate between two Fock states, we are interested in the maximal information extraction rate about the storage state in the multiplexed measurement scheme. In particular, we will discuss how this maximal total measurement rate depends on the maximum number of photon $N_\mathrm{max}$ that is probed by the multiplexed scheme. In the following, we assume a perfect measurement apparatus, giving us access to all the information extracted by the measurement process, which is not necessarily the case of heterodyne measurement on each peak of the comb. In Sec.~\ref{sec:constantmeastime}, we propose such a measurement apparatus. 

We assume the number resolved regime $2\pi\chi \gg \Gamma_2,\Gamma_1$. Thus the decoherence rate between two Fock states is independent of the Fock state numbers and is equal to $\Gamma_1/2$ (see Fig.~\ref{fig:Opti3}). In the following, we show that under these assumptions, the total measurement rate does not depend on $N_\mathrm{max}$.

Since the multiplexed measurement operates by entangling the storage mode with $N_\mathrm{max}+1$ frequency modes of the transmission line, we can describe the system and the extraction of information without the multiplexing qubit and only consider its effect, which is the entanglement operation. To each Fock state $|n\rangle$ in the storage mode ($0\leq n \leq N_\mathrm{max}$) is associated one out of the $N_\mathrm{max}+1$ modes of the transmission line at $f_\mathrm{mp}-n\chi$. Every mode is driven so that at the input it is in a coherent state (it can even be the vacuum as in the gedanken experiment). At the output, if we change of reference frame by displacing the outgoing modes $\hat{a}_{\mathrm{out},n}$ by the opposite of the input coherent state, a single mode will be excited and all the non resonant modes will be in the vacuum state. 
Therefore, any quantum state of the outgoing modes can be expressed as a superposition of $N_\mathrm{max}+2$ states only. States $|n\rangle_m$ correspond to all modes in the vacuum except the one at frequency $f_\mathrm{mp}-n\chi$ and $|\perp\rangle_m$ is the vacuum state.




Thus we can describe the system using two modes only: the storage mode and a simplified measurement mode. The storage mode is described using the Fock state basis $\{ |n\rangle_\mathrm{s} \}_{0\leq n\leq N_\mathrm{max}}$. The measurement mode has the $N_\mathrm{max}+2$ states discussed above. 
The bipartite system starts in the state 
\begin{equation}
    |\Psi(0)\rangle =  |\Psi_\mathrm{storage}\rangle_s \otimes |\Psi_\mathrm{meas}\rangle_m =  \left( \sum_{n=0}^{N_\mathrm{max}}  \psi_n |n\rangle_\mathrm{s} \right) \otimes |\perp\rangle_\mathrm{m}.
\end{equation}
After a measurement time $t$, the storage mode and the measurement mode become entangled and the readout of the measurement mode extracts information about the storage photon number. As the decoherence rate between every storage Fock state pair is $\Gamma_1/2$, one can write the state of the bipartite system as 
\begin{eqnarray}
    \begin{array}{rcl}
      |\Psi(t)\rangle &=&  \sqrt{e^{-\Gamma_1 t/2}}|\Psi(0)\rangle \\
      &&+ \sqrt{1-e^{-\Gamma_1 t/2}} \sum_{n=0}^{N_\mathrm{max}}  \psi_n |n\rangle_\mathrm{s}\otimes|n\rangle_\mathrm{m}.
    \end{array}
\end{eqnarray}
\noindent As expected, if we trace over the measurement mode, the diagonal of the density matrix of the storage mode remains unchanged while the off-diagonal terms decrease at a rate $\Gamma_1/2$.

\begin{figure}
    \centering
    \includegraphics[width=\columnwidth]{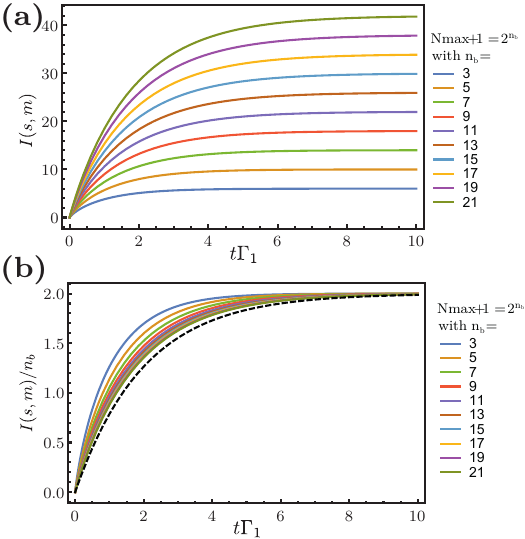}
    \caption{Mutual information between the storage and measurement modes for various maximum photon number $N_\mathrm{max}=2^{n_b}-1$ as a function of measurement time. a. Mutual information in bits. b. Mutual information divided by the number of bits $n_b$  as a function of time. The dashed black curve is the lower bound.}
    \label{fig:mutualinfo}
\end{figure}

One can calculate the mutual information~\cite{Chuang2010} of the bipartite system in order to obtain the information extraction rate and study how it scales with $N_\mathrm{max}$. The mutual information is defined as 
\begin{equation}
    I(s,m) = S(\rho_\mathrm{s})+S(\rho_\mathrm{m}) - S(\rho_\mathrm{s,m}),
\end{equation}
\noindent where $S(\rho)= -\mathrm{Tr}(\rho \log_2(\rho)) $ is the Von Neumann entropy, $\rho_\mathrm{s,m}$ is the bipartite density matrix and $\rho_\mathrm{s}$ ( respectively $\rho_\mathrm{m}$) is the density matrix of the storage mode ( resp. measurement mode) obtained by tracing out on the other mode. If the state of the bipartite system is pure then its entropy of is zero and the entropy of the storage mode is equal to the entropy of the measurement mode. Thus the mutual information can be written as
\begin{equation}
    I(s,m) = 2 S(\rho_\mathrm{s}).
\end{equation}
The mutual information is twice the amount of information that we can acquire about the storage mode using the measurement mode. Thus it is twice the amount of information extracted by the measurement process. 
The density matrix $\rho_\mathrm{s}$ of the storage mode depends on the initial photon number distribution. As our goal is to measure the probability to have $n$ photons for all $n$ simultaneously, we consider a initially uniform photon number distribution, such as when $\psi_n=(N_\mathrm{max}+1)^{-1/2}$ for all $n$. For this photon number distribution, the storage density matrix becomes
\begin{equation}
    \rho_\mathrm{s} = \dfrac{e^{-\Gamma_1 t/2}}{N_\mathrm{max}+1} \sum_{n,l=0}^{N_\mathrm{max}} |n\rangle_\mathrm{s}\langle l|_\mathrm{s} + \dfrac{1-e^{-\Gamma_1 t/2}}{N_\mathrm{max}+1}\sum_{n=0}^{N_\mathrm{max}} |n\rangle_\mathrm{s}\langle n|_\mathrm{s}.
\end{equation}
The eigenvalues of $\rho_\mathrm{s}$ are $e^{-\Gamma_1 t/2}+(1-e^{-\Gamma_1 t/2})/(N_\mathrm{max}+1)$ with degeneracy 1 and $(1-e^{-\Gamma_1 t/2})/(N_\mathrm{max}+1)$ with degeneracy $N_\mathrm{max}$. Thus one can derive the mutual information of the bipartite system
\begin{equation}
    \begin{array}{rcl}
I(s,m) & = & -2\left(r+\dfrac{1-r}{N_\mathrm{max}+1}\right) \log_2\left(r+\dfrac{1-r}{N_\mathrm{max}+1}\right) \\
&&-2 \dfrac{N_\mathrm{max}(1-r)}{N_\mathrm{max}+1}\log_2\left(\dfrac{1-r}{N_\mathrm{max}+1}\right),
    \end{array}
\end{equation}
     
\noindent with $r = e^{-\Gamma_1 t/2}$. The time evolution of the mutual information $I(s,m)$ is shown in Fig.~\ref{fig:mutualinfo}. At short times, the mutual information increases with time $t$ at a speed which depends on $N_\mathrm{max}+1$. As the mutual information is the information we extract out of the system, the time derivative of the mutual information at short times gives the rate at which information is extracted (e.g. the total measurement rate). In order to determine how the total measurement rate depends on the maximum photon number $N_\mathrm{max}$, we look at the mutual information per bit of information $n_b=\mathrm{log}_2(N_\mathrm{max}+1)$. The mutual information per bit decreases with $n_b$ for small photon numbers but converges to a lower bound when $n_b$ goes to infinity 
\begin{equation}
    \lim\limits_{n_b \to +\infty} \dfrac{I(s,m)}{n_b} = 2(1-e^{-\Gamma_1 t/2}).
\end{equation}
The total measurement time is of the order of the number of bits $n_b$ we have to measure divided by the total measurement rate in bits per second. As the mutual information per bit is always bigger than $2(1-r)$, the total measurement time will be at least about $1/\Gamma_1$. 
Because the mutual information per bit is always bigger than the the lower bound $2(1-r)$ which does not depend on $n_b$, the total measurement time for the multiplexing protocol is independent of the maximum photon number $N_\mathrm{max}$. Fig.~\ref{fig:mutualinfo}b shows the mutual information per bit as a function of time and its lower bound. 


\section{Measurement bandwidth}

The asymptotic former results are possible as long as the total measurement bandwidth increases with the photon number. We consider two main limitations to the maximal number of photons that can be measured.
First, above a certain number of photons, the qubit frequency overlaps the transition between the first and second excited states of the transmon. Indeed the $|e\rangle$ to $|f\rangle$ transition at zero photon in the resonator becomes resonant with the qubit $|g\rangle$ to $|e\rangle$ transition for $\chi_\mathrm{mp,mp}/\chi_{s,mp}$ photons, which complicates the analysis. In our experiment, it would set a limit to about 20 photons. However, this limit can be bypassed using a qubit with a much larger anharmonicity, such as a fluxonium qubit. 

Second, the higher order terms in the Hamiltonian tend to reduce the cross-Kerr rate $\chi_\mathrm{s,mp}$ when the photon number increases. Beyond some critical photon number $n_\mathrm{crit}$, the dispersive shift $\chi_\mathrm{s,mp}$ will be smaller than the decoherence rate of the qubit, which will escape from the number resolved regime. This limit occurs around 1000 photons in our device, which is therefore not the dominant limitation. Indeed one can show that
\begin{equation}
    \chi_\mathrm{s,mp} = p_\mathrm{mp}p_\mathrm{s}\dfrac{h f_\mathrm{mp} f_\mathrm{s}}{4 E_J}
\end{equation}
\begin{equation}
    \chi_{s,s,mp} = p_\mathrm{mp}p_\mathrm{s}^2\dfrac{h^2 f_\mathrm{mp} f_\mathrm{s}^2}{32 E_J^2}
\end{equation}
where $\chi_{s,s,mp}$ is the decrease of the cross Kerr rate $\chi_\mathrm{s,mp}$ when a photon is added to the storage mode, $E_J$ is the junction energy of the multiplexing qubit and $p_\mathrm{mp/s}$ is the fraction of the energy of the multiplexing qubit mode (resp. the storage mode) stored in the junction of the multiplexing qubit \cite{Minev2020}. Thus the critical photon number $n_\mathrm{crit}$ in the storage mode is given by
\begin{equation}
    \chi_\mathrm{s,mp}-n_\mathrm{crit}\chi_\mathrm{s,s,mp} = \Gamma_\mathrm{2,mp}
\end{equation}
which leads to 
\begin{equation}
    n_\mathrm{crit} = \left(1-\dfrac{\Gamma_\mathrm{2,mp}}{\chi_\mathrm{s,mp}} \right)\dfrac{8 E_J}{p_\mathrm{s} h f_\mathrm{s}}
\end{equation}
and a upper bound of $8 E_J/(p_\mathrm{s} h f_\mathrm{s}) \approx 10^3$ photons for $E_J/h\approx15\mathrm{~GHz}$, $p_\mathrm{s} \approx 10^{-2}$ and $f_\mathrm{s}\approx 5 \mathrm{~GHz}$.

Therefore, we do not believe that these limitations are hard enough to prevent the multiplexing scheme to count very large photon numbers.

\section{Modeling of the measurement operators}
 \label{sec:measurementoperators}
  
We introduce here a simple model to characterize the measurement and its backaction on the resonator. The measurement uses a phase preserving amplifier in order to amplify the signal at each frequency $f_\mathrm{mp}-k\chi_\mathrm{s,mp}$ in the comb and record a complex amplitude $I_m^{(k)}+iQ_m^{(k)}$. We assume that each of these measurement records only extracts the information on the occupation of the Fock state $|k\rangle$, which is experimentally valid in the limit $2 \pi \chi_\mathrm{s,mp} \gg \Gamma_\mathrm{2,mp}$. Without decoherence and in the limit of long measurement time, its backaction on the storage mode would project the storage state on Fock state $|k\rangle$ or on the complementary subspace $\Pi^{(k)}_\perp\mathcal{H}_s$, where $\Pi^{(k)}_\perp=\mathbf{1}-|k\rangle\langle k|$ and $\mathcal{H}_s$ is the Hilbert space of the storage resonator.

In practice, the measurement proceeds by first entangling the resonator, which is in a state $|\psi\rangle$, and the signal mode of the phase preserving amplifier. For the measurement channel $k$, the entangled state reads
$|\alpha,0\rangle\otimes \Pi^{(k)}|\psi\rangle + |\alpha_\perp,0 \rangle \otimes \Pi^{(k)}_\perp|\psi\rangle,$
where  $\Pi^{(k)}=|k\rangle\langle k|$, and states denoted as $|\alpha,\beta\rangle$ are the coherent states of the signal and the idler modes at the input of the amplifier. We distinguish two cases: the case where the probe is resonant with the multiplexing qubit, leading to a reflected amplitude $\alpha$, and the case where it is off resonant leading to a reflected amplitude $\alpha_\perp$. The resonance frequency of the qubit depends on the number of photons in the resonator so that the reflected amplitude $\alpha$ indicates $k$ photons while $\alpha_\perp$ indicates that there are not $k$ photons. For an incoming amplitude $\alpha_\mathrm{in}$ onto the multiplexing qubit, we get
\begin{equation}\left\{
    \begin{aligned}
        &\alpha_\perp  = \alpha_\mathrm{in}\\
        &\alpha = \alpha_\mathrm{in}\left(1-\frac{\Gamma_{1,mp}}{\pi\Omega} \frac{\langle\sigma_\mathrm{-,mp}\rangle_\mathrm{ss}}{i} \right)
    \end{aligned}
    \right.
\end{equation}
where $ \langle\sigma_\mathrm{-,mp}\rangle_\mathrm{ss}$ is the steady state mean value of the multiplexing qubit lowering operator.

  If the qubit is driven by a single tone, the maximum of $| \langle \sigma _\mathrm{-,mp} \rangle _\mathrm{ss}|$ is reached for $2\pi\Omega = \Gamma_\mathrm{1,mp}/\sqrt{2}$. However, in the case of a qubit driven by an infinite frequency comb, the time average of the lowering operator is
  \begin{equation}
        \langle\sigma_\mathrm{-,mp}\rangle_\mathrm{comb} = i\overline{y} \dfrac{1-e^{-\Gamma_\mathrm{1,mp}/2\chi_\mathrm{s,mp}}}{\Gamma_\mathrm{1,mp}/\chi_\mathrm{s,mp}} 
  \end{equation}
  
 \noindent with $\overline{y}$ the steady state solution defined in Eq.~(\ref{eq:ybar}). The fraction on the right corresponds to the average emission between two jumps in the time domain version of the comb.

 The measurement operator (Kraus operator) corresponding to the heterodyne detection of a propagating field encoding the information on the $|k \rangle$ state thus reads
 \begin{equation}
 \begin{array}{rcl}
M^{(k)}(I^{(k)}_m,Q^{(k)}_m) &=& 
     \langle\Psi_\mathrm{I^{(k)}_m,Q^{(k)}_m}|\alpha,0\rangle \otimes \Pi^{(k)}\\
     &&+\langle\Psi_\mathrm{I^{(k)}_m,Q^{(k)}_m}|\alpha_\perp,0\rangle \otimes \Pi^{(k)}_\perp
 \end{array}
 \end{equation}
 where $|\Psi_\mathrm{I^{(k)}_m,Q^{(k)}_m}\rangle$ is the state on which the propagating field is projected after the heterodyne measurement performed by the phase preserving amplifier followed by a heterodyne detection setup.
 
 
 Following the supplementary information of Ref.~\cite{Hatridge2013}, in the case of a phase preserving amplifier the inner product $\xi(\beta,I_m,Q_m)=\langle\Psi_\mathrm{I_m,Q_m}|\beta,0\rangle$ is given up to a global phase factor (independent on $\beta$, $I_m$ and $Q_m$) by
 \begin{equation}
 \begin{array}{rcl}
\xi(\beta,I_m,Q_m)&=&\frac{1}{\sqrt{\pi}2\sigma_0}e^{-\dfrac{|\beta|^2}{2(2\sigma_0)^2}}  \\
&& \times e^{-\dfrac{(I_m-\beta)^2}{2(2\sigma_0)^2}}  e^{-\dfrac{(Q_m+i\beta)^2}{2(2\sigma_0)^2}}  
 \end{array}
 \end{equation}
  \noindent where $\sigma_0$ is the amplitude of the zero-point fluctuations (the variance of the measured $I_m$ is $2\sigma_0^2$ in the quantum limit of phase preserving amplification).
   Therefore, we finally get the following analytical expression of the measurement operators for each channel $k$, in the case of $\Gamma_\mathrm{2,mp} \ll 2\pi \chi_\mathrm{s,mp}$
   \begin{widetext}
  \begin{equation}
  \begin{array}{rcl}
M^{(k)}(I^{(k)}_m,Q^{(k)}_m)& =& \frac{1}{\sqrt{\pi}2\sigma_0}e^{-\dfrac{|\alpha_\perp|^2}{2(2\sigma_0)^2}} e^{-\dfrac{(I^{(k)}_m-\alpha_\perp)^2}{2(2\sigma_0)^2}}  e^{-\dfrac{(Q^{(k)}_m+i \alpha_\perp)^2}{2(2\sigma_0)^2}} \Pi_\perp^{(k)}\\& &+ \frac{1}{\sqrt{\pi}2\sigma_0}e^{-\dfrac{|\alpha|^2}{2(2\sigma_0)^2}} e^{-\dfrac{(I^{(k)}_m- \alpha)^2}{2(2\sigma_0)^2}}  e^{-\dfrac{(Q^{(k)}_m+i \alpha)^2}{2(2\sigma_0)^2}} \Pi^{(k)}.
  \end{array}
 \end{equation}  
   \end{widetext}

\section{How fast can the multiplexed measurement be?}
 \label{sec:measurement_time}

In principle, multiplexing the measurement enables to determine the photon number in a constant time, no matter the maximal number of photons one wants to resolve. Note that this measurement rate is not in contradiction with the amount of information that would be contained in a qubit. Indeed, in the present scheme, we are using the two-level system not as a memory, but rather as a device whose frequency is changed by the Fock state to be measured. We are thus somehow replacing a communication channel faithfully sending a qubit state, by a communication channel faithfully sending a two-level system in one out of many propagating microwave modes. In Sec.~\ref{sec:constantmeastime}, we showed how to obtain the photon number in a time set by the lifetime of the qubit. Here, we investigate the measurement time in the case of our experiment.

 \subsection{Multiplexing with quadrature measurements}
 \label{sec:multiplexed_time}

The gedanken experiment requires components that are out of reach with current technologies. Yet, the experiment we perform and present in the main text implements a similar experiment that replaces the multiplexer and photodetector array by signal processing of propagating microwave modes.
 
 It is out of scope of this work to derive an exact expression of the photon number measurement time in our experiment. However, we can determine whether the measurement time depends on the maximum photon number $N_\mathrm{max}$ one wants to measure. In contrast with the photodetectors, quadrature measurements are inherently noisy. Identifying the photon number in the storage mode consists in determining which channel contains an amplitude $\alpha$ while all the others contain an amplitude $\alpha_\perp$ (see Sec.~\ref{sec:measurementoperators}). The measurement records $\{I^{(k)}_m,Q^{(k)}_m\}_k$ are stochastic processes centered on $\alpha_\perp$ (except for one value of $k$, where it is centered on $\alpha$). Determining the photon number $k$ comes down to discriminating which record is centered on $\alpha$ only using the ensemble of noisy records $\{I^{(k)}_m,Q^{(k)}_m\}_k$. 

After a measurement time $t$, the measurement records $\{I^{(k)}_m,Q^{(k)}_m\}_k$ are averaged along a duration $t$. Thus the time averaged intrinsic noise contained in the measurement records scales as $1/\sqrt{t}$ and the time average value is independent on $t$. The problem can be mapped onto the following game. $N_\mathrm{max}$ stochastic variables $\{u_i\}_{1\leq i\leq N_\mathrm{max}}$ are each randomly chosen using a Gaussian distribution centered on $0$ with a width $1/\sqrt{t}$. Another stochastic variable  $u_0$ is  randomly chosen using a Gaussian distribution centered on $1$ with a width $1/\sqrt{t}$. The list $\{u_i\}_{0\leq i\leq N_\mathrm{max}}$ is scrambled randomly into a list $l$ and the goal consists in identifying the variable $u_0$ using only the list $l$. The optimal strategy is to pick the highest element of the list $l$. The probability to make an error and lose the game is then given by the probability that the maximum of the $\{u_i\}_{1\leq i\leq N_\mathrm{max}}$ are higher than $u_0$
 \begin{equation}
     \mathcal{P}_\mathrm{error} = \mathcal{P}\left(\max_{N_\mathrm{max}\geq i\geq 1}(u_i) > u_0 \right).
 \end{equation}
We can rescale all the distribution by $\sqrt{t}$, thus $u_0$ are chosen randomly using a Gaussian distribution centered on $\sqrt{t}$ with a width of 1 and each of the $\{u_i\}_{1\leq i\leq N_\mathrm{max}}$ using a Gaussian distribution centered on 0 with a width of 1. One can show that the mean of the maximum of $\{u_i\}_{1\leq i\leq N_\mathrm{max}}$ tends towards $\sqrt{2 \ln(N_\mathrm{max})}$~\cite{DasGupta2014}. 
 \begin{equation}
     \mathrm{mean}\left[\max_{N_\mathrm{max}\geq i\geq 1}(u_i)\right]\sim\sqrt{2 \ln(N_\mathrm{max})}
 \end{equation}
 Besides, the median of the max of $\{u_i\}_{1\leq i\leq N_\mathrm{max}}$ is equal to the mean value within an error scaling as $1/\sqrt{N_\mathrm{max}}$:
 \begin{equation}
\begin{array}{rcl}
\mathrm{median}\left[\max_{N_\mathrm{max}\geq i\geq 1}(u_i)\right]&=&\mathrm{mean}\left[\max_{N_\mathrm{max}\geq i\geq 1}(u_i)\right]\\
&&+O\left(1/\sqrt{N_\mathrm{max}}\right).
\end{array}
 \end{equation}
 Since the error probability is between $1/4$ and $3/4$ if the median of $u_0$ is equal to the median of the maximum of $\{u_i\}_{1\leq i\leq N_\mathrm{max}}$, it leads to
 \begin{equation}
 \begin{array}{cc}
1/4<\mathcal{P}_\mathrm{error}<3/4 \Rightarrow \sqrt{t} \sim \sqrt{2\log{N_\mathrm{max}}} \\\Rightarrow t \sim \log(N_\mathrm{max})
 \end{array}
 \end{equation}
 
 From this expression, we understand that the measurement time for a fixed error probability scales as $\log(N_\mathrm{max})$. 
 
 \subsection{Comparison between measurement schemes}
  \label{sec:comparison}

 In this section we compare the following various photon number measurement schemes using a qubit of frequency $f_q$ that is dispersively coupled to a storage mode of frequency $f_s$. We assume the cross Kerr rate $\chi$ between the storage mode and the qubit to be greater than the decoherence rate of the qubit $\Gamma_2$. The goal is to measure the photon number $N$ assuming it is smaller than $N_\mathrm{max}$.
 
 
 
 \textbf{Sequential brute force}
 
 The brute force approach consists in measuring whether or not there are $k$ photons in the storage mode for all possible values of $k$ from $0$ to $N_\mathrm{max}$~\cite{Johnson2010}. For each $k = 0,1,2,3, ...$, we apply a photon number conditional $\pi$ pulse to the qubit at frequency $f_q - k\chi$ so that the qubit is  excited only if there are $k$ photons in the storage mode. Reading out the qubit state gives the answer to the question 'Are there $k$ photons?'. The full measurement stops as soon as this binary answer is positive so that it takes a time given by $(T_\pi+T_\mathrm{ro})(N+1)$. The time $T_\pi$ is the time of an conditional $\pi$ pulse, hence it is at least $1/\chi$, while the qubit readout time $T_\mathrm{ro}$ is limited by other parameters in order to get a single shot readout.

\onecolumngrid

\begin{table}[]
\begin{tabular}{|c|c|c|c|c|c|}
\hline
Protocol                                                                        & $t_\mathrm{meas}\propto$                                       & complexity                                                                                     & check $N>N_\mathrm{max}$ & qubit role & error propagation \\ \hline
\begin{tabular}[c]{@{}c@{}}Sequential \\ brute force\end{tabular}               & $(N+1)/\chi$                                                   & $N+1$  gates                                                                                   & yes                      & pointer    & yes               \\ \hline
\begin{tabular}[c]{@{}c@{}}Passive\\ decimation\end{tabular}                    & $N\mathrm{max}/\chi$                                           & \begin{tabular}[c]{@{}c@{}}$N_\mathrm{max}^2$ gates\\ and complex analysis\end{tabular}        & no                       & pointer    & no                \\ \hline
\begin{tabular}[c]{@{}c@{}}Binary code\\ feedback\end{tabular}                  & $T_\mathrm{fb}\log_2(N_\mathrm{max}+1)$                        & feedback                                                                                       & no                       & pointer    & yes               \\ \hline
\begin{tabular}[c]{@{}c@{}}Binary code\\ optimal control\end{tabular}           & $( T_\mathrm{reset} + 1/\chi ) \log_2(N_\mathrm{max}+1)$ & optimal control                                                                                & no                       & pointer    & no                \\ \hline
\begin{tabular}[c]{@{}c@{}}Multiplexed\\ quadrature \\ measurement\end{tabular} & $\ln(N_\mathrm{max})/\chi$                                     & \begin{tabular}[c]{@{}c@{}}relies on near\\ quantum limited\\ broadband amplifier\end{tabular} & yes                      & encoder    & no                \\ \hline
\begin{tabular}[c]{@{}c@{}}Gedanken\\ multiplexed\\ measurement\end{tabular}    & $1/\chi$                                                       & hardware to develop                                                                            & yes                      & encoder    & gedanken!         \\ \hline
\end{tabular}
\caption{Protocols for photocounting using a qubit}
\end{table}
\null
\twocolumngrid

 \textbf{Passive photon number decimation using weak measurement}

This approach, which was implemented with Rydberg atoms in cavity~\cite{Guerlin2007}, consists in encoding the photon number in the phase of the qubit by waiting a time $1/(N_\mathrm{max}\chi)$ after the qubit has been prepared in state $(|g\rangle+|e\rangle)/\sqrt{2}$. The protocol is composed of a series of $p$ sequences, where each sequence encodes the photon number into the phase of the qubit and realizes a $\pi/2$ pulse on the qubit with a phase $2\pi p/N_\mathrm{max}$ followed by a qubit readout. Using generalized measurement theory, one infers the probability that the cavity is in a given Fock state.

After $p$ qubit readouts the variance on the photon number is $\sigma=N_\mathrm{max}/(\sqrt{p}\pi)$ (see appendix A in Ref.~\cite{PeaudecerfThesis}). Therefore, the required number of repetitions to get a fixed error probability on the photon number scales as $p\propto N_\mathrm{max}^2$. Since each measurement takes at least $1/(N_\mathrm{max}\chi)$, the total measurement time scales at least as $N_\mathrm{max}/\chi$.
 
 \textbf{Active photon number decimation}
 
 The previous protocol can be improved by optimizing the phase of the final $\pi/2$ pulse to maximize the amount of information extracted on the cavity photon number. It was realized in Ref.~\cite{Peaudecerf2014} using Rydberg atoms in cavity. Because of the use of feedback on a weak measurement, we could not find a closed form for the measurement time in this case~\cite{PeaudecerfThesis}. However it was shown that the total time is larger than the total time taken by a binary decimation with feedback (see below).

 
  \textbf{Binary decimation with feedback} 
  
This method was shown to provide the least number of steps for sequential photocounting~\cite{Haroche1992}. Each step consists in applying an unconditional $\pi/2$ pulse to the qubit, wait a time $1/2^{k+1}\chi$, apply a new unconditional $\pi/2$ pulse with a phase $\phi_k$ that encodes the least significant\footnote{here least significant is to be understood as the last bit in the binary decomposition, and not in terms of amount of information} $k^\mathrm{th}$ bit $b_k$ of the photon number $N=\sum_k b_k 2^k$ into the qubit state. Importantly, the phase $\phi_k$ depends on the results of the $k-1$ former measurements. The sequence needs to be repeated $p = \log_2(N_\mathrm{max}+1)$ times with $k$ going from $0$ to $p-1$. This procedure was recently implemented in Ref.~\cite{Dassonneville2020}.

The measurement time is at least given by the sum of the total interaction time between qubit and cavity and of the total feedback latency. The total interaction time is bounded by $\sum_p 1/(2^{p+1}\chi)=1/\chi$. However the feedback latency scales as $p$ and can be written as $T_\mathrm{fb}\log_2(N_\mathrm{max}+1)$.
 
\textbf{Binary decimation with optimal pulse control} 

  An optimal binary decimation can also be implemented without using a feedback loop by measuring a series of generalized parity operators which yields the bit values of the binary decomposition of the photon number in the storage mode. The $k^\mathrm{th}$ generalized parity measurement consists in an optimal pulse that excites the qubit conditioned on the value of the $k^\mathrm{th}$ bit. The $p= \log_2(N_\mathrm{max}+1)$ parity measurements are performed in a time sequence. A subsequent measurement and dynamic reset of the qubit state completes the sequence~\cite{Wang2020,Curtis2020}. Such an optimal pulse can only be performed in a time of the order of the dispersive interaction time $1/\chi$. It leads to a total measurement time scaling as $\log_2(N_\mathrm{max}+1)(1/\chi+T_\mathrm{reset})$ where $T_\mathrm{reset}$ is the duration of the active reset protocol.

  In the table, we provide a summary of the various advantages and drawbacks of the photocounting methods. No time sequence measurement is able to provide a measurement time that does not depend on the photon number. Using a multiplexed quantum measurement thus enables a qualitative improvement on the measurement time. In practice, this drastic improvement in the scaling with $N_\mathrm{max}$ requires a detection setup that is out of reach currently. Our experiment demonstrates that multiplexing is possible using an heterodyne detection setup instead. The scaling of the measurement time is then in $\ln(N_\mathrm{max})$ as in state-of-the-art sequential measurements. Besides, we deport the complexity of optimal control or feedback into the challenge to reach large measurement efficiencies on a large frequency band (many $\chi$'s).

\section{Master Equation simulations}
In this section, we briefly describe the master equation simulations used to understand our experimental results. We simulated the main photon counting experiments with both qubits as well as the photon number calibration of the storage mode, and the dephasing rate induced by the multiplexed measurement of the storage mode.

All simulations were performed using Python package \emph{QuTiP}~\cite{Johansson2013}. We simulated the complete system composed of the storage mode, the yes-no qubit and the multiplexing qubit with all couplings, except in the simulation of the measurement induced dephasing rate for which we only took into account the storage mode and the multiplexing qubit. The storage mode was modeled as an harmonic oscillator while the transmons were replaced by two level systems. The Hilbert space of the storage mode was truncated at a photon number ranging from $10$ to $25$ photons depending on the simulation. In this section we will use Pauli matrices to describe operators acting on qubits.

\subsection{Photocounting simulations}
\label{sec:Photon_counting_sim}
\subsubsection{Photocounting with the yes-no qubit}
Both photon counting approaches are simulated in a very similar manner. The first simulation (yes-no simulation) describes the use of conditional operations on the yes-no qubit. This experiment serves as a calibration of the number of photons in the storage mode and of all relevant parameters. This experiment starts with a displacement of the storage mode followed by a conditional $\pi$ pulse on the yes-no qubit at frequency $f_\mathrm{yn}-\delta\!f_\mathrm{yn}$ before detecting the expectation value of the Pauli operator $\sigma_\mathrm{z,yn}$.

    
    
    
    
    
    

We write the Hamiltonian of the system in a frame rotating at $f_\mathrm{s}-\chi_\mathrm{s,mp}/2-\chi_\mathrm{s,yn}/2$ for storage mode, $f_\mathrm{yn}-\delta\!f_\mathrm{yn}$ for yes-no qubit mode and $f_\mathrm{mp}$ for multiplexing qubit mode as follows
\begin{equation}
    \begin{array}{rcl}
  \hat{H_1}/h &=& \delta\!f_{\mathrm{yn}} \dfrac{\hat{\sigma}_\mathrm{z,yn}}{2}
    -  \chi_\mathrm{s,yn} \hat{n}_\mathrm{s}\dfrac{\hat{\sigma}_\mathrm{z,yn}}{2} -  \chi_\mathrm{s,mp} \hat{n}_\mathrm{s}\dfrac{\hat{\sigma}_\mathrm{z,mp}}{2} \\&&- \chi_\mathrm{s,s} \hat{n}_\mathrm{s}(\hat{n}_\mathrm{s}-1) 
    - \chi_\mathrm{s,s,yn} \hat{n}_\mathrm{s} (\hat{n}_\mathrm{s}-1) \dfrac{\hat{\sigma}_\mathrm{z,yn}}{2} \\&&- \chi_\mathrm{s,s,mp} \hat{n}_\mathrm{s} (\hat{n}_\mathrm{s}-1) \dfrac{\hat{\sigma}_\mathrm{z,mp}}{2} + \frac{\epsilon_\mathrm{yn}(t)}{h} \hat{\sigma}_\mathrm{x,yn} 
    \\&&+ \frac{\lambda(t)}{2 \pi} (\epsilon_\mathrm{max}e^{i\pi(\chi_\mathrm{s,mp}+\chi_\mathrm{s,yn})t} \hat{a}_\mathbf{s}\\&&+\epsilon_\mathrm{max}^\ast e^{-i\pi(\chi_\mathrm{s,mp}+\chi_\mathrm{s,yn})t} \hat{a}^\dag_\mathbf{s})
\end{array},
\label{hamiltonianH_1}
\end{equation}where $\lambda(t)$ is a Gaussian function with duration $100 \mathrm{~ns}$, width $25 \mathrm{~ns}$ and a maximum of 1 so that the storage mode displacement pulse reads $\epsilon_\mathrm{s}(t)=\lambda(t)\epsilon_\mathrm{max}$ and $\epsilon_\mathrm{yn}(t)$ is the time envelope of a Gaussian pulse with duration $1.9 \mathrm{~\mu s}$ and width $475 \mathrm{~ns}$. The amplitude of the pulse is chosen to obtain a $\pi$ rotation on the yes-no qubit. The term $- \delta\!f_{\mathrm{yn}} \dfrac{\hat{\sigma}_\mathrm{z,yn}}{2}$ takes into account the detuning between the $\pi$ pulse and the yes-no qubit frequency. $\epsilon_\mathrm{yn}(t)$ is delayed with respect to $\lambda(t)$ to match the experimental pulse sequence. In comparison with Hamiltonian (Eq.~(1) in Methods), this simulation adds cross-Kerr interactions between each qubit and the storage mode, a self-Kerr term on the storage mode but it does not take into account the readout resonator.

In addition to the Hamiltonian (\ref{hamiltonianH_1}), we supply the solver with eight collapse operators to simulate the dynamics of the following master equation
\begin{equation}
    \begin{array}{rcl}
        \dot{\rho} &=& - \dfrac{i}{\hbar} [\hat{H_1},\rho] + 2 \Gamma_\mathrm{\phi,s}\mathcal{L}(\hat{n}_\mathrm{s})\rho
        \\&&+ (1+n_\mathrm{th,s})\Gamma_\mathrm{1,s}\mathcal{L}(\hat{a}_\mathrm{s})\rho 
        + n_\mathrm{th,s}\Gamma_\mathrm{1,s}\mathcal{L}(\hat{a}^\dag_\mathrm{s})\rho
        \\&&+ \dfrac{1}{2} \Gamma_\mathrm{\phi,yn}\mathcal{L}(\hat{\sigma}_\mathrm{z,yn})\rho+  \Gamma_\mathrm{1,yn}\mathcal{L}(\hat{\sigma}^-_\mathrm{yn})\rho \\&&+ \dfrac{1}{2} \Gamma_\mathrm{\phi,mp}\mathcal{L}(\hat{\sigma}_\mathrm{z,mp})\rho + \Gamma_\mathrm{1,mp}\mathcal{L}(\hat{\sigma}^-_\mathrm{mp})\rho
    \end{array},
    \label{eq:Lindblad_simu_photon_counting}
\end{equation}
with $n_\mathrm{th,s}$ the expectation values of $\hat{n}_\mathrm{s}$ when the system is at rest due to thermal occupation. All decoherence and relaxation rates are measured using previously explained calibration.

The master equation is solved using the function ``mesolve" of QuTiP starting from a thermal state with $n_\mathrm{th,s}$ photon in storage mode, the yes-no qubit in the ground state $|g\rangle$ and the multiplexing qubit also in the ground state $|g\rangle$. The solver iteratively computes the density matrix with a $10 \mathrm{~ns}$ time step during the displacement pulse and the $\pi$ pulse. We compute the expectation value $\langle\hat{\sigma}_\mathrm{z,yn}\rangle$ at the end of the sequence and convert it into a probability $\mathbb{P}_e$ of finding the yes-no qubit in the $|e\rangle$ state.

This simulation can be used to reproduce the experiment in Fig.~2A,B of the main text by adjusting the following parameters $\left\{ \mu=\epsilon_\mathrm{max}/V_\mathrm{max,s}, \chi_\mathrm{s,yn}, \chi_\mathrm{s,s}, \chi_\mathrm{s,s,yn}, n_\mathrm{th,s} \right\}$. Note that we need to run the simulation for every couple of parameters ($V_\mathrm{max,s}$,$\delta\!f_\mathrm{yn}$). Table~\ref{tab:fit_parameter_sim} compiles the values of fitted parameters.

\subsubsection{Photocounting with the multiplexing qubit}
A second simulation (fluorescence simulation) was carried out to compare the photon counting experiment in Fig.~3A,B using a single drive on the multiplexing qubit with theory. This experiment also starts with a storage mode displacement but it is followed by a $2 \mathrm{~\mu s}$ Gaussian pulse on the multiplexing qubit at the frequency $f_\mathrm{mp}-\delta\!f_\mathrm{mp}$ with an amplitude expressed as a Rabi frequency $\Omega=\chi_\mathrm{s,mp}/4$. The measured reflection coefficient of the multiplexing qubit $r(\delta\!f_\mathrm{mp})$ can be expressed using input-output theory as~\cite{Cottet2018}

\begin{equation}
    \begin{array}{rcl}
r(\delta\!f_\mathrm{mp})& =& \frac{\langle a_\mathrm{out}\rangle}{\langle a_\mathrm{in}\rangle}=\frac{\langle a_\mathrm{in}\rangle-\sqrt{\Gamma_\mathrm{1,mp}}\langle\sigma_{-,\mathrm{mp}}\rangle}{\langle a_\mathrm{in}\rangle}\\&=& 1-\frac{\sqrt{\Gamma_\mathrm{1,mp}}}{\langle a_\mathrm{in}\rangle} \langle \hat{\sigma}_\mathrm{-,mp}\rangle.
    \end{array}
\end{equation}

And since the Rabi frequency is given by $\Omega=\sqrt{\Gamma_\mathrm{1,mp}}|\langle a_\mathrm{in}\rangle|/\pi$ we get an emission coefficient
\[ 1-\mathcal{R}e\left(r(\delta\!f_\mathrm{mp})\right)=\frac{\Gamma_\mathrm{1,mp}}{\pi\Omega} \mathcal{R}e\left(e^{-i\arg(\langle a_\mathrm{in}\rangle)} \langle\hat{\sigma}_\mathrm{-,mp}\rangle\right)\]
in the frame rotating at $f_\mathrm{mp}-\delta\!f_\mathrm{mp}$ for the multiplexing qubit. If we set the phase of the drive so that $i\langle a_\mathrm{in}\rangle\geq 0$, meaning we drive the qubit along $\sigma_\mathrm{x,mp}$, the emission coefficient becomes
\[ 1-\mathcal{R}e\left(r(\delta\!f_\mathrm{mp})\right)=\frac{\Gamma_\mathrm{1,mp}}{2\pi\Omega} \langle\hat{\sigma}_\mathrm{y,mp}\rangle\]

The Hamiltonian of the problem in the frame rotating at $f_\mathrm{s}-\chi_\mathrm{s,mp}/2-\chi_\mathrm{s,yn}/2$ for storage mode, $f_\mathrm{yn}$ for yes-no qubit and $f_\mathrm{mp}-\delta\!f_\mathrm{mp}$ for multiplexing qubit reads
\begin{equation}
    \begin{array}{rcl}
  \hat{H_2}/h &=&  \delta\!f_{\mathrm{mp}} \dfrac{\hat{\sigma}_\mathrm{z,mp}}{2}
    - \chi_\mathrm{s,yn} \hat{n}_\mathrm{s}\dfrac{\hat{\sigma}_\mathrm{z,yn}}{2} - \chi_\mathrm{s,mp} \hat{n}_\mathrm{s}\dfrac{\hat{\sigma}_\mathrm{z,mp}}{2} \\
    &&- \chi_\mathrm{s,s} \hat{n}_\mathrm{s}(\hat{n}_\mathrm{s}-1)
    -\chi_\mathrm{s,s,yn} \hat{n}_\mathrm{s} (\hat{n}_\mathrm{s}-1) \dfrac{\hat{\sigma}_\mathrm{z,yn}}{2} \\
    &&- \chi_\mathrm{s,s,mp} \hat{n}_\mathrm{s} (\hat{n}_\mathrm{s}-1) \dfrac{\hat{\sigma}_\mathrm{z,mp}}{2} + \dfrac{ \Omega}{2}\epsilon_\mathrm{mp}(t) \hat{\sigma}_\mathrm{x,mp} \\
    &&+ \frac{\lambda(t)}{2 \pi} (\epsilon_\mathrm{max}e^{i\pi(\chi_\mathrm{s,mp}+\chi_\mathrm{s,yn})t} \hat{a}_\mathbf{s}\\&&+\epsilon_\mathrm{max}^\ast e^{-i\pi(\chi_\mathrm{s,mp}+\chi_\mathrm{s,yn})t} \hat{a}^\dag_\mathbf{s})
\end{array},
\end{equation}
where $\epsilon_\mathrm{mp}(t)\geq 0$ is a Gaussian function of duration 2 $\mu$s, width 250 ns and amplitude 1. $\epsilon_\mathrm{mp}(t)$ is delayed compare to $\lambda(t)$ to reproduce the experimental pulse sequence. We add to this Hamiltonian the same relaxation and decoherence channels as for the yes-no simulation (see Eq.~(\ref{eq:Lindblad_simu_photon_counting})) for which the decoherence and relaxation rates were measured independently. The resulting master equation only differs from the yes-no simulation by the Rabi drive that addresses the multiplexing qubit instead of the yes-no qubit. The master equation is solved using the "mesolve" function of QuTiP with a time step of $5.25 \mathrm{~ns}$ starting from a thermal state with $n_\mathrm{th,s}$ photons for storage and the yes-no qubit and the multiplexing qubit in the ground state $|g\rangle$. Finally, the expectation value $\langle\hat{\sigma}_\mathrm{y,mp}\rangle$ is computed and integrated during the $2~\mathrm{\mu s}$ of the pulse.

We compare the measured emission coefficient in Fig.~2B,D to the simulated signal $A \langle\hat{\sigma}_\mathrm{y,mp}\rangle$ where $A$ is left as a free parameter due to a small parasitic reflection in the measurement setup and thermal population. The parameters $\left\{ \mu, \chi_\mathrm{s,t}, \chi_\mathrm{s,s}, \chi_\mathrm{s,s,t}, n_\mathrm{th,s} \right\}$ is already set by the calibration above using the simulation of the yes-no qubit. From the fluorescence simulation, we thus extract the parameters $\left\{\chi_\mathrm{s,mp}, \chi_\mathrm{s,s,mp}, A \right\}$ by comparing the experimental observations in Fig.~2B,D with the simulation for various $V_\mathrm{max,s}$ and $\delta\!f_\mathrm{mp}$. Fitted values are given in Tab.~\ref{tab:fit_parameter_sim}. Finally, we ran the yes-no simulation again taking into account the updated  multiplexing qubit parameters. As expected only small changes in the results of the yes-no qubit simulation are observed.

\begin{table}[h!]
    \centering
    \begin{tabular}{|c|c|}
    \hline
    parameter & fitted values\\
    \hline
    $\mu$& 1.45 (mV~$\mu$s)$^{-1}$ \\
    $\chi_\mathrm{s,yn}$& 1.42 MHz\\
    $\chi_\mathrm{s,mp}$& 4.9 MHz \\
    $\chi_\mathrm{s,s}$& -0.02 MHz\\
    $\chi_\mathrm{s,s,yn}$& -0.003 MHz \\
    $\chi_\mathrm{s,s,mp}$& -0.08 MHz \\
    $n_\mathrm{th,s}$& 0.03 \\
    \hline
    \end{tabular}
    \caption{\textbf{Parameters extracted from the photocounting simulations using the multiplexing or yes-no qubit.} All parameters except those related to the multiplexing qubit are determined using a fit of the yes-no qubit simulation to the Fig.~2A,C. Parameters related to multiplexing qubit are obtained using a fit of the simulation to the Fig.~2B,D.}
    \label{tab:fit_parameter_sim}
\end{table}

\subsection{Evolution of the average photon number in the storage mode}
\label{sec:photon_number_sim}

We simulated the filling of the storage mode by a displacement pulse on the resonator. We simulated the same master equation used for the photocounting simulations with parameters obtained from the photocounting simulations (see Tab.~\ref{tab:fit_parameter_sim}) but without applying any drive on the qubits. Only the displacement pulse on the storage mode is modeled i.e.  $\epsilon_\mathrm{mp}(t)=0$, $\delta\!f_\mathrm{mp}=0$,  $\epsilon_\mathrm{yn}(t)=0$, and $\delta\!f_\mathrm{yn}=0$.

The "mesolve" function of QuTiP computes the density matrix with a time step of $10~\mathrm{ns}$ and returns the mean number of photons in the storage mode at the end of the displacement pulse for various drive amplitudes. Fig.~\ref{fig:photon_number} shows the square root of the expected mean photon number as a function of the amplitude $\epsilon_\mathrm{max}$. We obtain a scaling factor $\sqrt{\left< n_\mathrm{s} \right>} = 85.9~\mathrm{V^{-1}} V_\mathrm{max,s}$ used in the photon number calibration of the storage mode.

\subsection{Simulation of multiplexed readout}
 \label{sec:sim_multiplexing}

In this subsection, we simulate how a frequency comb reflects off the multiplexing qubit. We write the Hamiltonian in the frame rotating at  $f_\mathrm{s}-\chi_\mathrm{s,mp}/2-\chi_\mathrm{s,yn}/2$ for the storage mode and at the qubit frequencies for the qubits as
\begin{equation}
    \begin{aligned}
  \hat{H_3}/h = & - \chi_\mathrm{s,yn} \hat{n}_\mathrm{s}\dfrac{\hat{\sigma}_\mathrm{z,yn}}{2} - \chi_\mathrm{s,mp} \hat{n}_\mathrm{s}\dfrac{\hat{\sigma}_\mathrm{z,mp}}{2} - \chi_\mathrm{s,s} \hat{n}_\mathrm{s}(\hat{n}_\mathrm{s}-1) \\& - \chi_\mathrm{s,s,yn} \hat{n}_\mathrm{s} (\hat{n}_\mathrm{s}-1) \dfrac{\hat{\sigma}_\mathrm{z,yn}}{2} 
  - \chi_\mathrm{s,s,mp} \hat{n}_\mathrm{s} (\hat{n}_\mathrm{s}-1) \dfrac{\hat{\sigma}_\mathrm{z,mp}}{2} \\& + \dfrac{\Omega}{2}(\epsilon_\mathrm{comb}(t) \hat{\sigma}^+_\mathrm{mp}+\epsilon_\mathrm{comb}^\ast(t) \hat{\sigma}^-_\mathrm{mp})  \\& + \frac{\lambda(t)}{2 \pi} (\epsilon_\mathrm{max}e^{i\pi(\chi_\mathrm{s,mp}+\chi_\mathrm{s,yn})t} \hat{a}_\mathbf{s}\\&+\epsilon_\mathrm{max}^\ast e^{-i\pi(\chi_\mathrm{s,mp}+\chi_\mathrm{s,yn})t} \hat{a}^\dag_\mathbf{s})
\end{aligned},
\end{equation}where $\Omega=\chi_\mathrm{s,mp}/2$ and $\epsilon_\mathrm{comb}(t)$ is the product of a Gaussian function with the sum of nine complex exponential $\sum_{k=0}^8 \mathrm{exp}(2i\pi\chi_\mathrm{s,mp} k t)$. The Gaussian envelope of $\epsilon_\mathrm{comb}(t)$ has a duration of $2 ~\mu s$, a width of $250\mathrm{~ns}$, and a maximum amplitude of $1$ and the delay between $\epsilon_\mathrm{comb}(t)$ and $\lambda(t)$ reproduces the experimental sequence. The master equation (\ref{eq:Lindblad_simu_photon_counting}) is used with a time step of $1 \mathrm{~ns}$ for various amplitude $\epsilon_\mathrm{max}$. We obtain the time evolution of $\langle\sigma_{y,mp}\rangle$ enabling us to compare the experimental measurements of Fig.~2E to the model. To do so, we integrate the simulated function $\langle\sigma_{y,mp}\rangle \times \mathrm{cos}(2 \pi \chi_\mathrm{s,mp} k t)$ for each integer $k$, similarly to the demultiplexing processing we perform on the multiplexed experimental signal. Note that, in the case $k=0$, we need to divide the integral by 2 in order to perform a proper demultiplexing. By combining this simulation with the photon number calibration, we get the expected values of the 9 multiplexing readout signals as a function of the mean number of photon in the storage mode used in Fig.~2E.

\subsection{Simulation of measurement induced dephasing on the storage mode}
 \label{sec:sim_MID}

\begin{figure}
    \centering
    \includegraphics[width=\columnwidth]{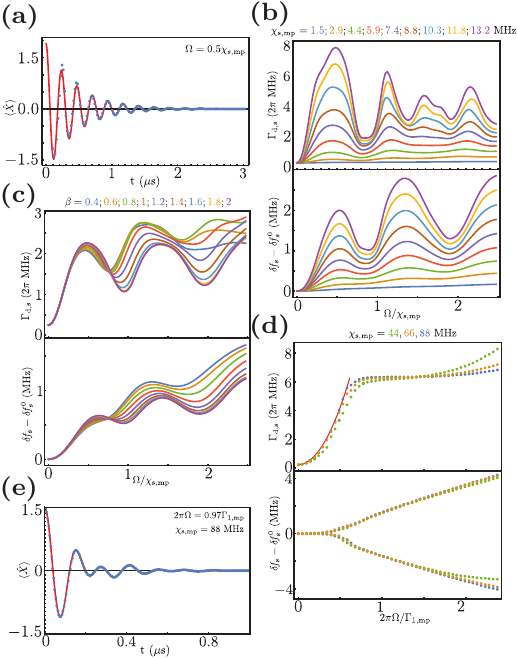}
    \caption{\textbf{Simulations of the measurement induced dephasing rate and of the ac-Stark shift induced by a frequency comb. a.} Ramsey-like oscillations of the storage mode for $\Omega =\chi_{s,mp}/2$ and an initial coherent field amplitude $\beta=-1.55$. Blue dots are the simulated expectation values of $\hat{X}$ and red line is the theory given by the Eq.~(\ref{eq:XPoftime}). \textbf{b.} Simulated measurement induced dephasing rate $\Gamma_\mathrm{d,s}$ and ac-Stark shift as a function of $\Omega/\chi_{s,mp}$ for various values of $\chi_\mathrm{s,mp}$. Simulations show the same pattern with maxima and minima for some specific values of $\Omega/\chi_{s,mp}$ as in the experiment in Fig.~3C. \textbf{c.} Simulated measurement induced dephasing rate and ac-Stark shift as a function of $\Omega/\chi_{s,mp}$ for various initial coherent state amplitudes $\beta$ in the storage mode. For $\Omega/\chi_{s,mp}>1$, we see a difference of about $20~\%$ between $\beta = 1.6$ and $\beta=1.2$. \textbf{d.} Simulations for $\chi_\mathrm{s,mp}\gg \Gamma_\mathrm{1,mp}$. The evolution of the measurement induced dephasing and ac-Stark shift with  $\Omega$ is different compared to the case of \textbf{b}. The evolution of the measurement induced dephasing rate and the ac-Stark shift seems to be given by the ratio $2\pi\Omega/ \Gamma_\mathrm{1,mp}$. The red line is a guide for eyes representing a quadratic function. It shows that the measurement induced dephasing rate increases linearly with $\Omega^2$ for small drive amplitudes. For the ac-Stark shift, on the contrary with \textbf{b}, the detuning is constant at the small drive amplitudes, then two frequencies appear with comparable contributions to the Ramsey oscillations. The two frequencies evolve linearly with $\Omega$. \textbf{e.} Example of simulated Ramsey oscillations exhibiting two frequencies.}
    \label{fig:simu_MID}
\end{figure}

In this part, we only simulate the multiplexing qubit and the storage mode to decrease the computational cost of the simulation. The Hamiltonian of the simulation in the frame rotating at the multiplexing qubit resonant frequency and at $f_\mathrm{s}+\delta\!f_\mathrm{s}^0$ for the storage mode is
\begin{equation}
\begin{aligned}
    \hat{H_4} /h= &-\chi_\mathrm{s,mp}\dfrac{\hat{\sigma}_\mathrm{z,mp}+\mathbbm{1}}{2}\hat{n}_\mathrm{s}- \delta\!f_\mathrm{s}^0\hat{n}_\mathrm{s}\\&-\chi_\mathrm{s,s,mp}\hat{n}_\mathrm{s}(\hat{n}_\mathrm{s}-1)\dfrac{\hat{\sigma}_\mathrm{z,mp}+\mathbbm{1}}{2}\\&+\dfrac{\Omega}{2}(\epsilon_\mathrm{comb}(t) \hat{\sigma}^+_\mathrm{mp}+\epsilon_\mathrm{comb}^\ast(t) \hat{\sigma}^-_\mathrm{mp}) ,
\end{aligned}
\end{equation}
where $\epsilon_\mathrm{comb}(t)$ is the product of a Gaussian function with the sum of nine complex exponential $\sum_{k=0}^8 \mathrm{exp}( 2i\pi\chi_\mathrm{s,mp} k \tau)$. The width of the Gaussian function is equal to one quarter of the duration $t$ of the pulse. We add four dephasing and relaxation channels to this Hamiltonian to obtain the master equation 
\begin{equation}
\begin{aligned}
        \dot{\rho} = &- \dfrac{i}{\hbar} [\hat{H_4},\rho] + 2 \Gamma_\mathrm{\phi,s}\mathcal{L}(\hat{n}_\mathrm{s})\rho+ \Gamma_\mathrm{1,s}\mathcal{L}(\hat{a}_\mathrm{s})\rho \\&+ \dfrac{1}{2} \Gamma_\mathrm{\phi,mp}\mathcal{L}(\hat{\sigma}_\mathrm{z,mp})\rho + \Gamma_\mathrm{1,mp}\mathcal{L}(\hat{\sigma}^-_\mathrm{mp})\rho.
\end{aligned}
\label{eq:Lindblad_MID_simu}
\end{equation}
The storage is initialized in a coherent state of amplitude $\beta = 1.55$ and the multiplexing qubit is initialized in state $|g\rangle$. We simulate the dynamics of the system for a pulse duration $t$ going from $100\mathrm{~ns}$ to $5 \mathrm{~\mu s}$ and for $\Omega$ ranging from $0$ to $2\chi_\mathrm{s,mp}$. We compute the expectation value of $\hat{X} = (\hat{a}_\mathrm{s}+\hat{a}^\dag_\mathrm{s})/2$ at the end of each simulation. For a given $\Omega$, we extract the time evolution of $\langle\hat{X}\rangle$ under the influence of the multiplexed measurement as shown on Fig.~\ref{fig:simu_MID}a. This decaying sinusoid is fitted using Eq.~(\ref{eq:XPoftime}) to obtain the oscillation frequency $\delta\!f_\mathrm{s}$ and the decay rate $\Gamma_\mathrm{d,s}$. Fig.~\ref{fig:MID_Storage_cal}c and d show the measurement induced dephasing and ac-Stark shift as a function of amplitude of the comb $\Omega$ for two sets of coherent state amplitudes $\beta$ and detuning $\delta\!f_\mathrm{s}^0$.

We identify three interesting features. The first one is the evolution of the shape of the curves $\delta\!f_\mathrm{s}(\Omega)$ and $\Gamma_\mathrm{d,s}(\Omega)$ with $\chi_\mathrm{s,mp}$. We repeat the simulation using a square pulse envelope instead of Gaussian pulse for $\epsilon_\mathrm{comb}$ to make the simulation faster for several values of $\chi_\mathrm{s,mp}$ from $1.5$ to $13.2 \mathrm{~MHz}$ by steps of $1.4 \mathrm{~MHz}$. We observe that $\delta\!f_\mathrm{s}(\Omega)$ and $\Gamma_\mathrm{d,s}(\Omega)$ increase as $\chi_\mathrm{s,mp}$ becomes larger but that the maxima and minima of the curve are always found for the same $\Omega/\chi_\mathrm{s,mp}$ ratio (Fig.~\ref{fig:simu_MID}b) as predicted (see Sec.~\ref{sec:theory}).

The second observation is that $\delta\!f_\mathrm{s}(\Omega)$ and $\Gamma_\mathrm{d,s}(\Omega)$ vary with the initial coherent state amplitude $\beta$ (Fig.~\ref{fig:simu_MID}c). As the frequency comb contains only a finite number of frequencies ${f_\mathrm{mp}-n\chi_\mathrm{s,mp}}$ with $0\leq k\leq 8$, the decoherence rate between two Fock states, $|i\rangle$ and $|j\rangle$, depends on $i$ and $j$. As the probability distribution of photon numbers depends on the amplitude of the coherent state, the mean decoherence rate also depends on the coherent state amplitude.

The third observation is that in the regime $\chi_\mathrm{s,mp}\gg \Gamma_\mathrm{1,mp}/2\pi$, the dephasing rate and ac-Stark shift are a function of the ratio $2\pi\Omega/\Gamma_\mathrm{1,mp}$ as shown on Fig.~\ref{fig:simu_MID}d. The dephasing rate increases as $\Omega^2$ until a plateau is reached for $2\pi\Omega/\Gamma_\mathrm{1,mp}=0.7$, this plateau is predicted (see Sec.~\ref{sec:theory}). In contrast, the Stark shift is constant for $2\pi\Omega/\Gamma_\mathrm{1,mp}<0.3$ and splits into two frequencies (two oscillations on top of each other in Ramsey interferometry) with a splitting proportional to $2\pi\Omega/\Gamma_\mathrm{1,mp}$. Since there are two frequencies, we use Eq.~(\ref{eq:XPlargeM}) to fit the simulated Ramsey oscillations for $\chi_\mathrm{s,mp}\gg \Gamma_\mathrm{1,mp}/2\pi$. In practice Eq.~(\ref{eq:XPlargeM}) is a good fit function because a Fourier analysis shows that the signal is composed of two frequencies with the same amplitude. Fig.~\ref{fig:simu_MID}e shows an example of simulated Ramsey oscillations for $\chi_\mathrm{s,mp}\gg \Gamma_\mathrm{1,mp}/2\pi$.

\section{Density Matrix Elements}
 \label{sec:density_matrix}

In this part, we explain how one can calculate the density matrix of the storage mode from the measured Wigner function. It is the recipe we used to produce the bottom part of Fig.~3A in the main text. We further present original results on the decay of density matrix elements when the multiplexing qubit is driven by a single tone or by the comb of frequencies. We characterize the quantum non-demolition nature of our photocounter. Finally, we present an experiment in which we show revivals of density matrix elements as a function of time and show simulations that reproduce them qualitatively. We discuss a new quantity called the mean coherence and show its measured evolution in various measurement configurations.

\subsection{Density matrix reconstruction}
The Wigner tomography contains all the information about the state of the storage mode. We explain below how we reconstruct the density matrix from the measured Wigner function. We compute the Wigner map for every operator $|n\rangle\langle m|$ with $|n\rangle$ and $|m\rangle$ two fock states with $n$ and $m$ photons. The mean value of those operators is equal to the $(n,m)$ element $\rho_\mathrm{nm}$ of the density matrix. Using the mathematical expression of $\langle x|n\rangle$
\begin{equation}
   \langle x|n\rangle=\psi_n(x) = \left(\dfrac{2}{\pi} \right)^{1/4} \dfrac{1}{\sqrt{2^n n!}} \mathcal{H}_n(\sqrt{2} x) e^{-x^2}
\end{equation}
with $\mathcal{H}_n(x)=(-1)^n e^{x^2}\frac{d^n}{dx^n}e^{-x^2}$ the Hermite polynomial function of order $n$ and $|x\rangle$ the eigenvector of the quadrature $(\hat{a}_\mathrm{s}+\hat{a}^\dagger_\mathrm{s})/2$ associated to the eigenvalue $x$. The Wigner map of the operator $|n\rangle$ and $|m\rangle$ becomes
\begin{equation}
    W_{|n\rangle\langle m|}(x,p) = \dfrac{1}{\pi} \int \mathrm{d}y e^{-2ipy} \psi_n(x+y/2)\overline{\psi_m}(x-y/2)
\end{equation}
and the matrix element $\rho_\mathrm{nm}$ of the storage mode is given by Eq.~(\ref{eq:Wigner_expect_value})
\begin{equation}
    \rho_\mathrm{nm} = \pi \iint \mathrm{d}x\mathrm{d}p W_{|n\rangle \langle m|}(x,p) W_{\rho}(x,p).
    \label{eq:density_matrix}
\end{equation}

\subsection{Accessing the measurement induced dephasing rate}

In order to characterize the decoherence due to the multiplexed measurement, we use a renormalization of the density matrix elements in order to remove most of the effects of the storage mode relaxation. Let us now show that in the absence of Hamiltonian evolution and measurement back-action, the quantity $|\rho_\mathrm{nm}|/\sqrt{\rho_\mathrm{nn}\rho_\mathrm{mm}}$ evolves only because of dephasing and that its dynamics is not affected by relaxation. We consider the storage mode alone under the influence of its relaxation and dephasing channels in a frame rotating at $f_\mathrm{s}$
\begin{equation}
    \dot{\rho} = \Gamma_{1,s}\mathcal{L}(\hat{a}_\mathrm{s})\rho + 2\Gamma_{\phi,s}\mathcal{L}(\hat{a}_\mathrm{s}^\dag \hat{a}_\mathrm{s})\rho.
\end{equation}
From this equation, we can compute the time derivative of the density matrix element 
\begin{equation}
\begin{aligned}
    \dot{\rho}_\mathrm{nm} = &\Gamma_\mathrm{1,s} \left(\rho_\mathrm{n+1m+1} \sqrt{(n+1)(m+1)} - \dfrac{n+m}{2} \rho_\mathrm{nm}\right)\\& - \Gamma_\mathrm{\phi,s} \rho_\mathrm{nm}(n-m)^2.
\end{aligned}
\end{equation}
If the storage mode is initialized in a coherent state $|\alpha_o\rangle$, the solution of the equation is  
\begin{widetext}
\begin{equation}
    \rho_\mathrm{nm}(t) = e^{-|\alpha_o|^2e^{-\Gamma_\mathrm{1,s}t}} \dfrac{\alpha_o^m e^{-m\Gamma_\mathrm{1,s}t/2} (\alpha_o^\ast)^n e^{-n\Gamma_\mathrm{1,s}t/2}}{\sqrt{n! m!}} e^{-\Gamma_\mathrm{\phi,s}(n-m)^2t},
\end{equation}
\end{widetext}
and we get
\begin{equation}
    \dfrac{|\rho_\mathrm{nm}|}{\sqrt{\rho_\mathrm{nn}\rho_\mathrm{mm}}}(t) = e^{-\Gamma_\mathrm{\phi,s}(n-m)^2t}.\label{eq:decoherence_natdephasing}
\end{equation}
Thus indeed, the renormalization removes the effect of the relaxation rate $\Gamma_{1,s}$ and only characterizes the dephasing rate.
Under the action of measurement, the dephasing rate $\Gamma_\mathrm{\phi,s}$ is increased by the measurement induced dephasing rate. We will use this property in the following sections and study the evolution of $\rho_{mn}/\sqrt{\rho_{nn} \rho_{mm}}$.

\subsection{Decoherence of the storage mode induced by a single measurement drive}
\label{sec:dephasing_single_drive}
Measuring whether there are $n$ photons spoils the coherence of the superposition between Fock state
$|n\rangle$ and Fock state $|m\ne n\rangle$. We evidence this dephasing by observing the evolution of $\rho_{n m}$

We prepare the storage mode state in a coherent state with an amplitude $\beta = -1.7$, and probe the multiplexing qubit during a time $t$ with a drive at the frequency $f_\mathrm{mp}-\Delta_\mathrm{mp}$ before doing a Wigner tomography. For various times $t$ and detunings $\Delta_\mathrm{mp}$, we compute the density matrix of the storage mode using Eq.~(\ref{eq:density_matrix}). One can fit the time evolution of $|\rho_\mathrm{nm}|/\sqrt{\rho_\mathrm{nn}\rho_\mathrm{mm}}$ with a decreasing exponential function. The extracted decoherence rate $\Gamma_\mathrm{d,s}^{nm}(\Delta_\mathrm{mp})$ is then compared to the theoretical value.

\begin{figure}
    \centering
    \includegraphics[width=\columnwidth]{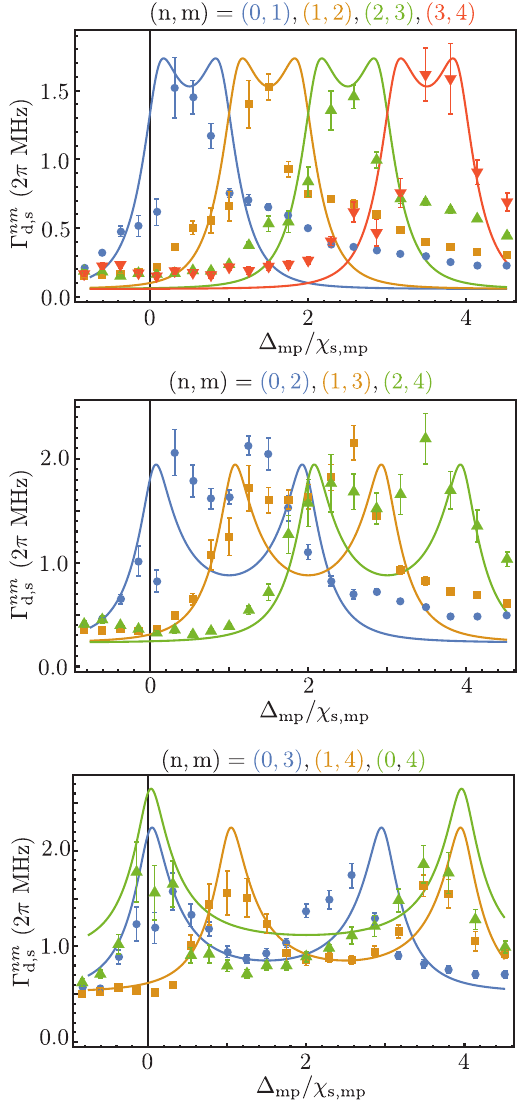}
    \caption{\textbf{Decoherence rate of superpositions between Fock states induced by a single drive on the multiplexing qubit.} In each panel, dots are obtained using Eq.~(\ref{eq:density_matrix}) on the measured Wigner function of the storage mode when driven by a single drive at $f_\mathrm{mp}-\Delta_\mathrm{mp}$ with an amplitude $\Omega=\chi_\mathrm{s,mp}/2$. Lines represent the highest eigenvalue of (\ref{eq:pmatrix}) without any free parameters. An offset equals to $\Gamma_\mathrm{\phi,s}(n-m)^2$, which is the intrinsic dephasing of the storage mode, is added to obtain the total dephasing rate.}
    \label{fig:single_drive}
\end{figure}

In Ref.~\cite{Sarlette2020a}, we show that an exact, infinite-order adiabatic elimination of the multiplexing qubit  probed with a single frequency drive is possible under the assumption that there is no photon loss in the storage mode. It shows that the decoherence rate between the Fock state $|n\rangle$ and $|m\rangle$ due to the measurement is given by the highest eigenvalue, which are all negatives, of the following matrix
\begin{widetext}
\begin{equation}
    \begin{pmatrix}
   -\Gamma_\mathrm{1,mp}/2 & -2\pi\Delta+\dfrac{n+m}{2}2\pi\chi_\mathrm{s,mp}&0&0 \\
   2\pi\Delta-\dfrac{n+m}{2}2\pi\chi_\mathrm{s,mp} & -\Gamma_\mathrm{1,mp}/2 & -2\pi\Omega&0\\
   0&2\pi\Omega&-\Gamma_\mathrm{1,mp}&-\Gamma_\mathrm{1,mp}-i\dfrac{n+m}{2}2\pi\chi_\mathrm{s,mp}\\
   0&0&-i\dfrac{n+m}{2}2\pi\chi_\mathrm{s,mp}&0
\end{pmatrix}.
\label{eq:pmatrix}
\end{equation}
\end{widetext}
\noindent Fig.~\ref{fig:single_drive} shows the measured density matrix decoherence rates $\Gamma_\mathrm{d,s}^{nm}$ and the above theory for $n$ and $m$ going from $0$ to $4$ (with an offset corresponding to the natural dephasing rate in Eq.~(\ref{eq:decoherence_natdephasing})). As expected in a regime with resolved resonance peaks $(2\pi \chi_\mathrm{s,mp} |m-n| > \Gamma_\mathrm{2,mp})$, the decoherence rate $\Gamma_\mathrm{d,s}^{nm}$ between Fock states $|n\rangle$ and $|m\rangle$ is larger when the single drive probes whether there are $n$ photons or $m$ photons with a moderate drive amplitude $\Omega$ (dependence on $\Omega$ not shown here). For much larger drive amplitude $\Omega$, one can increase the decoherence rate $\Gamma_\mathrm{d,s}^{nm}$ by driving with a detuning $\Delta_\mathrm{mp}=(n+m)\chi_\mathrm{s,mp}/2$, similarly to dispersive qubit readout which is optimal for information extraction at large drive power and for a drive frequency detuned by $\chi_\mathrm{s,mp}/2$. In fact this regime would become particularly attractive for poorly resolved resonances as a function of photon number $(2\pi \chi_\mathrm{s,mp} |m-n| < \Gamma_\mathrm{2,mp})$. Premises of this effect are visible on Fig.~\ref{fig:single_drive}, as the maximal decoherence rate occurs at a detuning slightly closer to $(n+m)\chi_\mathrm{s,mp}/2$, with a stronger effect for small $|m-n|$, both in theory and in the experimental observations. The small discrepancy between theory and experiment, in particular the asymmetry as a function of $n$ and $m$, may be explained by the photon loss rate of the storage mode, which is not captured in the simplified theoretical model.

\subsection{Multiplexed measurement vs single tone measurement}
\label{sec:density_vs_frequency}
In Fig.~3A of the main text, one sees that the dephasing of the storage mode induced by the measurement is stronger for multiplexed measurement than for single tone measurement. This conclusion is based on the Wigner tomography of the storage mode in three distinct cases. The storage mode is initialized in a coherent state of amplitude $\beta = -1.55$, then, before performing the tomography of the storage mode, we either (i) wait for a time $t$, (ii) probe the multiplexing qubit for a time $t$ at a single frequency $f_\mathrm{mp}-\chi_\mathrm{s,mp}$  corresponding to $1$ photon or (iii) with a frequency comb.

\begin{figure}
    \centering
    \includegraphics{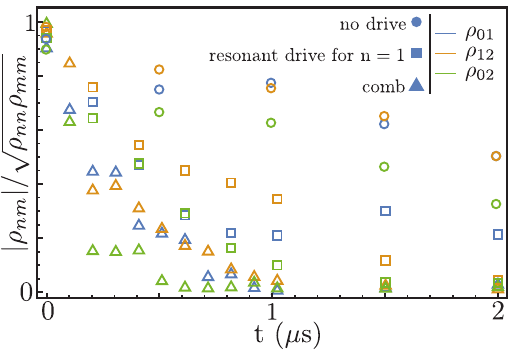}
    \caption{\textbf{Dynamics of the storage mode coherences under various measurement schemes.} Normalized off-diagonal elements of the density matrix extracted from the measured Wigner function of the storage mode as a function of time. The figure focuses on three elements $\rho_{01}$ (blue), $\rho_{12}$ (orange) and $\rho_{02}$ (green). Circles: case (i) without driving the multiplexing qubit. Squares: case (ii) where a single tone at $f_\mathrm{mp}-\chi_\mathrm{s,mp}$ drives the multiplexing qubit with a strength $\Omega=\chi_\mathrm{s,mp}/2$. Triangles: case (iii) where the multiplexing qubit is driven by a comb of 9 peaks with the same strength $\Omega$ each.}
    \label{fig:coherencevsmeas}
\end{figure}

From the measured Wigner functions, we compute the density matrix of the storage mode $\rho(t)$ for various times $t$ for the three cases and compare the evolution of the normalized elements $\rho_\mathrm{nm}(t)$ (see Fig.~\ref{fig:coherencevsmeas}). Without any drive on the multiplexing qubit (circles and case (i)), the density matrix elements decay due to natural dephasing only. Clearly, the drive on the multiplexing qubit induces a decay of the coherences, with a stronger effect when the comb is turned on than when a singe drive is turned on. We conclude that a multiplexing measurement extracts more information than a single measurement.

The effect on $\rho_{02}$ when probing with a resonant drive for $n=1$, is consistent with the significant measurement-induced detuning that can be read off the top right plot of Fig.~\ref{fig:single_drive} (blue, value $1$ on the horizontal axis). Apparently, when driving with a comb, such an effect combines with the ones on $n=0$ and $n=2$ resonances, and other components, to induce a stronger overall measurement rate.
We will investigate this comb effect more precisely in section~\ref{sec:revivals}.

\begin{figure}
    \centering
    \includegraphics[width=\columnwidth]{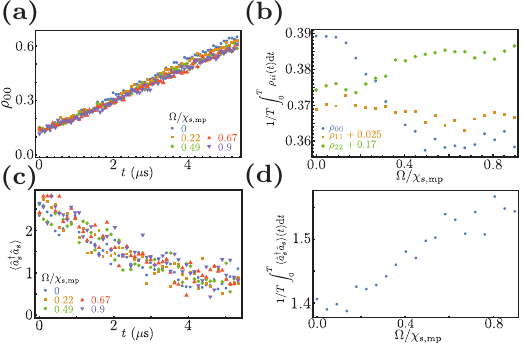}
    \caption{\textbf{Impact of multiplexed measurement on the occupation of the Fock states.} \textbf{a.} Measured probability to find the storage mode in Fock state $\left| 0 \right>$ as a function of time $t$ for various comb drive amplitudes $\Omega/\chi_\mathrm{s,mp}$. \textbf{b.} Measured diagonal elements of the density matrix integrated during $5\mathrm{~\mu s}$ as a function of drive amplitude $\Omega/\chi_\mathrm{s,mp}$. \textbf{c.} Decay of the average photon number in time for various drive amplitudes. \textbf{d.} Average photon number evolution as a function of $\Omega/\chi_\mathrm{s,mp}$.
    }
    \label{fig:QND}
\end{figure}

\subsection{Quantum Non Demolition nature of the multiplexed measurement}

The goal of this subsection is to quantify the \emph{Quantum Non Demolition} (QND) nature of our multiplexed measurement. A measurement is said to be QND if
\begin{itemize}
\item the measurement time is very short compared to the timescale of evolution of the system under study,
\item the interaction with the probe does not disturb the quantum state of the system if it belongs to the measurement basis.
\end{itemize}

If a photocounter is QND, the diagonal elements of the density matrix, i.e. the average Fock state populations, of the resonator are unchanged (on average on all measurements) by the measurement process. In our experiment, we observe that the diagonal elements of the density matrix in the energy basis evolve owing to the decay of the storage mode but do not strongly depend on the measurement strength. 
To be more accurate, we notice that for large probe amplitude (red and purple points in Fig.~\ref{fig:QND}a), the probability of finding the storage mode with $0$ photon is slightly lower. This dependence on the amplitude of the drive $\Omega$ is best characterized by extracting the populations (Fig.~\ref{fig:QND}b) and photon number (Fig.~\ref{fig:QND}d) integrated during $T=5 \mathrm{~\mu s}$ as a function of $\Omega/\chi_\mathrm{s,mp}$. For small drive amplitude $\Omega/\chi_\mathrm{s,mp}<0.1$, the probability to find a given number of photon does not change with $\Omega/\chi_\mathrm{s,mp}$ but for larger drive amplitude the resonator gets populated probably because of Zeno effect due to the non-Markovian environment originating from the multiplexing qubit.

In practice, for small drive amplitude and a measurement time of $5 \mathrm{~\mu s}$, the relaxation dynamics of the system during the measurement process increases the probability of having $0$ photon at the end of the measurement by approximately $10~\%$. We find that the mean photon number is decreased by the same percentage.

\subsection{Off-diagonal density matrix elements and revivals of the coherences}

\label{sec:revivals}
In the main text, we use Wigner tomography in order to observe Ramsey like oscillations of the storage mode. In fact, using Eq.~(\ref{eq:density_matrix}), the Wigner function allows us to visualize the dynamics of every off-diagonal elements of the storage density matrix to gain insight into the physics of the dephasing process.

\begin{figure}
    \centering
    \includegraphics[width=0.8\columnwidth]{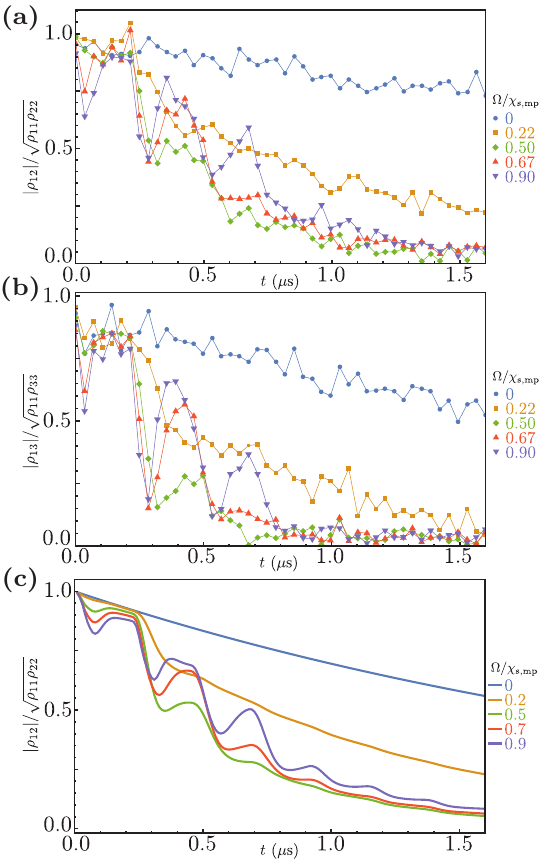}
    \caption{\textbf{Normalized off-diagonal elements of the storage matrix density. a.} Measured normalised coherence $|\rho_{12}|/\sqrt{\rho_{11} \rho_{22}}$ between Fock states $|1\rangle$ and $|2\rangle$ of the storage mode as a function of time and for various amplitudes $\Omega$ of the driving frequency comb. \textbf{b.} Similar plot for $|\rho_{13}|/\sqrt{\rho_{11} \rho_{33}}$. \textbf{c.} Results of the simulation of the master equation Eq.~(\ref{eq:Lindblad_MID_simu}).
    }
    \label{fig:off_diagonal_1}
\end{figure}

Fig.~\ref{fig:off_diagonal_1}a and b show the decay of off-diagonal elements $\rho_{12}$ and $\rho_{13}$ as a function of time. For small drive amplitudes $\Omega < 0.5 \chi_\mathrm{s,mp}$, off-diagonal elements decay faster when $\Omega$ is increased since more and more information is extracted per unit time by the drive. As larger drive amplitudes are reached, off-diagonal elements start oscillating. The contrast of these coherence revivals become more pronounced as the drive amplitude becomes larger and they exhibit a quasi periodicity. Note that the $10~\%$ deviation to exact periodicity may originate from the Gaussian pulse shaping of the comb. This behavior is qualitatively reproduced by our simulations (Fig.~\ref{fig:off_diagonal_1}c).

\begin{figure}
    \centering
    \includegraphics[width=0.8\columnwidth]{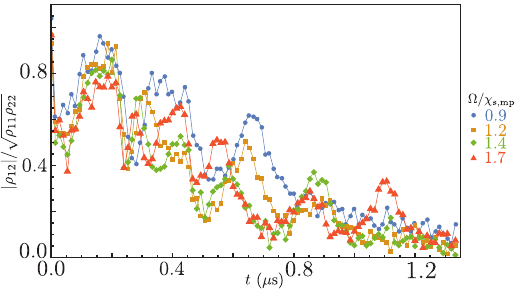}
    \caption{\textbf{Normalized off-diagonal elements of the storage matrix density for the largest measurement amplitudes.} Measured normalized coherence $|\rho_{12}|/\sqrt{\rho_{11} \rho_{22}}$ between Fock states $|1\rangle$ and $|2\rangle$ of the storage mode as a function of time and for various amplitudes $\Omega>0.9\chi_\mathrm{s,mp}$ of the driving frequency comb. }
    \label{fig:off_diagonal_2}
\end{figure}

\newpage
\subsection{Mean coherence between Fock states}

Since the driving frequency comb holds the promise to probe how many photons are in the storage mode, it should affect all coherences $\rho_{nm}$. In this section, we introduce two ways of characterizing the impact of the multiplexed photocounting on the global coherence of the storage mode.

The first one, shown in the main text in Fig.~3B, is the quadrature of the storage mode in the frame rotating at the frequency of this mode when the qubit is probed by a comb. It can be expressed as $\mathrm{Re}[(\langle\hat{X}\rangle+ i \langle\hat{P}\rangle)e^{-2i\pi \delta\!f_\mathrm{s} t}]$ with $\langle \hat{X}\rangle$ and $\langle \hat{P}\rangle$ the expectation values of the quadratures in the frame rotating at the frequency of the storage drive. This quantity is related to the first off-diagonal of the density matrix.

\begin{figure}[t]
    \centering
    \includegraphics[width=\columnwidth]{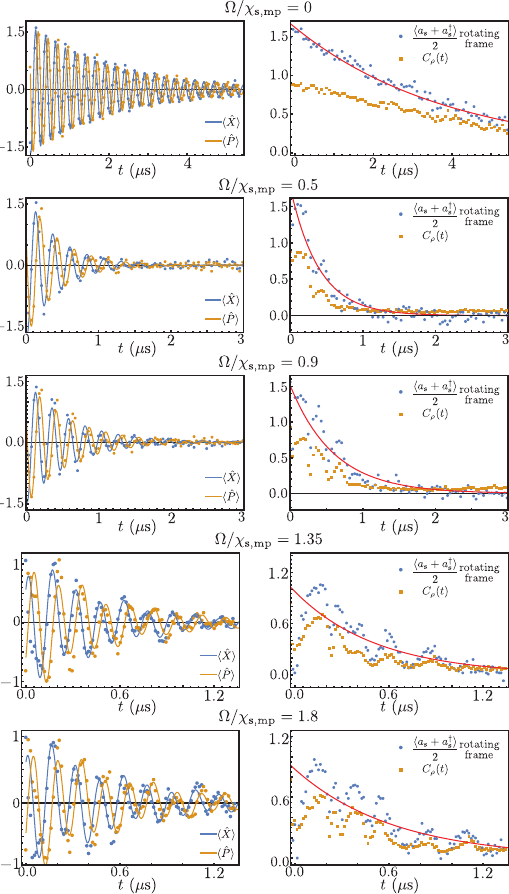}
    \caption{\textbf{Impact of the multiplexed drive on storage mode quadrature and mean coherence between Fock states.} Left column: Ramsey oscillations measured (dots) and fits (line) using Eq.~\ref{eq:XPoftime} and \ref{eq:XPlargeM} for five multiplexing drive amplitudes. Right column: quadrature of storage mode in the frame rotating at its resonance frequency (blue dots) and mean coherence between Fock states $ C_\rho (t)$ (orange squares) defined in Eq.~(\ref{eq:mean_coherence}) for five different drive amplitudes. Incertiture are about $0.07$ for the quadrature and $0.015$ for the mean coherence. The red solid lines are exponential fits of the quadrature decay.}
    \label{fig:Ramey_vs_density_matrix}
\end{figure}

We introduce a second quantity: the mean coherence $C_\rho$ between Fock states $0$ to $4$. It is defined as
\begin{center}
\begin{equation}
    C_\rho  = \underset{4\geq i>j\geq 0}{\mathrm{Mean}}\left[\dfrac{|\rho_{ij}|}{\sqrt{\rho_{ii}\rho_{jj}}}\right] .
    \label{eq:mean_coherence}
\end{equation}
\end{center}$C_{\rho}(t)$ contains the information about the dephasing between every different Fock states.
The left part of Fig.~\ref{fig:Ramey_vs_density_matrix} shows oscillations of the storage mode quadratures in the frame of the drive on the storage mode (for state preparation and Wigner tomography) for various multiplexing qubit drive amplitudes. On the right part of Fig.~\ref{fig:Ramey_vs_density_matrix}, we display the mean quadrature $\langle \hat{a}_\mathrm{s}+\hat{a}_\mathrm{s}^\dag \rangle=\mathrm{Re}[(\langle\hat{X}\rangle+ i \langle\hat{P}\rangle)e^{-2i\pi \delta\!f_\mathrm{s} t}]$ in the frame rotating at the storage mode frequency and the mean coherence $C_\rho$. Those two quantities show the same dynamics, leading to the same dephasing rate and both quantities can be used to characterize it. 
The revivals that can be seen on each of the density matrix off-diagonal elements (see Fig.~\ref{fig:off_diagonal_1}) also appear in the evolution of the quadrature and of the mean coherence between Fock states. 

\begin{thebibliography}{50}%
\makeatletter
\providecommand \@ifxundefined [1]{%
 \@ifx{#1\undefined}
}%
\providecommand \@ifnum [1]{%
 \ifnum #1\expandafter \@firstoftwo
 \else \expandafter \@secondoftwo
 \fi
}%
\providecommand \@ifx [1]{%
 \ifx #1\expandafter \@firstoftwo
 \else \expandafter \@secondoftwo
 \fi
}%
\providecommand \natexlab [1]{#1}%
\providecommand \enquote  [1]{``#1''}%
\providecommand \bibnamefont  [1]{#1}%
\providecommand \bibfnamefont [1]{#1}%
\providecommand \citenamefont [1]{#1}%
\providecommand \href@noop [0]{\@secondoftwo}%
\providecommand \href [0]{\begingroup \@sanitize@url \@href}%
\providecommand \@href[1]{\@@startlink{#1}\@@href}%
\providecommand \@@href[1]{\endgroup#1\@@endlink}%
\providecommand \@sanitize@url [0]{\catcode `\\12\catcode `\$12\catcode
  `\&12\catcode `\#12\catcode `\^12\catcode `\_12\catcode `\%12\relax}%
\providecommand \@@startlink[1]{}%
\providecommand \@@endlink[0]{}%
\providecommand \url  [0]{\begingroup\@sanitize@url \@url }%
\providecommand \@url [1]{\endgroup\@href {#1}{\urlprefix }}%
\providecommand \urlprefix  [0]{URL }%
\providecommand \Eprint [0]{\href }%
\providecommand \doibase [0]{https://doi.org/}%
\providecommand \selectlanguage [0]{\@gobble}%
\providecommand \bibinfo  [0]{\@secondoftwo}%
\providecommand \bibfield  [0]{\@secondoftwo}%
\providecommand \translation [1]{[#1]}%
\providecommand \BibitemOpen [0]{}%
\providecommand \bibitemStop [0]{}%
\providecommand \bibitemNoStop [0]{.\EOS\space}%
\providecommand \EOS [0]{\spacefactor3000\relax}%
\providecommand \BibitemShut  [1]{\csname bibitem#1\endcsname}%
\let\auto@bib@innerbib\@empty
\bibitem [{\citenamefont {Guerlin}\ \emph {et~al.}(2007)\citenamefont
  {Guerlin}, \citenamefont {Bernu}, \citenamefont {Deleglise}, \citenamefont
  {Sayrin}, \citenamefont {Gleyzes}, \citenamefont {Kuhr}, \citenamefont
  {Brune}, \citenamefont {Raimond},\ and\ \citenamefont
  {Haroche}}]{Guerlin2007}%
  \BibitemOpen
  \bibfield  {author} {\bibinfo {author} {\bibfnamefont {C.}~\bibnamefont
  {Guerlin}}, \bibinfo {author} {\bibfnamefont {J.}~\bibnamefont {Bernu}},
  \bibinfo {author} {\bibfnamefont {S.}~\bibnamefont {Deleglise}}, \bibinfo
  {author} {\bibfnamefont {C.}~\bibnamefont {Sayrin}}, \bibinfo {author}
  {\bibfnamefont {S.}~\bibnamefont {Gleyzes}}, \bibinfo {author} {\bibfnamefont
  {S.}~\bibnamefont {Kuhr}}, \bibinfo {author} {\bibfnamefont {M.}~\bibnamefont
  {Brune}}, \bibinfo {author} {\bibfnamefont {J.-M.}\ \bibnamefont {Raimond}},\
  and\ \bibinfo {author} {\bibfnamefont {S.}~\bibnamefont {Haroche}},\
  }\bibfield  {title} {\bibinfo {title} {{Progressive field-state collapse and
  quantum non-demolition photon counting.}},\ }\href
  {https://doi.org/10.1038/nature06057} {\bibfield  {journal} {\bibinfo
  {journal} {Nature}\ }\textbf {\bibinfo {volume} {448}},\ \bibinfo {pages}
  {889} (\bibinfo {year} {2007})}\BibitemShut {NoStop}%
\bibitem [{\citenamefont {Johnson}\ \emph {et~al.}(2010)\citenamefont
  {Johnson}, \citenamefont {Reed}, \citenamefont {Houck}, \citenamefont
  {Schuster}, \citenamefont {Bishop}, \citenamefont {Ginossar}, \citenamefont
  {Gambetta}, \citenamefont {DiCarlo}, \citenamefont {Frunzio}, \citenamefont
  {Girvin},\ and\ \citenamefont {Schoelkopf}}]{Johnson2010}%
  \BibitemOpen
  \bibfield  {author} {\bibinfo {author} {\bibfnamefont {B.~R.}\ \bibnamefont
  {Johnson}}, \bibinfo {author} {\bibfnamefont {M.~D.}\ \bibnamefont {Reed}},
  \bibinfo {author} {\bibfnamefont {A.~A.}\ \bibnamefont {Houck}}, \bibinfo
  {author} {\bibfnamefont {D.~I.}\ \bibnamefont {Schuster}}, \bibinfo {author}
  {\bibfnamefont {L.~S.}\ \bibnamefont {Bishop}}, \bibinfo {author}
  {\bibfnamefont {E.}~\bibnamefont {Ginossar}}, \bibinfo {author}
  {\bibfnamefont {J.~M.}\ \bibnamefont {Gambetta}}, \bibinfo {author}
  {\bibfnamefont {L.}~\bibnamefont {DiCarlo}}, \bibinfo {author} {\bibfnamefont
  {L.}~\bibnamefont {Frunzio}}, \bibinfo {author} {\bibfnamefont {S.~M.}\
  \bibnamefont {Girvin}},\ and\ \bibinfo {author} {\bibfnamefont {R.~J.}\
  \bibnamefont {Schoelkopf}},\ }\bibfield  {title} {\bibinfo {title} {{Quantum
  non-demolition detection of single microwave photons in a circuit}},\ }\href
  {https://doi.org/10.1038/nphys1710} {\bibfield  {journal} {\bibinfo
  {journal} {Nature Physics}\ }\textbf {\bibinfo {volume} {6}},\ \bibinfo
  {pages} {663} (\bibinfo {year} {2010})}\BibitemShut {NoStop}%
\bibitem [{\citenamefont {Peaudecerf}\ \emph {et~al.}(2014)\citenamefont
  {Peaudecerf}, \citenamefont {Rybarczyk}, \citenamefont {Gerlich},
  \citenamefont {Gleyzes}, \citenamefont {Raimond}, \citenamefont {Haroche},
  \citenamefont {Dotsenko},\ and\ \citenamefont {Brune}}]{Peaudecerf2014}%
  \BibitemOpen
  \bibfield  {author} {\bibinfo {author} {\bibfnamefont {B.}~\bibnamefont
  {Peaudecerf}}, \bibinfo {author} {\bibfnamefont {T.}~\bibnamefont
  {Rybarczyk}}, \bibinfo {author} {\bibfnamefont {S.}~\bibnamefont {Gerlich}},
  \bibinfo {author} {\bibfnamefont {S.}~\bibnamefont {Gleyzes}}, \bibinfo
  {author} {\bibfnamefont {J.}~\bibnamefont {Raimond}}, \bibinfo {author}
  {\bibfnamefont {S.}~\bibnamefont {Haroche}}, \bibinfo {author} {\bibfnamefont
  {I.}~\bibnamefont {Dotsenko}},\ and\ \bibinfo {author} {\bibfnamefont
  {M.}~\bibnamefont {Brune}},\ }\bibfield  {title} {\bibinfo {title} {{Adaptive
  Quantum Nondemolition Measurement of a Photon Number}},\ }\href
  {https://doi.org/10.1103/PhysRevLett.112.080401} {\bibfield  {journal}
  {\bibinfo  {journal} {Physical Review Letters}\ }\textbf {\bibinfo {volume}
  {112}},\ \bibinfo {pages} {080401} (\bibinfo {year} {2014})}\BibitemShut
  {NoStop}%
\bibitem [{\citenamefont {{S. Haroche}}\ \emph {et~al.}(1992)\citenamefont {{S.
  Haroche}}, \citenamefont {{M. Brune}},\ and\ \citenamefont {{J.M.
  Raimond}}}]{Haroche1992}%
  \BibitemOpen
  \bibfield  {author} {\bibinfo {author} {\bibnamefont {{S. Haroche}}},
  \bibinfo {author} {\bibnamefont {{M. Brune}}},\ and\ \bibinfo {author}
  {\bibnamefont {{J.M. Raimond}}},\ }\bibfield  {title} {\bibinfo {title}
  {{Measuring photon numbers in a cavity by atomic interferometry: optimizing
  the convergence procedure}},\ }\href {https://doi.org/10.1051/jp2:1992157}
  {\bibfield  {journal} {\bibinfo  {journal} {J. Phys. II France}\ }\textbf
  {\bibinfo {volume} {2}},\ \bibinfo {pages} {659} (\bibinfo {year}
  {1992})}\BibitemShut {NoStop}%
\bibitem [{\citenamefont {Dassonneville}\ \emph {et~al.}(2020)\citenamefont
  {Dassonneville}, \citenamefont {Assouly}, \citenamefont {Peronnin},
  \citenamefont {Rouchon},\ and\ \citenamefont {Huard}}]{Dassonneville2020}%
  \BibitemOpen
  \bibfield  {author} {\bibinfo {author} {\bibfnamefont {R.}~\bibnamefont
  {Dassonneville}}, \bibinfo {author} {\bibfnamefont {R.}~\bibnamefont
  {Assouly}}, \bibinfo {author} {\bibfnamefont {T.}~\bibnamefont {Peronnin}},
  \bibinfo {author} {\bibfnamefont {P.}~\bibnamefont {Rouchon}},\ and\ \bibinfo
  {author} {\bibfnamefont {B.}~\bibnamefont {Huard}},\ }\bibfield  {title}
  {\bibinfo {title} {{Number-Resolved Photocounter for Propagating Microwave
  Mode}},\ }\bibfield  {journal} {\bibinfo  {journal} {Physical Review
  Applied}\ }\textbf {\bibinfo {volume} {14}},\ \href
  {https://doi.org/10.1103/PhysRevApplied.14.044022}
  {10.1103/PhysRevApplied.14.044022} (\bibinfo {year} {2020})\BibitemShut
  {NoStop}%
\bibitem [{\citenamefont {Curtis}\ \emph {et~al.}(2021)\citenamefont {Curtis},
  \citenamefont {Hann}, \citenamefont {Elder}, \citenamefont {Wang},
  \citenamefont {Frunzio}, \citenamefont {Jiang},\ and\ \citenamefont
  {Schoelkopf}}]{Curtis2020}%
  \BibitemOpen
  \bibfield  {author} {\bibinfo {author} {\bibfnamefont {J.~C.}\ \bibnamefont
  {Curtis}}, \bibinfo {author} {\bibfnamefont {C.~T.}\ \bibnamefont {Hann}},
  \bibinfo {author} {\bibfnamefont {S.~S.}\ \bibnamefont {Elder}}, \bibinfo
  {author} {\bibfnamefont {C.~S.}\ \bibnamefont {Wang}}, \bibinfo {author}
  {\bibfnamefont {L.}~\bibnamefont {Frunzio}}, \bibinfo {author} {\bibfnamefont
  {L.}~\bibnamefont {Jiang}},\ and\ \bibinfo {author} {\bibfnamefont {R.~J.}\
  \bibnamefont {Schoelkopf}},\ }\bibfield  {title} {\bibinfo {title}
  {{Single-shot number-resolved detection of microwave photons with error
  mitigation}},\ }\href {https://doi.org/10.1103/PhysRevA.103.023705}
  {\bibfield  {journal} {\bibinfo  {journal} {Physical Review A}\ }\textbf
  {\bibinfo {volume} {103}},\ \bibinfo {pages} {023705} (\bibinfo {year}
  {2021})},\ \Eprint {https://arxiv.org/abs/2010.04817} {arXiv:2010.04817}
  \BibitemShut {NoStop}%
\bibitem [{\citenamefont {Wang}\ \emph {et~al.}(2020)\citenamefont {Wang},
  \citenamefont {Curtis}, \citenamefont {Lester}, \citenamefont {Zhang},
  \citenamefont {Gao}, \citenamefont {Freeze}, \citenamefont {Batista},
  \citenamefont {Vaccaro}, \citenamefont {Chuang}, \citenamefont {Frunzio},
  \citenamefont {Jiang}, \citenamefont {Girvin},\ and\ \citenamefont
  {Schoelkopf}}]{Wang2020}%
  \BibitemOpen
  \bibfield  {author} {\bibinfo {author} {\bibfnamefont {C.~S.}\ \bibnamefont
  {Wang}}, \bibinfo {author} {\bibfnamefont {J.~C.}\ \bibnamefont {Curtis}},
  \bibinfo {author} {\bibfnamefont {B.~J.}\ \bibnamefont {Lester}}, \bibinfo
  {author} {\bibfnamefont {Y.}~\bibnamefont {Zhang}}, \bibinfo {author}
  {\bibfnamefont {Y.~Y.}\ \bibnamefont {Gao}}, \bibinfo {author} {\bibfnamefont
  {J.}~\bibnamefont {Freeze}}, \bibinfo {author} {\bibfnamefont {V.~S.}\
  \bibnamefont {Batista}}, \bibinfo {author} {\bibfnamefont {P.~H.}\
  \bibnamefont {Vaccaro}}, \bibinfo {author} {\bibfnamefont {I.~L.}\
  \bibnamefont {Chuang}}, \bibinfo {author} {\bibfnamefont {L.}~\bibnamefont
  {Frunzio}}, \bibinfo {author} {\bibfnamefont {L.}~\bibnamefont {Jiang}},
  \bibinfo {author} {\bibfnamefont {S.~M.}\ \bibnamefont {Girvin}},\ and\
  \bibinfo {author} {\bibfnamefont {R.~J.}\ \bibnamefont {Schoelkopf}},\
  }\bibfield  {title} {\bibinfo {title} {{Efficient Multiphoton Sampling of
  Molecular Vibronic Spectra on a Superconducting Bosonic Processor}},\ }\href
  {https://doi.org/10.1103/PhysRevX.10.021060} {\bibfield  {journal} {\bibinfo
  {journal} {Physical Review X}\ }\textbf {\bibinfo {volume} {10}},\ \bibinfo
  {pages} {021060} (\bibinfo {year} {2020})}\BibitemShut {NoStop}%
\bibitem [{\citenamefont {Schuster}\ \emph {et~al.}(2007)\citenamefont
  {Schuster}, \citenamefont {Houck}, \citenamefont {Schreier}, \citenamefont
  {Wallraff}, \citenamefont {Gambetta}, \citenamefont {Blais}, \citenamefont
  {Frunzio}, \citenamefont {Majer}, \citenamefont {Johnson}, \citenamefont
  {Devoret}, \citenamefont {Girvin},\ and\ \citenamefont
  {Schoelkopf}}]{Schuster2007}%
  \BibitemOpen
  \bibfield  {author} {\bibinfo {author} {\bibfnamefont {D.~I.}\ \bibnamefont
  {Schuster}}, \bibinfo {author} {\bibfnamefont {A.~A.}\ \bibnamefont {Houck}},
  \bibinfo {author} {\bibfnamefont {J.~A.}\ \bibnamefont {Schreier}}, \bibinfo
  {author} {\bibfnamefont {A.}~\bibnamefont {Wallraff}}, \bibinfo {author}
  {\bibfnamefont {J.~M.}\ \bibnamefont {Gambetta}}, \bibinfo {author}
  {\bibfnamefont {A.}~\bibnamefont {Blais}}, \bibinfo {author} {\bibfnamefont
  {L.}~\bibnamefont {Frunzio}}, \bibinfo {author} {\bibfnamefont
  {J.}~\bibnamefont {Majer}}, \bibinfo {author} {\bibfnamefont
  {B.}~\bibnamefont {Johnson}}, \bibinfo {author} {\bibfnamefont {M.~H.}\
  \bibnamefont {Devoret}}, \bibinfo {author} {\bibfnamefont {S.~M.}\
  \bibnamefont {Girvin}},\ and\ \bibinfo {author} {\bibfnamefont {R.~J.}\
  \bibnamefont {Schoelkopf}},\ }\bibfield  {title} {\bibinfo {title}
  {{Resolving photon number states in a superconducting circuit}},\ }\href
  {http://www.nature.com/nature/journal/v445/n7127/abs/nature05461.html
  file://localhost/Users/benjaminhuard/Documents/Papers/2007/Schuster/Nature
  2007 Schuster.pdf papers://1032888b-409c-43f0-add6-7d88c315be36/Paper/p418}
  {\bibfield  {journal} {\bibinfo  {journal} {Nature}\ }\textbf {\bibinfo
  {volume} {445}},\ \bibinfo {pages} {515} (\bibinfo {year}
  {2007})}\BibitemShut {NoStop}%
\bibitem [{\citenamefont {Gely}\ \emph {et~al.}(2019)\citenamefont {Gely},
  \citenamefont {Kounalakis}, \citenamefont {Dickel}, \citenamefont {Dalle},
  \citenamefont {Vatr{\'{e}}}, \citenamefont {Baker}, \citenamefont {Jenkins},\
  and\ \citenamefont {Steele}}]{Gely2019}%
  \BibitemOpen
  \bibfield  {author} {\bibinfo {author} {\bibfnamefont {M.~F.}\ \bibnamefont
  {Gely}}, \bibinfo {author} {\bibfnamefont {M.}~\bibnamefont {Kounalakis}},
  \bibinfo {author} {\bibfnamefont {C.}~\bibnamefont {Dickel}}, \bibinfo
  {author} {\bibfnamefont {J.}~\bibnamefont {Dalle}}, \bibinfo {author}
  {\bibfnamefont {R.}~\bibnamefont {Vatr{\'{e}}}}, \bibinfo {author}
  {\bibfnamefont {B.}~\bibnamefont {Baker}}, \bibinfo {author} {\bibfnamefont
  {M.~D.}\ \bibnamefont {Jenkins}},\ and\ \bibinfo {author} {\bibfnamefont
  {G.~A.}\ \bibnamefont {Steele}},\ }\bibfield  {title} {\bibinfo {title}
  {{Observation and stabilization of photonic Fock states in a hot
  radio-frequency resonator}},\ }\href
  {https://doi.org/10.1126/science.aaw3101} {\bibfield  {journal} {\bibinfo
  {journal} {Science}\ }\textbf {\bibinfo {volume} {363}},\ \bibinfo {pages}
  {1072} (\bibinfo {year} {2019})}\BibitemShut {NoStop}%
\bibitem [{\citenamefont {Proakis}\ and\ \citenamefont
  {Salehi}(2002)}]{ProakisSaheli2002}%
  \BibitemOpen
  \bibfield  {author} {\bibinfo {author} {\bibfnamefont {J.~G.}\ \bibnamefont
  {Proakis}}\ and\ \bibinfo {author} {\bibfnamefont {M.}~\bibnamefont
  {Salehi}},\ }\href@noop {} {\emph {\bibinfo {title} {{Communication Systems
  Engineering}}}},\ \bibinfo {edition} {2nd}\ ed.\ (\bibinfo  {publisher}
  {Pentice Hall},\ \bibinfo {year} {2002})\BibitemShut {NoStop}%
\bibitem [{\citenamefont {Macklin}\ \emph {et~al.}(2015)\citenamefont
  {Macklin}, \citenamefont {Hover}, \citenamefont {Schwartz}, \citenamefont
  {Zhang}, \citenamefont {Oliver},\ and\ \citenamefont
  {Siddiqi}}]{Macklin2015}%
  \BibitemOpen
  \bibfield  {author} {\bibinfo {author} {\bibfnamefont {C.}~\bibnamefont
  {Macklin}}, \bibinfo {author} {\bibfnamefont {D.}~\bibnamefont {Hover}},
  \bibinfo {author} {\bibfnamefont {M.~E.}\ \bibnamefont {Schwartz}}, \bibinfo
  {author} {\bibfnamefont {X.}~\bibnamefont {Zhang}}, \bibinfo {author}
  {\bibfnamefont {W.~D.}\ \bibnamefont {Oliver}},\ and\ \bibinfo {author}
  {\bibfnamefont {I.}~\bibnamefont {Siddiqi}},\ }\bibfield  {title} {\bibinfo
  {title} {{A near – quantum-limited Josephson traveling-wave parametric
  amplifier}},\ }\href@noop {} {\bibfield  {journal} {\bibinfo  {journal}
  {Science}\ }\textbf {\bibinfo {volume} {350}},\ \bibinfo {pages} {307}
  (\bibinfo {year} {2015})}\BibitemShut {NoStop}%
\bibitem [{\citenamefont {Houck}\ \emph {et~al.}(2007)\citenamefont {Houck},
  \citenamefont {Schuster}, \citenamefont {Gambetta}, \citenamefont {Schreier},
  \citenamefont {Johnson}, \citenamefont {Chow}, \citenamefont {Frunzio},
  \citenamefont {Majer}, \citenamefont {Devoret}, \citenamefont {Girvin},\ and\
  \citenamefont {Schoelkopf}}]{Houck2007}%
  \BibitemOpen
  \bibfield  {author} {\bibinfo {author} {\bibfnamefont {A.~A.}\ \bibnamefont
  {Houck}}, \bibinfo {author} {\bibfnamefont {D.~I.}\ \bibnamefont {Schuster}},
  \bibinfo {author} {\bibfnamefont {J.~M.}\ \bibnamefont {Gambetta}}, \bibinfo
  {author} {\bibfnamefont {J.~A.}\ \bibnamefont {Schreier}}, \bibinfo {author}
  {\bibfnamefont {B.~R.}\ \bibnamefont {Johnson}}, \bibinfo {author}
  {\bibfnamefont {J.~M.}\ \bibnamefont {Chow}}, \bibinfo {author}
  {\bibfnamefont {L.}~\bibnamefont {Frunzio}}, \bibinfo {author} {\bibfnamefont
  {J.}~\bibnamefont {Majer}}, \bibinfo {author} {\bibfnamefont {M.~H.}\
  \bibnamefont {Devoret}}, \bibinfo {author} {\bibfnamefont {S.~M.}\
  \bibnamefont {Girvin}},\ and\ \bibinfo {author} {\bibfnamefont {R.~J.}\
  \bibnamefont {Schoelkopf}},\ }\bibfield  {title} {\bibinfo {title}
  {{Generating single microwave photons in a circuit}},\ }\href
  {http://www.nature.com/nature/journal/v449/n7160/abs/nature06126.html
  file://localhost/Users/benjaminhuard/Documents/Papers/2007/Houck/Nature 2007
  Houck.pdf papers://1032888b-409c-43f0-add6-7d88c315be36/Paper/p405}
  {\bibfield  {journal} {\bibinfo  {journal} {Nature}\ }\textbf {\bibinfo
  {volume} {449}},\ \bibinfo {pages} {328} (\bibinfo {year}
  {2007})}\BibitemShut {NoStop}%
\bibitem [{\citenamefont {Astafiev}\ \emph {et~al.}(2010)\citenamefont
  {Astafiev}, \citenamefont {Zagoskin}, \citenamefont {Abdumalikov},
  \citenamefont {Pashkin}, \citenamefont {Yamamoto}, \citenamefont {Inomata},
  \citenamefont {Nakamura},\ and\ \citenamefont {Tsai}}]{Astafiev2010a}%
  \BibitemOpen
  \bibfield  {author} {\bibinfo {author} {\bibfnamefont {O.}~\bibnamefont
  {Astafiev}}, \bibinfo {author} {\bibfnamefont {A.~M.}\ \bibnamefont
  {Zagoskin}}, \bibinfo {author} {\bibfnamefont {A.~A.}\ \bibnamefont
  {Abdumalikov}}, \bibinfo {author} {\bibfnamefont {Y.~A.}\ \bibnamefont
  {Pashkin}}, \bibinfo {author} {\bibfnamefont {T.}~\bibnamefont {Yamamoto}},
  \bibinfo {author} {\bibfnamefont {K.}~\bibnamefont {Inomata}}, \bibinfo
  {author} {\bibfnamefont {Y.}~\bibnamefont {Nakamura}},\ and\ \bibinfo
  {author} {\bibfnamefont {J.~S.}\ \bibnamefont {Tsai}},\ }\bibfield  {title}
  {\bibinfo {title} {{Resonance fluorescence of a single artificial atom.}},\
  }\href {https://doi.org/10.1126/science.1181918} {\bibfield  {journal}
  {\bibinfo  {journal} {Science (New York, N.Y.)}\ }\textbf {\bibinfo {volume}
  {327}},\ \bibinfo {pages} {840} (\bibinfo {year} {2010})}\BibitemShut
  {NoStop}%
\bibitem [{\citenamefont {Abdumalikov}\ \emph {et~al.}(2011)\citenamefont
  {Abdumalikov}, \citenamefont {Astafiev}, \citenamefont {Pashkin},
  \citenamefont {Nakamura},\ and\ \citenamefont
  {Tsai}}]{PhysRevLett.107.043604}%
  \BibitemOpen
  \bibfield  {author} {\bibinfo {author} {\bibfnamefont {A.~A.}\ \bibnamefont
  {Abdumalikov}}, \bibinfo {author} {\bibfnamefont {O.~V.}\ \bibnamefont
  {Astafiev}}, \bibinfo {author} {\bibfnamefont {Y.~A.}\ \bibnamefont
  {Pashkin}}, \bibinfo {author} {\bibfnamefont {Y.}~\bibnamefont {Nakamura}},\
  and\ \bibinfo {author} {\bibfnamefont {J.~S.}\ \bibnamefont {Tsai}},\
  }\bibfield  {title} {\bibinfo {title} {{Dynamics of Coherent and Incoherent
  Emission from an Artificial Atom in a 1D Space}},\ }\href
  {https://doi.org/10.1103/PhysRevLett.107.043604} {\bibfield  {journal}
  {\bibinfo  {journal} {Phys. Rev. Lett.}\ }\textbf {\bibinfo {volume} {107}},\
  \bibinfo {pages} {43604} (\bibinfo {year} {2011})}\BibitemShut {NoStop}%
\bibitem [{\citenamefont {Campagne-Ibarcq}\ \emph {et~al.}(2014)\citenamefont
  {Campagne-Ibarcq}, \citenamefont {Bretheau}, \citenamefont {Flurin},
  \citenamefont {Auff{\`{e}}ves}, \citenamefont {Mallet},\ and\ \citenamefont
  {Huard}}]{PhysRevLett.112.180402}%
  \BibitemOpen
  \bibfield  {author} {\bibinfo {author} {\bibfnamefont {P.}~\bibnamefont
  {Campagne-Ibarcq}}, \bibinfo {author} {\bibfnamefont {L.}~\bibnamefont
  {Bretheau}}, \bibinfo {author} {\bibfnamefont {E.}~\bibnamefont {Flurin}},
  \bibinfo {author} {\bibfnamefont {A.}~\bibnamefont {Auff{\`{e}}ves}},
  \bibinfo {author} {\bibfnamefont {F.}~\bibnamefont {Mallet}},\ and\ \bibinfo
  {author} {\bibfnamefont {B.}~\bibnamefont {Huard}},\ }\bibfield  {title}
  {\bibinfo {title} {{Observing Interferences between Past and Future Quantum
  States in Resonance Fluorescence}},\ }\href
  {https://doi.org/10.1103/PhysRevLett.112.180402} {\bibfield  {journal}
  {\bibinfo  {journal} {Phys. Rev. Lett.}\ }\textbf {\bibinfo {volume} {112}},\
  \bibinfo {pages} {180402} (\bibinfo {year} {2014})}\BibitemShut {NoStop}%
\bibitem [{\citenamefont {Cohen-Tannoudji}\ \emph {et~al.}(2001)\citenamefont
  {Cohen-Tannoudji}, \citenamefont {Dupont-Roc},\ and\ \citenamefont
  {Grynberg}}]{Cohen-Tannoudji2001en}%
  \BibitemOpen
  \bibfield  {author} {\bibinfo {author} {\bibfnamefont {C.}~\bibnamefont
  {Cohen-Tannoudji}}, \bibinfo {author} {\bibfnamefont {J.}~\bibnamefont
  {Dupont-Roc}},\ and\ \bibinfo {author} {\bibfnamefont {G.}~\bibnamefont
  {Grynberg}},\ }\href@noop {} {\emph {\bibinfo {title} {{Atom-Photon
  Interactions: Basic Processes and Applications}}}}\ (\bibinfo {year}
  {2001})\BibitemShut {NoStop}%
\bibitem [{\citenamefont {Besse}\ \emph {et~al.}(2018)\citenamefont {Besse},
  \citenamefont {Gasparinetti}, \citenamefont {Collodo}, \citenamefont
  {Walter}, \citenamefont {Kurpiers}, \citenamefont {Pechal}, \citenamefont
  {Eichler},\ and\ \citenamefont {Wallraff}}]{Besse2018}%
  \BibitemOpen
  \bibfield  {author} {\bibinfo {author} {\bibfnamefont {J.-C.}\ \bibnamefont
  {Besse}}, \bibinfo {author} {\bibfnamefont {S.}~\bibnamefont {Gasparinetti}},
  \bibinfo {author} {\bibfnamefont {M.~C.}\ \bibnamefont {Collodo}}, \bibinfo
  {author} {\bibfnamefont {T.}~\bibnamefont {Walter}}, \bibinfo {author}
  {\bibfnamefont {P.}~\bibnamefont {Kurpiers}}, \bibinfo {author}
  {\bibfnamefont {M.}~\bibnamefont {Pechal}}, \bibinfo {author} {\bibfnamefont
  {C.}~\bibnamefont {Eichler}},\ and\ \bibinfo {author} {\bibfnamefont
  {A.}~\bibnamefont {Wallraff}},\ }\bibfield  {title} {\bibinfo {title}
  {{Single-Shot Quantum Nondemolition Detection of Individual Itinerant
  Microwave Photons}},\ }\href {https://doi.org/10.1103/PhysRevX.8.021003}
  {\bibfield  {journal} {\bibinfo  {journal} {Physical Review X}\ }\textbf
  {\bibinfo {volume} {8}},\ \bibinfo {pages} {021003} (\bibinfo {year}
  {2018})}\BibitemShut {NoStop}%
\bibitem [{\citenamefont {Chen}\ \emph {et~al.}(2011)\citenamefont {Chen},
  \citenamefont {Hover}, \citenamefont {Sendelbach}, \citenamefont {Maurer},
  \citenamefont {Merkel}, \citenamefont {Pritchett}, \citenamefont {Wilhelm},\
  and\ \citenamefont {McDermott}}]{Chen2011}%
  \BibitemOpen
  \bibfield  {author} {\bibinfo {author} {\bibfnamefont {Y.-F.}\ \bibnamefont
  {Chen}}, \bibinfo {author} {\bibfnamefont {D.}~\bibnamefont {Hover}},
  \bibinfo {author} {\bibfnamefont {S.}~\bibnamefont {Sendelbach}}, \bibinfo
  {author} {\bibfnamefont {L.}~\bibnamefont {Maurer}}, \bibinfo {author}
  {\bibfnamefont {S.~T.}\ \bibnamefont {Merkel}}, \bibinfo {author}
  {\bibfnamefont {E.~J.}\ \bibnamefont {Pritchett}}, \bibinfo {author}
  {\bibfnamefont {F.~K.}\ \bibnamefont {Wilhelm}},\ and\ \bibinfo {author}
  {\bibfnamefont {R.}~\bibnamefont {McDermott}},\ }\bibfield  {title} {\bibinfo
  {title} {{Microwave Photon Counter Based on Josephson Junctions}},\ }\href
  {https://doi.org/10.1103/PhysRevLett.107.217401} {\bibfield  {journal}
  {\bibinfo  {journal} {Physical Review Letters}\ }\textbf {\bibinfo {volume}
  {107}},\ \bibinfo {pages} {217401} (\bibinfo {year} {2011})}\BibitemShut
  {NoStop}%
\bibitem [{\citenamefont {Inomata}\ \emph {et~al.}(2016)\citenamefont
  {Inomata}, \citenamefont {Lin}, \citenamefont {Koshino}, \citenamefont
  {Oliver}, \citenamefont {Tsai}, \citenamefont {Yamamoto},\ and\ \citenamefont
  {Nakamura}}]{Inomata2016}%
  \BibitemOpen
  \bibfield  {author} {\bibinfo {author} {\bibfnamefont {K.}~\bibnamefont
  {Inomata}}, \bibinfo {author} {\bibfnamefont {Z.}~\bibnamefont {Lin}},
  \bibinfo {author} {\bibfnamefont {K.}~\bibnamefont {Koshino}}, \bibinfo
  {author} {\bibfnamefont {W.~D.}\ \bibnamefont {Oliver}}, \bibinfo {author}
  {\bibfnamefont {J.-S.}\ \bibnamefont {Tsai}}, \bibinfo {author}
  {\bibfnamefont {T.}~\bibnamefont {Yamamoto}},\ and\ \bibinfo {author}
  {\bibfnamefont {Y.}~\bibnamefont {Nakamura}},\ }\bibfield  {title} {\bibinfo
  {title} {{Single microwave-photon detector using an artificial $\Lambda$-type
  three-level system}},\ }\href {https://doi.org/10.1038/ncomms12303}
  {\bibfield  {journal} {\bibinfo  {journal} {Nature Communications}\ }\textbf
  {\bibinfo {volume} {7}},\ \bibinfo {pages} {12303} (\bibinfo {year}
  {2016})},\ \Eprint {https://arxiv.org/abs/1601.05513} {arXiv:1601.05513}
  \BibitemShut {NoStop}%
\bibitem [{\citenamefont {Narla}\ \emph {et~al.}(2016)\citenamefont {Narla},
  \citenamefont {Shankar}, \citenamefont {Hatridge}, \citenamefont {Leghtas},
  \citenamefont {Sliwa}, \citenamefont {Zalys-Geller}, \citenamefont
  {Mundhada}, \citenamefont {Pfaff}, \citenamefont {Frunzio}, \citenamefont
  {Schoelkopf},\ and\ \citenamefont {Devoret}}]{Narla2016}%
  \BibitemOpen
  \bibfield  {author} {\bibinfo {author} {\bibfnamefont {A.}~\bibnamefont
  {Narla}}, \bibinfo {author} {\bibfnamefont {S.}~\bibnamefont {Shankar}},
  \bibinfo {author} {\bibfnamefont {M.}~\bibnamefont {Hatridge}}, \bibinfo
  {author} {\bibfnamefont {Z.}~\bibnamefont {Leghtas}}, \bibinfo {author}
  {\bibfnamefont {K.~M.}\ \bibnamefont {Sliwa}}, \bibinfo {author}
  {\bibfnamefont {E.}~\bibnamefont {Zalys-Geller}}, \bibinfo {author}
  {\bibfnamefont {S.~O.}\ \bibnamefont {Mundhada}}, \bibinfo {author}
  {\bibfnamefont {W.}~\bibnamefont {Pfaff}}, \bibinfo {author} {\bibfnamefont
  {L.}~\bibnamefont {Frunzio}}, \bibinfo {author} {\bibfnamefont {R.~J.}\
  \bibnamefont {Schoelkopf}},\ and\ \bibinfo {author} {\bibfnamefont {M.~H.}\
  \bibnamefont {Devoret}},\ }\bibfield  {title} {\bibinfo {title} {{Robust
  Concurrent Remote Entanglement Between Two Superconducting Qubits}},\ }\href
  {https://doi.org/10.1103/PhysRevX.6.031036} {\bibfield  {journal} {\bibinfo
  {journal} {Physical Review X}\ }\textbf {\bibinfo {volume} {6}},\ \bibinfo
  {pages} {031036} (\bibinfo {year} {2016})},\ \Eprint
  {https://arxiv.org/abs/1603.03742} {arXiv:1603.03742} \BibitemShut {NoStop}%
\bibitem [{\citenamefont {Kono}\ \emph {et~al.}(2018)\citenamefont {Kono},
  \citenamefont {Koshino}, \citenamefont {Tabuchi}, \citenamefont {Noguchi},\
  and\ \citenamefont {Nakamura}}]{Kono2018}%
  \BibitemOpen
  \bibfield  {author} {\bibinfo {author} {\bibfnamefont {S.}~\bibnamefont
  {Kono}}, \bibinfo {author} {\bibfnamefont {K.}~\bibnamefont {Koshino}},
  \bibinfo {author} {\bibfnamefont {Y.}~\bibnamefont {Tabuchi}}, \bibinfo
  {author} {\bibfnamefont {A.}~\bibnamefont {Noguchi}},\ and\ \bibinfo {author}
  {\bibfnamefont {Y.}~\bibnamefont {Nakamura}},\ }\bibfield  {title} {\bibinfo
  {title} {{Quantum non-demolition detection of an itinerant microwave
  photon}},\ }\href {https://doi.org/10.1038/s41567-018-0066-3} {\bibfield
  {journal} {\bibinfo  {journal} {Nature Physics}\ }\textbf {\bibinfo {volume}
  {14}},\ \bibinfo {pages} {546} (\bibinfo {year} {2018})}\BibitemShut
  {NoStop}%
\bibitem [{\citenamefont {Lescanne}\ \emph {et~al.}(2020)\citenamefont
  {Lescanne}, \citenamefont {Del{\'{e}}glise}, \citenamefont {Albertinale},
  \citenamefont {R{\'{e}}glade}, \citenamefont {Capelle}, \citenamefont
  {Ivanov}, \citenamefont {Jacqmin}, \citenamefont {Leghtas},\ and\
  \citenamefont {Flurin}}]{Lescanne2019}%
  \BibitemOpen
  \bibfield  {author} {\bibinfo {author} {\bibfnamefont {R.}~\bibnamefont
  {Lescanne}}, \bibinfo {author} {\bibfnamefont {S.}~\bibnamefont
  {Del{\'{e}}glise}}, \bibinfo {author} {\bibfnamefont {E.}~\bibnamefont
  {Albertinale}}, \bibinfo {author} {\bibfnamefont {U.}~\bibnamefont
  {R{\'{e}}glade}}, \bibinfo {author} {\bibfnamefont {T.}~\bibnamefont
  {Capelle}}, \bibinfo {author} {\bibfnamefont {E.}~\bibnamefont {Ivanov}},
  \bibinfo {author} {\bibfnamefont {T.}~\bibnamefont {Jacqmin}}, \bibinfo
  {author} {\bibfnamefont {Z.}~\bibnamefont {Leghtas}},\ and\ \bibinfo {author}
  {\bibfnamefont {E.}~\bibnamefont {Flurin}},\ }\bibfield  {title} {\bibinfo
  {title} {{Irreversible Qubit-Photon Coupling for the Detection of Itinerant
  Microwave Photons}},\ }\href {https://doi.org/10.1103/PhysRevX.10.021038}
  {\bibfield  {journal} {\bibinfo  {journal} {Physical Review X}\ }\textbf
  {\bibinfo {volume} {10}},\ \bibinfo {pages} {021038} (\bibinfo {year}
  {2020})},\ \Eprint {https://arxiv.org/abs/1902.05102} {arXiv:1902.05102}
  \BibitemShut {NoStop}%
\bibitem [{\citenamefont {Schmitt}\ \emph {et~al.}(2014)\citenamefont
  {Schmitt}, \citenamefont {Zhou}, \citenamefont {Juliusson}, \citenamefont
  {Royer}, \citenamefont {Blais}, \citenamefont {Bertet}, \citenamefont
  {Vion},\ and\ \citenamefont {Esteve}}]{Schmitt2014}%
  \BibitemOpen
  \bibfield  {author} {\bibinfo {author} {\bibfnamefont {V.}~\bibnamefont
  {Schmitt}}, \bibinfo {author} {\bibfnamefont {X.}~\bibnamefont {Zhou}},
  \bibinfo {author} {\bibfnamefont {K.}~\bibnamefont {Juliusson}}, \bibinfo
  {author} {\bibfnamefont {B.}~\bibnamefont {Royer}}, \bibinfo {author}
  {\bibfnamefont {A.}~\bibnamefont {Blais}}, \bibinfo {author} {\bibfnamefont
  {P.}~\bibnamefont {Bertet}}, \bibinfo {author} {\bibfnamefont
  {D.}~\bibnamefont {Vion}},\ and\ \bibinfo {author} {\bibfnamefont
  {D.}~\bibnamefont {Esteve}},\ }\bibfield  {title} {\bibinfo {title}
  {{Multiplexed readout of transmon qubits with Josephson bifurcation
  amplifiers}},\ }\href {https://doi.org/10.1103/PhysRevA.90.062333} {\bibfield
   {journal} {\bibinfo  {journal} {Physical Review A}\ }\textbf {\bibinfo
  {volume} {90}},\ \bibinfo {pages} {062333} (\bibinfo {year}
  {2014})}\BibitemShut {NoStop}%
\bibitem [{\citenamefont {Heinsoo}\ \emph {et~al.}(2018)\citenamefont
  {Heinsoo}, \citenamefont {Andersen}, \citenamefont {Remm}, \citenamefont
  {Krinner}, \citenamefont {Walter}, \citenamefont {Salath{\'{e}}},
  \citenamefont {Gasparinetti}, \citenamefont {Besse}, \citenamefont
  {Poto{\v{c}}nik}, \citenamefont {Wallraff},\ and\ \citenamefont
  {Eichler}}]{Heinsoo2018}%
  \BibitemOpen
  \bibfield  {author} {\bibinfo {author} {\bibfnamefont {J.}~\bibnamefont
  {Heinsoo}}, \bibinfo {author} {\bibfnamefont {C.~K.}\ \bibnamefont
  {Andersen}}, \bibinfo {author} {\bibfnamefont {A.}~\bibnamefont {Remm}},
  \bibinfo {author} {\bibfnamefont {S.}~\bibnamefont {Krinner}}, \bibinfo
  {author} {\bibfnamefont {T.}~\bibnamefont {Walter}}, \bibinfo {author}
  {\bibfnamefont {Y.}~\bibnamefont {Salath{\'{e}}}}, \bibinfo {author}
  {\bibfnamefont {S.}~\bibnamefont {Gasparinetti}}, \bibinfo {author}
  {\bibfnamefont {J.-C.}\ \bibnamefont {Besse}}, \bibinfo {author}
  {\bibfnamefont {A.}~\bibnamefont {Poto{\v{c}}nik}}, \bibinfo {author}
  {\bibfnamefont {A.}~\bibnamefont {Wallraff}},\ and\ \bibinfo {author}
  {\bibfnamefont {C.}~\bibnamefont {Eichler}},\ }\bibfield  {title} {\bibinfo
  {title} {{Rapid High-fidelity Multiplexed Readout of Superconducting
  Qubits}},\ }\href {https://doi.org/10.1103/PhysRevApplied.10.034040}
  {\bibfield  {journal} {\bibinfo  {journal} {Physical Review Applied}\
  }\textbf {\bibinfo {volume} {10}},\ \bibinfo {pages} {034040} (\bibinfo
  {year} {2018})}\BibitemShut {NoStop}%
\bibitem [{\citenamefont {Kundu}\ \emph {et~al.}(2019)\citenamefont {Kundu},
  \citenamefont {Gheeraert}, \citenamefont {Hazra}, \citenamefont {Roy},
  \citenamefont {Salunkhe}, \citenamefont {Patankar},\ and\ \citenamefont
  {Vijay}}]{Kundu2019}%
  \BibitemOpen
  \bibfield  {author} {\bibinfo {author} {\bibfnamefont {S.}~\bibnamefont
  {Kundu}}, \bibinfo {author} {\bibfnamefont {N.}~\bibnamefont {Gheeraert}},
  \bibinfo {author} {\bibfnamefont {S.}~\bibnamefont {Hazra}}, \bibinfo
  {author} {\bibfnamefont {T.}~\bibnamefont {Roy}}, \bibinfo {author}
  {\bibfnamefont {K.~V.}\ \bibnamefont {Salunkhe}}, \bibinfo {author}
  {\bibfnamefont {M.~P.}\ \bibnamefont {Patankar}},\ and\ \bibinfo {author}
  {\bibfnamefont {R.}~\bibnamefont {Vijay}},\ }\bibfield  {title} {\bibinfo
  {title} {{Multiplexed readout of four qubits in 3D circuit QED architecture
  using a broadband Josephson parametric amplifier}},\ }\href
  {https://doi.org/10.1063/1.5089729} {\bibfield  {journal} {\bibinfo
  {journal} {Applied Physics Letters}\ }\textbf {\bibinfo {volume} {114}},\
  \bibinfo {pages} {172601} (\bibinfo {year} {2019})}\BibitemShut {NoStop}%
\bibitem [{\citenamefont {Arute}\ \emph {et~al.}(2019)\citenamefont {Arute},
  \citenamefont {Arya}, \citenamefont {Babbush}, \citenamefont {Bacon},
  \citenamefont {Bardin}, \citenamefont {Barends}, \citenamefont {Biswas},
  \citenamefont {Boixo}, \citenamefont {Brandao}, \citenamefont {Buell},
  \citenamefont {Burkett}, \citenamefont {Chen}, \citenamefont {Chen},
  \citenamefont {Chiaro}, \citenamefont {Collins}, \citenamefont {Courtney},
  \citenamefont {Dunsworth}, \citenamefont {Farhi}, \citenamefont {Foxen},
  \citenamefont {Fowler}, \citenamefont {Gidney}, \citenamefont {Giustina},
  \citenamefont {Graff}, \citenamefont {Guerin}, \citenamefont {Habegger},
  \citenamefont {Harrigan}, \citenamefont {Hartmann}, \citenamefont {Ho},
  \citenamefont {Hoffmann}, \citenamefont {Huang}, \citenamefont {Humble},
  \citenamefont {Isakov}, \citenamefont {Jeffrey}, \citenamefont {Jiang},
  \citenamefont {Kafri}, \citenamefont {Kechedzhi}, \citenamefont {Kelly},
  \citenamefont {Klimov}, \citenamefont {Knysh}, \citenamefont {Korotkov},
  \citenamefont {Kostritsa}, \citenamefont {Landhuis}, \citenamefont
  {Lindmark}, \citenamefont {Lucero}, \citenamefont {Lyakh}, \citenamefont
  {Mandr{\`{a}}}, \citenamefont {McClean}, \citenamefont {McEwen},
  \citenamefont {Megrant}, \citenamefont {Mi}, \citenamefont {Michielsen},
  \citenamefont {Mohseni}, \citenamefont {Mutus}, \citenamefont {Naaman},
  \citenamefont {Neeley}, \citenamefont {Neill}, \citenamefont {Niu},
  \citenamefont {Ostby}, \citenamefont {Petukhov}, \citenamefont {Platt},
  \citenamefont {Quintana}, \citenamefont {Rieffel}, \citenamefont {Roushan},
  \citenamefont {Rubin}, \citenamefont {Sank}, \citenamefont {Satzinger},
  \citenamefont {Smelyanskiy}, \citenamefont {Sung}, \citenamefont
  {Trevithick}, \citenamefont {Vainsencher}, \citenamefont {Villalonga},
  \citenamefont {White}, \citenamefont {Yao}, \citenamefont {Yeh},
  \citenamefont {Zalcman}, \citenamefont {Neven},\ and\ \citenamefont
  {Martinis}}]{Arute2019}%
  \BibitemOpen
  \bibfield  {author} {\bibinfo {author} {\bibfnamefont {F.}~\bibnamefont
  {Arute}}, \bibinfo {author} {\bibfnamefont {K.}~\bibnamefont {Arya}},
  \bibinfo {author} {\bibfnamefont {R.}~\bibnamefont {Babbush}}, \bibinfo
  {author} {\bibfnamefont {D.}~\bibnamefont {Bacon}}, \bibinfo {author}
  {\bibfnamefont {J.~C.}\ \bibnamefont {Bardin}}, \bibinfo {author}
  {\bibfnamefont {R.}~\bibnamefont {Barends}}, \bibinfo {author} {\bibfnamefont
  {R.}~\bibnamefont {Biswas}}, \bibinfo {author} {\bibfnamefont
  {S.}~\bibnamefont {Boixo}}, \bibinfo {author} {\bibfnamefont {F.~G. S.~L.}\
  \bibnamefont {Brandao}}, \bibinfo {author} {\bibfnamefont {D.~A.}\
  \bibnamefont {Buell}}, \bibinfo {author} {\bibfnamefont {B.}~\bibnamefont
  {Burkett}}, \bibinfo {author} {\bibfnamefont {Y.}~\bibnamefont {Chen}},
  \bibinfo {author} {\bibfnamefont {Z.}~\bibnamefont {Chen}}, \bibinfo {author}
  {\bibfnamefont {B.}~\bibnamefont {Chiaro}}, \bibinfo {author} {\bibfnamefont
  {R.}~\bibnamefont {Collins}}, \bibinfo {author} {\bibfnamefont
  {W.}~\bibnamefont {Courtney}}, \bibinfo {author} {\bibfnamefont
  {A.}~\bibnamefont {Dunsworth}}, \bibinfo {author} {\bibfnamefont
  {E.}~\bibnamefont {Farhi}}, \bibinfo {author} {\bibfnamefont
  {B.}~\bibnamefont {Foxen}}, \bibinfo {author} {\bibfnamefont
  {A.}~\bibnamefont {Fowler}}, \bibinfo {author} {\bibfnamefont
  {C.}~\bibnamefont {Gidney}}, \bibinfo {author} {\bibfnamefont
  {M.}~\bibnamefont {Giustina}}, \bibinfo {author} {\bibfnamefont
  {R.}~\bibnamefont {Graff}}, \bibinfo {author} {\bibfnamefont
  {K.}~\bibnamefont {Guerin}}, \bibinfo {author} {\bibfnamefont
  {S.}~\bibnamefont {Habegger}}, \bibinfo {author} {\bibfnamefont {M.~P.}\
  \bibnamefont {Harrigan}}, \bibinfo {author} {\bibfnamefont {M.~J.}\
  \bibnamefont {Hartmann}}, \bibinfo {author} {\bibfnamefont {A.}~\bibnamefont
  {Ho}}, \bibinfo {author} {\bibfnamefont {M.}~\bibnamefont {Hoffmann}},
  \bibinfo {author} {\bibfnamefont {T.}~\bibnamefont {Huang}}, \bibinfo
  {author} {\bibfnamefont {T.~S.}\ \bibnamefont {Humble}}, \bibinfo {author}
  {\bibfnamefont {S.~V.}\ \bibnamefont {Isakov}}, \bibinfo {author}
  {\bibfnamefont {E.}~\bibnamefont {Jeffrey}}, \bibinfo {author} {\bibfnamefont
  {Z.}~\bibnamefont {Jiang}}, \bibinfo {author} {\bibfnamefont
  {D.}~\bibnamefont {Kafri}}, \bibinfo {author} {\bibfnamefont
  {K.}~\bibnamefont {Kechedzhi}}, \bibinfo {author} {\bibfnamefont
  {J.}~\bibnamefont {Kelly}}, \bibinfo {author} {\bibfnamefont {P.~V.}\
  \bibnamefont {Klimov}}, \bibinfo {author} {\bibfnamefont {S.}~\bibnamefont
  {Knysh}}, \bibinfo {author} {\bibfnamefont {A.}~\bibnamefont {Korotkov}},
  \bibinfo {author} {\bibfnamefont {F.}~\bibnamefont {Kostritsa}}, \bibinfo
  {author} {\bibfnamefont {D.}~\bibnamefont {Landhuis}}, \bibinfo {author}
  {\bibfnamefont {M.}~\bibnamefont {Lindmark}}, \bibinfo {author}
  {\bibfnamefont {E.}~\bibnamefont {Lucero}}, \bibinfo {author} {\bibfnamefont
  {D.}~\bibnamefont {Lyakh}}, \bibinfo {author} {\bibfnamefont
  {S.}~\bibnamefont {Mandr{\`{a}}}}, \bibinfo {author} {\bibfnamefont {J.~R.}\
  \bibnamefont {McClean}}, \bibinfo {author} {\bibfnamefont {M.}~\bibnamefont
  {McEwen}}, \bibinfo {author} {\bibfnamefont {A.}~\bibnamefont {Megrant}},
  \bibinfo {author} {\bibfnamefont {X.}~\bibnamefont {Mi}}, \bibinfo {author}
  {\bibfnamefont {K.}~\bibnamefont {Michielsen}}, \bibinfo {author}
  {\bibfnamefont {M.}~\bibnamefont {Mohseni}}, \bibinfo {author} {\bibfnamefont
  {J.}~\bibnamefont {Mutus}}, \bibinfo {author} {\bibfnamefont
  {O.}~\bibnamefont {Naaman}}, \bibinfo {author} {\bibfnamefont
  {M.}~\bibnamefont {Neeley}}, \bibinfo {author} {\bibfnamefont
  {C.}~\bibnamefont {Neill}}, \bibinfo {author} {\bibfnamefont {M.~Y.}\
  \bibnamefont {Niu}}, \bibinfo {author} {\bibfnamefont {E.}~\bibnamefont
  {Ostby}}, \bibinfo {author} {\bibfnamefont {A.}~\bibnamefont {Petukhov}},
  \bibinfo {author} {\bibfnamefont {J.~C.}\ \bibnamefont {Platt}}, \bibinfo
  {author} {\bibfnamefont {C.}~\bibnamefont {Quintana}}, \bibinfo {author}
  {\bibfnamefont {E.~G.}\ \bibnamefont {Rieffel}}, \bibinfo {author}
  {\bibfnamefont {P.}~\bibnamefont {Roushan}}, \bibinfo {author} {\bibfnamefont
  {N.~C.}\ \bibnamefont {Rubin}}, \bibinfo {author} {\bibfnamefont
  {D.}~\bibnamefont {Sank}}, \bibinfo {author} {\bibfnamefont {K.~J.}\
  \bibnamefont {Satzinger}}, \bibinfo {author} {\bibfnamefont {V.}~\bibnamefont
  {Smelyanskiy}}, \bibinfo {author} {\bibfnamefont {K.~J.}\ \bibnamefont
  {Sung}}, \bibinfo {author} {\bibfnamefont {M.~D.}\ \bibnamefont
  {Trevithick}}, \bibinfo {author} {\bibfnamefont {A.}~\bibnamefont
  {Vainsencher}}, \bibinfo {author} {\bibfnamefont {B.}~\bibnamefont
  {Villalonga}}, \bibinfo {author} {\bibfnamefont {T.}~\bibnamefont {White}},
  \bibinfo {author} {\bibfnamefont {Z.~J.}\ \bibnamefont {Yao}}, \bibinfo
  {author} {\bibfnamefont {P.}~\bibnamefont {Yeh}}, \bibinfo {author}
  {\bibfnamefont {A.}~\bibnamefont {Zalcman}}, \bibinfo {author} {\bibfnamefont
  {H.}~\bibnamefont {Neven}},\ and\ \bibinfo {author} {\bibfnamefont {J.~M.}\
  \bibnamefont {Martinis}},\ }\bibfield  {title} {\bibinfo {title} {{Quantum
  supremacy using a programmable superconducting processor}},\ }\href
  {https://doi.org/10.1038/s41586-019-1666-5} {\bibfield  {journal} {\bibinfo
  {journal} {Nature}\ }\textbf {\bibinfo {volume} {574}},\ \bibinfo {pages}
  {505} (\bibinfo {year} {2019})}\BibitemShut {NoStop}%
\bibitem [{\citenamefont {Clerk}\ \emph {et~al.}(2010)\citenamefont {Clerk},
  \citenamefont {Devoret}, \citenamefont {Girvin}, \citenamefont {Marquardt},\
  and\ \citenamefont {Schoelkopf}}]{Clerk2008}%
  \BibitemOpen
  \bibfield  {author} {\bibinfo {author} {\bibfnamefont {A.~A.}\ \bibnamefont
  {Clerk}}, \bibinfo {author} {\bibfnamefont {M.~H.}\ \bibnamefont {Devoret}},
  \bibinfo {author} {\bibfnamefont {S.~M.}\ \bibnamefont {Girvin}}, \bibinfo
  {author} {\bibfnamefont {F.}~\bibnamefont {Marquardt}},\ and\ \bibinfo
  {author} {\bibfnamefont {R.~J.}\ \bibnamefont {Schoelkopf}},\ }\bibfield
  {title} {\bibinfo {title} {{Introduction to quantum noise, measurement, and
  amplification}},\ }\href {https://doi.org/10.1103/RevModPhys.82.1155}
  {\bibfield  {journal} {\bibinfo  {journal} {Reviews of Modern Physics}\
  }\textbf {\bibinfo {volume} {82}},\ \bibinfo {pages} {1155} (\bibinfo {year}
  {2010})},\ \Eprint {https://arxiv.org/abs/0810.4729} {arXiv:0810.4729}
  \BibitemShut {NoStop}%
\bibitem [{\citenamefont {Boissonneault}\ \emph {et~al.}(2009)\citenamefont
  {Boissonneault}, \citenamefont {Gambetta},\ and\ \citenamefont
  {Blais}}]{Boissonneault2009}%
  \BibitemOpen
  \bibfield  {author} {\bibinfo {author} {\bibnamefont {Boissonneault}},
  \bibinfo {author} {\bibnamefont {Gambetta}},\ and\ \bibinfo {author}
  {\bibnamefont {Blais}},\ }\bibfield  {title} {\bibinfo {title} {{Dispersive
  regime of circuit QED: Photon-dependent qubit dephasing and relaxation
  rates}},\ }\href {http://dx.doi.org/10.1103/PhysRevA.79.013819
  papers://b942896d-313a-49f0-995e-1484170e0fc9/Paper/p396} {\bibfield
  {journal} {\bibinfo  {journal} {Physical Review A}\ }\textbf {\bibinfo
  {volume} {79}},\ \bibinfo {pages} {13819} (\bibinfo {year}
  {2009})}\BibitemShut {NoStop}%
\bibitem [{\citenamefont {Lutterbach}\ and\ \citenamefont
  {Davidovich}(1997)}]{Lutterbach1997}%
  \BibitemOpen
  \bibfield  {author} {\bibinfo {author} {\bibfnamefont {L.}~\bibnamefont
  {Lutterbach}}\ and\ \bibinfo {author} {\bibfnamefont {L.}~\bibnamefont
  {Davidovich}},\ }\bibfield  {title} {\bibinfo {title} {{Method for Direct
  Measurement of the Wigner Function in Cavity QED and Ion Traps}},\ }\href
  {https://doi.org/10.1103/PhysRevLett.78.2547} {\bibfield  {journal} {\bibinfo
   {journal} {Physical Review Letters}\ }\textbf {\bibinfo {volume} {78}},\
  \bibinfo {pages} {2547} (\bibinfo {year} {1997})}\BibitemShut {NoStop}%
\bibitem [{\citenamefont {Bertet}\ \emph {et~al.}(2002)\citenamefont {Bertet},
  \citenamefont {Auffeves}, \citenamefont {Maioli}, \citenamefont {Osnaghi},
  \citenamefont {Meunier}, \citenamefont {Brune}, \citenamefont {Raimond},\
  and\ \citenamefont {Haroche}}]{Bertet2002}%
  \BibitemOpen
  \bibfield  {author} {\bibinfo {author} {\bibfnamefont {P.}~\bibnamefont
  {Bertet}}, \bibinfo {author} {\bibfnamefont {A.}~\bibnamefont {Auffeves}},
  \bibinfo {author} {\bibfnamefont {P.}~\bibnamefont {Maioli}}, \bibinfo
  {author} {\bibfnamefont {S.}~\bibnamefont {Osnaghi}}, \bibinfo {author}
  {\bibfnamefont {T.}~\bibnamefont {Meunier}}, \bibinfo {author} {\bibfnamefont
  {M.}~\bibnamefont {Brune}}, \bibinfo {author} {\bibfnamefont
  {J.}~\bibnamefont {Raimond}},\ and\ \bibinfo {author} {\bibfnamefont
  {S.}~\bibnamefont {Haroche}},\ }\bibfield  {title} {\bibinfo {title} {{Direct
  Measurement of the Wigner Function of a One-Photon Fock State in a Cavity}},\
  }\href {https://doi.org/10.1103/PhysRevLett.89.200402} {\bibfield  {journal}
  {\bibinfo  {journal} {Physical Review Letters}\ }\textbf {\bibinfo {volume}
  {89}},\ \bibinfo {pages} {200402} (\bibinfo {year} {2002})}\BibitemShut
  {NoStop}%
\bibitem [{\citenamefont {Vlastakis}\ \emph {et~al.}(2013)\citenamefont
  {Vlastakis}, \citenamefont {Kirchmair}, \citenamefont {Leghtas},
  \citenamefont {Nigg}, \citenamefont {Frunzio}, \citenamefont {Girvin},
  \citenamefont {Mirrahimi}, \citenamefont {Devoret},\ and\ \citenamefont
  {Schoelkopf}}]{Vlastakis2013a}%
  \BibitemOpen
  \bibfield  {author} {\bibinfo {author} {\bibfnamefont {B.}~\bibnamefont
  {Vlastakis}}, \bibinfo {author} {\bibfnamefont {G.}~\bibnamefont
  {Kirchmair}}, \bibinfo {author} {\bibfnamefont {Z.}~\bibnamefont {Leghtas}},
  \bibinfo {author} {\bibfnamefont {S.~E.}\ \bibnamefont {Nigg}}, \bibinfo
  {author} {\bibfnamefont {L.}~\bibnamefont {Frunzio}}, \bibinfo {author}
  {\bibfnamefont {S.~M.}\ \bibnamefont {Girvin}}, \bibinfo {author}
  {\bibfnamefont {M.}~\bibnamefont {Mirrahimi}}, \bibinfo {author}
  {\bibfnamefont {M.~H.}\ \bibnamefont {Devoret}},\ and\ \bibinfo {author}
  {\bibfnamefont {R.~J.}\ \bibnamefont {Schoelkopf}},\ }\bibfield  {title}
  {\bibinfo {title} {{Deterministically Encoding Quantum Information Using
  100-Photon Schrodinger Cat States}},\ }\href
  {https://doi.org/10.1126/science.1243289} {\bibfield  {journal} {\bibinfo
  {journal} {Science}\ }\textbf {\bibinfo {volume} {342}},\ \bibinfo {pages}
  {607} (\bibinfo {year} {2013})}\BibitemShut {NoStop}%
\bibitem [{\citenamefont {Campagne-Ibarcq}\ \emph {et~al.}(2020)\citenamefont
  {Campagne-Ibarcq}, \citenamefont {Eickbusch}, \citenamefont {Touzard},
  \citenamefont {Zalys-Geller}, \citenamefont {Frattini}, \citenamefont
  {Sivak}, \citenamefont {Reinhold}, \citenamefont {Puri}, \citenamefont
  {Shankar}, \citenamefont {Schoelkopf}, \citenamefont {Frunzio}, \citenamefont
  {Mirrahimi},\ and\ \citenamefont {Devoret}}]{Campagne-Ibarcq2020}%
  \BibitemOpen
  \bibfield  {author} {\bibinfo {author} {\bibfnamefont {P.}~\bibnamefont
  {Campagne-Ibarcq}}, \bibinfo {author} {\bibfnamefont {A.}~\bibnamefont
  {Eickbusch}}, \bibinfo {author} {\bibfnamefont {S.}~\bibnamefont {Touzard}},
  \bibinfo {author} {\bibfnamefont {E.}~\bibnamefont {Zalys-Geller}}, \bibinfo
  {author} {\bibfnamefont {N.~E.}\ \bibnamefont {Frattini}}, \bibinfo {author}
  {\bibfnamefont {V.~V.}\ \bibnamefont {Sivak}}, \bibinfo {author}
  {\bibfnamefont {P.}~\bibnamefont {Reinhold}}, \bibinfo {author}
  {\bibfnamefont {S.}~\bibnamefont {Puri}}, \bibinfo {author} {\bibfnamefont
  {S.}~\bibnamefont {Shankar}}, \bibinfo {author} {\bibfnamefont {R.~J.}\
  \bibnamefont {Schoelkopf}}, \bibinfo {author} {\bibfnamefont
  {L.}~\bibnamefont {Frunzio}}, \bibinfo {author} {\bibfnamefont
  {M.}~\bibnamefont {Mirrahimi}},\ and\ \bibinfo {author} {\bibfnamefont
  {M.~H.}\ \bibnamefont {Devoret}},\ }\bibfield  {title} {\bibinfo {title}
  {{Quantum error correction of a qubit encoded in grid states of an
  oscillator}},\ }\href {https://doi.org/10.1038/s41586-020-2603-3} {\bibfield
  {journal} {\bibinfo  {journal} {Nature}\ }\textbf {\bibinfo {volume} {584}},\
  \bibinfo {pages} {368} (\bibinfo {year} {2020})},\ \Eprint
  {https://arxiv.org/abs/1907.12487} {arXiv:1907.12487} \BibitemShut {NoStop}%
\bibitem [{\citenamefont {Wiseman}\ and\ \citenamefont
  {Milburn}(2009)}]{Wiseman2009}%
  \BibitemOpen
  \bibfield  {author} {\bibinfo {author} {\bibfnamefont {H.~M.}\ \bibnamefont
  {Wiseman}}\ and\ \bibinfo {author} {\bibfnamefont {G.~J.}\ \bibnamefont
  {Milburn}},\ }\href
  {http://books.google.fr/books?id=7n0wHQAACAAJ{\&}dq=wiseman+milburn+quantum{\&}hl=en{\&}sa=X{\&}ei=xD7tUOP7OM25hAeck4H4DQ{\&}ved=0CC8Q6AEwAA
  papers://b942896d-313a-49f0-995e-1484170e0fc9/Paper/p504} {\emph {\bibinfo
  {title} {{Quantum Measurement and Control}}}}\ (\bibinfo  {publisher}
  {Cambridge University Press},\ \bibinfo {year} {2009})\BibitemShut {NoStop}%
\bibitem [{\citenamefont {Haroche}\ and\ \citenamefont
  {Raimond}(2006)}]{Haroche2006}%
  \BibitemOpen
  \bibfield  {author} {\bibinfo {author} {\bibfnamefont {S.}~\bibnamefont
  {Haroche}}\ and\ \bibinfo {author} {\bibfnamefont {J.}~\bibnamefont
  {Raimond}},\ }\bibfield  {title} {\bibinfo {title} {{Exploring the Quantum:
  Atoms, Cavities, and Photons}},\ }\href
  {http://books.google.com/books?id=FWcZAAAACAAJ{\&}printsec=frontcover
  papers://1032888b-409c-43f0-add6-7d88c315be36/Paper/p3984} {\bibfield
  {journal} {\bibinfo  {journal} {Oxford Graduated Text}\ ,\ \bibinfo {pages}
  {616}} (\bibinfo {year} {2006})}\BibitemShut {NoStop}%
\bibitem [{\citenamefont {Bretheau}\ \emph {et~al.}(2015)\citenamefont
  {Bretheau}, \citenamefont {Campagne-Ibarcq}, \citenamefont {Flurin},
  \citenamefont {Mallet},\ and\ \citenamefont {Huard}}]{Bretheau2015}%
  \BibitemOpen
  \bibfield  {author} {\bibinfo {author} {\bibfnamefont {L.}~\bibnamefont
  {Bretheau}}, \bibinfo {author} {\bibfnamefont {P.}~\bibnamefont
  {Campagne-Ibarcq}}, \bibinfo {author} {\bibfnamefont {E.}~\bibnamefont
  {Flurin}}, \bibinfo {author} {\bibfnamefont {F.}~\bibnamefont {Mallet}},\
  and\ \bibinfo {author} {\bibfnamefont {B.}~\bibnamefont {Huard}},\ }\bibfield
   {title} {\bibinfo {title} {{Quantum dynamics of an electromagnetic mode that
  cannot have N photons}},\ }\href {https://doi.org/10.1126/science.1259345}
  {\bibfield  {journal} {\bibinfo  {journal} {Science}\ }\textbf {\bibinfo
  {volume} {348}},\ \bibinfo {pages} {776} (\bibinfo {year}
  {2015})}\BibitemShut {NoStop}%
\bibitem [{\citenamefont {Cai}\ \emph {et~al.}(2021)\citenamefont {Cai},
  \citenamefont {Ma}, \citenamefont {Wang}, \citenamefont {Zou},\ and\
  \citenamefont {Sun}}]{Cai2021}%
  \BibitemOpen
  \bibfield  {author} {\bibinfo {author} {\bibfnamefont {W.}~\bibnamefont
  {Cai}}, \bibinfo {author} {\bibfnamefont {Y.}~\bibnamefont {Ma}}, \bibinfo
  {author} {\bibfnamefont {W.}~\bibnamefont {Wang}}, \bibinfo {author}
  {\bibfnamefont {C.-L.}\ \bibnamefont {Zou}},\ and\ \bibinfo {author}
  {\bibfnamefont {L.}~\bibnamefont {Sun}},\ }\bibfield  {title} {\bibinfo
  {title} {{Bosonic quantum error correction codes in superconducting quantum
  circuits}},\ }\href {https://doi.org/10.1016/j.fmre.2020.12.006} {\bibfield
  {journal} {\bibinfo  {journal} {Fundamental Research}\ }\textbf {\bibinfo
  {volume} {1}},\ \bibinfo {pages} {50} (\bibinfo {year} {2021})},\ \Eprint
  {https://arxiv.org/abs/2010.08699} {arXiv:2010.08699} \BibitemShut {NoStop}%
\bibitem [{\citenamefont {Gertler}\ \emph {et~al.}(2021)\citenamefont
  {Gertler}, \citenamefont {Baker}, \citenamefont {Li}, \citenamefont {Shirol},
  \citenamefont {Koch},\ and\ \citenamefont {Wang}}]{Gertler2020}%
  \BibitemOpen
  \bibfield  {author} {\bibinfo {author} {\bibfnamefont {J.~M.}\ \bibnamefont
  {Gertler}}, \bibinfo {author} {\bibfnamefont {B.}~\bibnamefont {Baker}},
  \bibinfo {author} {\bibfnamefont {J.}~\bibnamefont {Li}}, \bibinfo {author}
  {\bibfnamefont {S.}~\bibnamefont {Shirol}}, \bibinfo {author} {\bibfnamefont
  {J.}~\bibnamefont {Koch}},\ and\ \bibinfo {author} {\bibfnamefont
  {C.}~\bibnamefont {Wang}},\ }\bibfield  {title} {\bibinfo {title}
  {{Protecting a bosonic qubit with autonomous quantum error correction}},\
  }\href {https://doi.org/10.1038/s41586-021-03257-0} {\bibfield  {journal}
  {\bibinfo  {journal} {Nature}\ }\textbf {\bibinfo {volume} {590}},\ \bibinfo
  {pages} {243} (\bibinfo {year} {2021})},\ \Eprint
  {https://arxiv.org/abs/2004.09322} {arXiv:2004.09322} \BibitemShut {NoStop}%
\bibitem [{\citenamefont {Hacohen-Gourgy}\ \emph {et~al.}(2016)\citenamefont
  {Hacohen-Gourgy}, \citenamefont {Martin}, \citenamefont {Flurin},
  \citenamefont {Ramasesh}, \citenamefont {Whaley},\ and\ \citenamefont
  {Siddiqi}}]{Hacohen-Gourgy2016}%
  \BibitemOpen
  \bibfield  {author} {\bibinfo {author} {\bibfnamefont {S.}~\bibnamefont
  {Hacohen-Gourgy}}, \bibinfo {author} {\bibfnamefont {L.~S.}\ \bibnamefont
  {Martin}}, \bibinfo {author} {\bibfnamefont {E.}~\bibnamefont {Flurin}},
  \bibinfo {author} {\bibfnamefont {V.~V.}\ \bibnamefont {Ramasesh}}, \bibinfo
  {author} {\bibfnamefont {K.~B.}\ \bibnamefont {Whaley}},\ and\ \bibinfo
  {author} {\bibfnamefont {I.}~\bibnamefont {Siddiqi}},\ }\bibfield  {title}
  {\bibinfo {title} {{Quantum dynamics of simultaneously measured non-commuting
  observables}},\ }\href {https://doi.org/10.1038/nature19762} {\bibfield
  {journal} {\bibinfo  {journal} {Nature}\ }\textbf {\bibinfo {volume} {538}},\
  \bibinfo {pages} {491} (\bibinfo {year} {2016})},\ \Eprint
  {https://arxiv.org/abs/1608.06652} {arXiv:1608.06652} \BibitemShut {NoStop}%
\bibitem [{\citenamefont {Ficheux}\ \emph {et~al.}(2018)\citenamefont
  {Ficheux}, \citenamefont {Jezouin}, \citenamefont {Leghtas},\ and\
  \citenamefont {Huard}}]{FICHEUX2018}%
  \BibitemOpen
  \bibfield  {author} {\bibinfo {author} {\bibfnamefont {Q.}~\bibnamefont
  {Ficheux}}, \bibinfo {author} {\bibfnamefont {S.}~\bibnamefont {Jezouin}},
  \bibinfo {author} {\bibfnamefont {Z.}~\bibnamefont {Leghtas}},\ and\ \bibinfo
  {author} {\bibfnamefont {B.}~\bibnamefont {Huard}},\ }\bibfield  {title}
  {\bibinfo {title} {{Dynamics of a qubit while simultaneously monitoring its
  relaxation and dephasing}},\ }\href
  {https://doi.org/10.1038/s41467-018-04372-9} {\bibfield  {journal} {\bibinfo
  {journal} {Nature Communications}\ }\textbf {\bibinfo {volume} {9}},\
  \bibinfo {pages} {1926} (\bibinfo {year} {2018})}\BibitemShut {NoStop}%
\bibitem [{\citenamefont {Ficheux}(2018)}]{Ficheux2018thesis}%
  \BibitemOpen
  \bibfield  {author} {\bibinfo {author} {\bibfnamefont {Q.}~\bibnamefont
  {Ficheux}},\ }\emph {\bibinfo {title} {{Quantum Trajectories with
  Incompatible Decoherence Channels}}},\ \href
  {https://tel.archives-ouvertes.fr/tel-02098804} {Ph.D. thesis},\ \bibinfo
  {school} {Universite Paris Sciences et Lettres} (\bibinfo {year}
  {2018})\BibitemShut {NoStop}%
\bibitem [{Note1()}]{Note1}%
  \BibitemOpen
  \bibinfo {note} {We find two solutions where $z/\eta $ is stationary, one
  stable and one unstable. Considering the stable solution only, we compute $R$
  such that $\eta (kT+T)=R\protect \tmspace +\thinmuskip {.1667em} \eta (kT)$.
  The computation boils down to solving order-two polynomials.}\BibitemShut
  {Stop}%
\bibitem [{\citenamefont {Chuang}\ and\ \citenamefont
  {Nielsen}(2010)}]{Chuang2010}%
  \BibitemOpen
  \bibfield  {author} {\bibinfo {author} {\bibfnamefont {I.~L.}\ \bibnamefont
  {Chuang}}\ and\ \bibinfo {author} {\bibfnamefont {M.~A.}\ \bibnamefont
  {Nielsen}},\ }\href@noop {} {\emph {\bibinfo {title} {Contemporary
  Physics}}},\ Vol.~\bibinfo {volume} {52}\ (\bibinfo  {publisher} {Cambridge
  University Press},\ \bibinfo {year} {2010})\BibitemShut {NoStop}%
\bibitem [{\citenamefont {Minev}\ \emph {et~al.}(2020)\citenamefont {Minev},
  \citenamefont {Leghtas}, \citenamefont {Mundhada}, \citenamefont
  {Christakis}, \citenamefont {Pop},\ and\ \citenamefont
  {Devoret}}]{Minev2020}%
  \BibitemOpen
  \bibfield  {author} {\bibinfo {author} {\bibfnamefont {Z.~K.}\ \bibnamefont
  {Minev}}, \bibinfo {author} {\bibfnamefont {Z.}~\bibnamefont {Leghtas}},
  \bibinfo {author} {\bibfnamefont {S.~O.}\ \bibnamefont {Mundhada}}, \bibinfo
  {author} {\bibfnamefont {L.}~\bibnamefont {Christakis}}, \bibinfo {author}
  {\bibfnamefont {I.~M.}\ \bibnamefont {Pop}},\ and\ \bibinfo {author}
  {\bibfnamefont {M.~H.}\ \bibnamefont {Devoret}},\ }\bibfield  {title}
  {\bibinfo {title} {{Energy-participation quantization of Josephson
  circuits}},\ }\href {http://arxiv.org/abs/2010.00620} {\bibfield  {journal}
  {\bibinfo  {journal} {arXiv:2010.00620}\ } (\bibinfo {year} {2020})},\
  \Eprint {https://arxiv.org/abs/2010.00620} {arXiv:2010.00620} \BibitemShut
  {NoStop}%
\bibitem [{\citenamefont {Hatridge}\ \emph {et~al.}(2013)\citenamefont
  {Hatridge}, \citenamefont {Shankar}, \citenamefont {Mirrahimi}, \citenamefont
  {Schackert}, \citenamefont {Geerlings}, \citenamefont {Brecht}, \citenamefont
  {Sliwa}, \citenamefont {Abdo}, \citenamefont {Frunzio}, \citenamefont
  {Girvin}, \citenamefont {Schoelkopf},\ and\ \citenamefont
  {Devoret}}]{Hatridge2013}%
  \BibitemOpen
  \bibfield  {author} {\bibinfo {author} {\bibfnamefont {M.}~\bibnamefont
  {Hatridge}}, \bibinfo {author} {\bibfnamefont {S.}~\bibnamefont {Shankar}},
  \bibinfo {author} {\bibnamefont {Mirrahimi}}, \bibinfo {author}
  {\bibfnamefont {F.}~\bibnamefont {Schackert}}, \bibinfo {author}
  {\bibfnamefont {K.}~\bibnamefont {Geerlings}}, \bibinfo {author}
  {\bibfnamefont {T.}~\bibnamefont {Brecht}}, \bibinfo {author} {\bibfnamefont
  {K.}~\bibnamefont {Sliwa}}, \bibinfo {author} {\bibfnamefont
  {B.}~\bibnamefont {Abdo}}, \bibinfo {author} {\bibfnamefont {L.}~\bibnamefont
  {Frunzio}}, \bibinfo {author} {\bibnamefont {Girvin}}, \bibinfo {author}
  {\bibnamefont {Schoelkopf}},\ and\ \bibinfo {author} {\bibfnamefont
  {M.}~\bibnamefont {Devoret}},\ }\bibfield  {title} {\bibinfo {title}
  {{Quantum Back-Action of an Individual Variable-Strength Measurement}},\
  }\href {http://www.sciencemag.org/content/339/6116/178.short
  papers://b942896d-313a-49f0-995e-1484170e0fc9/Paper/p699} {\bibfield
  {journal} {\bibinfo  {journal} {Science}\ }\textbf {\bibinfo {volume}
  {339}},\ \bibinfo {pages} {178} (\bibinfo {year} {2013})}\BibitemShut
  {NoStop}%
\bibitem [{\citenamefont {DasGupta}\ \emph {et~al.}(2014)\citenamefont
  {DasGupta}, \citenamefont {Lahiri},\ and\ \citenamefont
  {Stoyanov}}]{DasGupta2014}%
  \BibitemOpen
  \bibfield  {author} {\bibinfo {author} {\bibfnamefont {A.}~\bibnamefont
  {DasGupta}}, \bibinfo {author} {\bibfnamefont {S.~N.}\ \bibnamefont
  {Lahiri}},\ and\ \bibinfo {author} {\bibfnamefont {J.}~\bibnamefont
  {Stoyanov}},\ }\bibfield  {title} {\bibinfo {title} {{Sharp fixed n bounds
  and asymptotic expansions for the mean and the median of a Gaussian sample
  maximum, and applications to the Donoho–Jin model}},\ }\href
  {https://doi.org/10.1016/j.stamet.2014.01.002} {\bibfield  {journal}
  {\bibinfo  {journal} {Statistical Methodology}\ }\textbf {\bibinfo {volume}
  {20}},\ \bibinfo {pages} {40} (\bibinfo {year} {2014})}\BibitemShut {NoStop}%
\bibitem [{\citenamefont {Peaudecerf}(2013)}]{PeaudecerfThesis}%
  \BibitemOpen
  \bibfield  {author} {\bibinfo {author} {\bibfnamefont {B.}~\bibnamefont
  {Peaudecerf}},\ }\emph {\bibinfo {title} {{Mesure adaptative non destructive
  du nombre de photons dans une cavit{\'{e}}.}}},\ \href@noop {} {Ph.D.
  thesis},\ \bibinfo  {school} {Paris 6} (\bibinfo {year} {2013})\BibitemShut
  {NoStop}%
\bibitem [{Note2()}]{Note2}%
  \BibitemOpen
  \bibinfo {note} {Here least significant is to be understood as the last bit
  in the binary decomposition, and not in terms of amount of
  information}\BibitemShut {NoStop}%
\bibitem [{\citenamefont {Johansson}\ \emph {et~al.}(2013)\citenamefont
  {Johansson}, \citenamefont {Nation},\ and\ \citenamefont
  {Nori}}]{Johansson2013}%
  \BibitemOpen
  \bibfield  {author} {\bibinfo {author} {\bibfnamefont {J.}~\bibnamefont
  {Johansson}}, \bibinfo {author} {\bibfnamefont {P.}~\bibnamefont {Nation}},\
  and\ \bibinfo {author} {\bibfnamefont {F.}~\bibnamefont {Nori}},\ }\bibfield
  {title} {\bibinfo {title} {{QuTiP 2: A Python framework for the dynamics of
  open quantum systems}},\ }\href {https://doi.org/10.1016/j.cpc.2012.11.019}
  {\bibfield  {journal} {\bibinfo  {journal} {Computer Physics Communications}\
  }\textbf {\bibinfo {volume} {184}},\ \bibinfo {pages} {1234} (\bibinfo {year}
  {2013})},\ \Eprint {https://arxiv.org/abs/1211.6518} {arXiv:1211.6518}
  \BibitemShut {NoStop}%
\bibitem [{\citenamefont {Cottet}(2018)}]{Cottet2018}%
  \BibitemOpen
  \bibfield  {author} {\bibinfo {author} {\bibfnamefont {N.}~\bibnamefont
  {Cottet}},\ }\emph {\bibinfo {title} {{Energy and Information in Fluorescence
  with Superconducting Circuits.}}},\ \href {https://doi.org/tel-02002463}
  {Ph.D. thesis},\ \bibinfo  {school} {PSL Research University; Ecole Normale
  Superieure} (\bibinfo {year} {2018})\BibitemShut {NoStop}%
\bibitem [{\citenamefont {Sarlette}\ \emph {et~al.}(2020)\citenamefont
  {Sarlette}, \citenamefont {Rouchon}, \citenamefont {Essig}, \citenamefont
  {Ficheux},\ and\ \citenamefont {Huard}}]{Sarlette2020a}%
  \BibitemOpen
  \bibfield  {author} {\bibinfo {author} {\bibfnamefont {A.}~\bibnamefont
  {Sarlette}}, \bibinfo {author} {\bibfnamefont {P.}~\bibnamefont {Rouchon}},
  \bibinfo {author} {\bibfnamefont {A.}~\bibnamefont {Essig}}, \bibinfo
  {author} {\bibfnamefont {Q.}~\bibnamefont {Ficheux}},\ and\ \bibinfo {author}
  {\bibfnamefont {B.}~\bibnamefont {Huard}},\ }\bibfield  {title} {\bibinfo
  {title} {{Quantum adiabatic elimination at arbitrary order for photon number
  measurement}},\ }\href {https://doi.org/10.1016/j.ifacol.2020.12.131}
  {\bibfield  {journal} {\bibinfo  {journal} {IFAC-PapersOnLine}\ }\textbf
  {\bibinfo {volume} {53}},\ \bibinfo {pages} {250} (\bibinfo {year}
  {2020})}\BibitemShut {NoStop}%
\end{thebibliography}
\end{document}